\documentclass[aps,pra,showpacs]{revtex4}
\usepackage{graphicx}
\usepackage{dcolumn}
\newcommand{\barz} {\bar{z}}

\newcommand{\barw} {\bar{w}}

\newcommand{\barpart}{\bar{\partial}}

\begin{document}
\title{Bosonizing one-dimensional  cold atomic gases}

\author{M. A. Cazalilla}
\affiliation{Donostia International Physics Center (DIPC),
Paseo Manuel de Lardizabal 4, 20018 Donostia, Spain.}
\begin{abstract}
We present results for the long-distance asymptotics of correlation functions of mesoscopic
one-dimensional systems with periodic and open (Dirichlet) boundary conditions, 
as well as at finite temperature in the thermodynamic limit.
The results are obtained using Haldane's harmonic-fluid approach (also known as ``bosonization''), 
and are valid for both bosons and fermions, in weakly and strongly interacting regimes. The harmonic-fluid
approach   and the method to  compute the correlation 
functions using conformal transformations are explained in great detail. As an application relevant to
one-dimensional systems of cold atomic gases, we consider the model 
of bosons interacting with a zero-range potential. The Luttinger-liquid parameters are obtained 
from the exact solution by solving the Bethe-ansatz equations in finite-size systems. The range
of applicability of the approach is discussed, and the prefactor of the one-body density matrix of bosons
is fixed by finding an appropriate parametrization of the weak-coupling result. The formula thus obtained
is shown to be accurate, when compared with recent diffusion Monte Carlo calculations, within less than $10\%$.  The experimental implications of these results for Bragg scattering experiments at low and high momenta
are also discussed.
 \end{abstract}
\maketitle

\section{Introduction}\label{intro}

 A number of recent experiments~\cite{Gorlitz01,Wada01,Schreck01,Hansel01,Greiner01} 
 have demonstrated  the possibility of confining 
 atoms to one dimension. This possibility  has produced an 
 outburst of theoretical activity over the last years~\cite{Olshanii98,Petrov00,Andersen02,Girardeau01,Dunjko01,Carusotto03,Papenbrock03,
 Gangardt03,Recati03,Buchler03,Cazalilla02,Luxat03,Olshanii02,Mora02,
 Giorgini02,Xianlong03,Cazalilla03}  in the field of cold atoms.  One important 
 focus of these studies has been on the correlation properties of  these new
 one-dimensional systems, with the goal of providing means to characterize 
 them experimentally. 
   
   Although one-dimensional models have been the favorite
 toy of mathematical physicists during the last century, they only became experimentally relevant
 in the late 1960's and 1970's in connection with a number of solid-state materials exhibiting 
 very anisotropic magnetic and electronic properties.  Later, in the 1980's 
 and 1990's, advances in chemical synthesis and nanotechnology made it 
 possible to manufacture materials and devices where electrons are mainly confined to move
 along one or a few conduction channels. The latest experimental 
 developments in the field of cold atoms, however, have several 
 distinct features. First of all, the constituent particles
 are not electrons but bosonic atoms (although in the future fermionic 
 atoms can also become available).  Second, the ``fundamental'' interaction is
 no longer of  Coulombic type  but a short range potential. Furthermore,
 in the case of dilute atomic vapors
 confined in magnetic and/or optical traps, the degree of controllability over
 parameters such density and  interaction strength seems unprecedented. This allows, 
 in principle, to exhaustively  explore phase diagrams, or to cleanly realize quantum-phase
 transitions, which so far are considered as theorists' extreme 
 idealizations of ``dirty'' solid-state phenomena. 
 
 The only {\it a priori}
  limitation offered by the  atomic systems, at least from the point of view of 
  a condensed matter theorist,  seems to be  their mesoscopic, rather than macroscopic, size.
 However, numerical calculations over the past decades (e.g.~\cite{Affleck01,DMRG}) 
 have taught us that many of the
 behaviors predicted for the thermodynamic limit, already manifest
 themselves at the mesoscopic scale, even when the system under study consists of just
 a few tens of particles. Furthermore, the study of mesoscopic systems is also a brach of
 modern condensed-matter physics as phenomena taking place 
 at the mesoscopic scale have an undeniable interest. Thus the main of motivation of this
 paper is to present a set of tools  that can be used to analyze the properties of these
 mesoscopic one-dimensional (1D) systems. Our main theoretical tool in this analysis
is the harmonic-fluid approach, which is nothing but the relevant quantum
hydrodynamics for 1D systems.

  The harmonic-fluid approach  has a 
long history~\cite{Tomonaga50,LiebMattis65,Luther74,EfetovLarkin75}, which in some 
respects culminated with the work of Haldane~\cite{Haldane81b,Haldane81a}.
He realized that many one-dimensional models exhibiting gapless
excitations with a linear spectrum can be described within the same framework.
This  framework defined a {\it universality class} of systems that Haldane termed 
``Luttinger liquids''. The name stems from an analogy with higher dimensional fermionic
systems, where the equivalent role is played by the (universality class of) Fermi liquids.
In the context of one-dimensional Fermi systems this approach has a second, more frequently used,
name: ``bosonization''. This refers to the fact that the method shows how to describe 
the low-energy degrees of freedom of the fermions 
in terms of a bosonic field which obeys a relativistic wave equation.
However, differently from the Fermi liquids, the class of Luttinger liquids 
also includes one-dimensional interacting boson systems. As we shall discuss below, this
has to do with absence of a well-defined concept of statistics in 1D. As a consequence, boson systems
can display fermion-like properties and vice-versa. One well-known example  in the field
of cold atoms is the behavior of the Tonks gas, where the bosons interact so strongly that they 
effectively behave as free fermions.

Besides blurring the line that separates bosons from fermions, confinement 
to one-dimension has another peculiarity that is worth discussing. Being all transverse
degrees of freedom frozen, fluctuations can only propagate longitudinally. This implies
that their effect is enormously enhanced. As a consequence, no long-range order
that breaks a continuous symmetry can exist in the thermodynamic limit, even at zero
temperature. 

 The harmonic-fluid approach has several advantages over other approaches that
are commonly employed to study the low-temperature behavior of 
1D systems. First of all, it is not a mean field theory,
and therefore does not break any symmetry. Furthermore, it can treat bosons and fermions
on equal footing, their difference being  manifested by the different structure of the 
correlation functions that describe them. It can also deal with strongly and weakly interacting
systems at the same time since the low-energy physics parametrized by
three {\it phenomenological} parameters (the particle density and two stiffnesses). These 
parameters are related to measurable properties of the system, which
makes  the approach conceptually simple.
Nonetheless, one has to be aware of its limitations, which essentially are related to its 
``effective field-theory'' character. Therefore, the description is a low-energy one, it comes
with a built-in cut-off, and it is unable to describe the high-energy structure or any model-specific (i.e.
non-universal) features. In some situations, when the interactions are strong, 
relating some of the phenomenological parameters to microscopic ones can be difficult. But in  
many such cases one can rely on an exact (i.e. Bethe-ansatz) solution to extract them. This will be illustrated here
for the case of  bosons interacting via a zero-range potential~\cite{Lieb63a,Lieb63b}. We 
will thus be able to show that many of the results obtained in the weakly interacting limit from the 
Bogoliubov-Popov~\cite{Popov,Popov80} approach and its modifications~\cite{Andersen02,Mora02}
can be recovered and understood within the harmonic-fluid approach. We shall also present
result for the correlation functions
of a number of ``toy models'' of 1D mesoscopic systems, such 
like a ring (i.e. a small system with periodic boundary
conditions) and a box (a system with open boundary conditions). Besides being analytically
tractable, these models are  important when comparing with numerical simulations using
Monte Carlo methods~\cite{Giorgini02,Carusotto03,Troyer02}, where periodic boundary conditions are used, 
or the density-matrix renormalization-group (DMRG)~\cite{DMRG,Kollath03}, for which open boundary conditions 
are best suited. Furthermore, in recent times, traps other than harmonic have become 
available~\cite{Hansel01,Sauer01}, which also makes relevant the study of these geometries.
 
 There already exist some studies of  the correlation functions in finite-size systems in the literature,
but they have mostly focused on  fermionic or spin  
correlations~\cite{Fabrizio95,Eggert96,Wang96,Affleck01}. 
The present approach allows to obtain the corrrelation functions  for both bosons and fermions at
the same time, and automatically includes all higher-harmonics, which have been omitted in 
previous treatments. Besides presenting in full detail the generalization of 
the harmonic-fluid approach to the box
case, in this paper we also address how to fix the prefactor of the one-body density
matrix for  bosons interacting with a zero-range potential (henceforth referred to
as ``delta-inteacting bosons''). We find that, when properly parametrized, the result obtained by
Popov in the weakly interacting limit~\cite{Popov80} is accurate even in the strong coupling
limit. This is demonstrated by comparing with exact results~\cite{Lenard72} and recent diffusion
Monte Carlo data~\cite{Giorgini02}.

 The organization of this paper is as follows: In the following section,
we shall review the harmonic-fluid approach following Ref.~\onlinecite{Haldane81b}. In
this section we also present its generalization to open boundary conditions.
In Sect.~\ref{correlators}, results for different correlation functions in the box and the ring,
as well as finite-temperature expressions in the thermodynamic limit, 
are presented.  We consider both bosons and spinless (i.e. spin-polarized) fermions.
Sect.~\ref{delta} specializes the discussion to a system of delta-interacting bosons.
This model is relevant for bosonic atoms confined to a one-dimensional channel.
We show how the parameters needed for the low-energy description can be extracted
from the exact (Bethe-ansatz) solution, and discuss how to fix the prefactor of the one-body density matrix. 
Some experimental consequences of the correlation functions obtained within
the harmonic-fluid approach are presented in this section. In particular, the 
line-shape of the momentum distribution at finite temperature is analyzed as the 
parameters of the system are varied. 
In Sect.~\ref{WFS}, we  discuss the asymptotic structure of the wave-function of
a bosonic Luttinger liquid, and obtain an expression for a system with open
boundary conditions   Finally, the appendices
contain a proof of a commutation relation used in the main text,
an alternative ``derivation'' of the low-energy Hamiltonian using
the path integral formalism, as well as a
detailed description of how to compute correlation
functions using conformal field theory methods. Some of
the results of this work have been briefly reported 
in Ref.~\onlinecite{Cazalilla02}.

\section{The harmonic-fluid approach}\label{sect1}
In this section we are going to review the 
harmonic-fluid approach in operator language, following the original
work of  Haldane~\cite{Haldane81b}. Some of the results of this
section can be also obtained  using a coherent-state 
path integral formulation (see Appendix~\ref{appb}).

\subsection{Haldane's construction}\label{ss1.1}
The discussion in this subsection will be independent of  boundary conditions, but
we shall assume the system to have a finite size $L$.
For the most part, the notation is similar to that of  Haldane
in Ref.~\onlinecite{Haldane81b}. However, we deviate
from it in a number of places, sometimes to agree
with more recent conventions. We shall work in second quantization 
most of the time. This means
that  a system  of bosons is described using field operators which obey 
$[\Psi(x), \Psi^{\dagger}(x')]  = \delta(x-x')$, and commute
otherwise;  $\rho(x) = \Psi^{\dagger}(x) \Psi(x)$ is the density operator. 
The mean ground state density, $\rho_{0}$, is fixed  either by 
the chemical potential, $\mu$, such that 
$\rho_{0} = \rho_{0}(\mu)$ (grand canonical ensemble), 
or by the total  particle number in the ground state, $N_{0}$, such that
$\rho_{0} = N_{0}/L$ (canonical ensemble).  Note that the discussion that follows applies
to uniform systems, the necessary modifications needed to deal with a non-uniform (but slowly varying)
ground state density  are discussed in Sect.~\ref{trap}.

 As pointed out  in the introduction, the effect of long wave-length
thermal and quantum fluctuations is enhanced by reduced dimensionality. 
In order to derive a low-temperature  description, we need  
to identify a set of variables that describe  the low-energy fluctuations of the system. 
For a bosonic system these variables are the  density 
and the phase.  At  low temperatures,  density 
and phase fluctuations are locally small in the sense 
to be defined below. To give a proper
definition of these variables, we need to introduce the phase-density 
representation of the bosonic field operator:
\begin{equation}\label{eq2.2}
\Psi^{\dagger}(x) = \sqrt{\rho(x)} \: e^{-i\phi(x)}.
\end{equation}
Consistently with  the bosonic commutation relations $\left[\Psi(x), \Psi^{\dag}(x)\right] = \delta(x-x')$,
phase and density operators obey
\begin{equation}\label{eq2.3}
e^{i \phi(x')} \rho(x) e^{-i\phi(x')} - \rho(x) = \delta(x-x').
\end{equation}
 A convenient way of describing the long wave-length density and phase fluctuations 
is to split $\rho(x) = \rho_{<}(x) + \rho_{>}(x)$ and $\phi(x) = \phi_{<}(x) + \phi_{>}(x)$, where
the $\rho_{<}(x)$ and $\phi_{<}(x)$ refer to the ``slow'' parts (i.e. coarse-grained over distances $\gg \rho^{-1}_0$)
of the operators, whereas $\rho_>(x)$ and $\phi_>(x)$ refer to the ``fast'' or short-wave length parts 
(see Appendix~\ref{appb} for a more careful definition). In the following
we  focus on the slow parts, and below, when there is not risk of confusion, we will denote as $\phi(x)$ the slow
part of the phase operator (i.e. $\phi_<(x)$).  Furthermore, since the density  fluctuates at low temperatures about the 
ground state value, $\rho_0$, it is convenient to introduce the operator $\Pi(x)$, defined by $\rho_<(x) = 
\rho_0 + \Pi(x)$.

\begin{figure}[t]
\centerline {
\includegraphics[width=5cm]{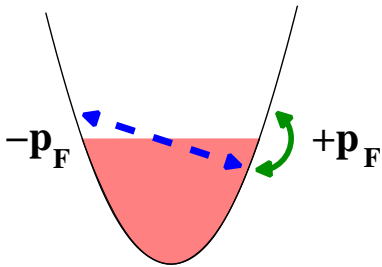}}
\caption{Low-energy excitations in a Tonks ($=$ free Fermi) gas. The continuous arrow represents
excitations about $+p_F$ (symmetrically, about $-p_F$) with momenta $q\approx 0$.
The dashed arrow represents low-energy excitations carrying momenta $q \approx \pm 2p_F = \pm 2 \pi \rho_0$.}
\label{fig1}
\end{figure}

  We  shall first consider the phase  fluctuations.  In Appendix~\ref{appa} we  
show that  the slow parts $\phi(x)$ and 
$\Pi(x)$  are  canonically conjugated fields, i.e.
\begin{equation}\label{eq2.4}
[\Pi(x), \phi(x')] = i \: \delta(x-x').
\end{equation}
It is  worth pointing out  that this commutation relation  holds,
for the slow parts of the density and phase, in arbitrary dimensions.
However, since in 1D fluctuations are constrained to 
propagate on a line, this causes the  aforementioned enhancement 
of their effect. This makes it impossible, in the thermodynamic limit,
to define of an order parameter. Thus we expect that  $\langle e^{i\phi(x)} \rangle = 0$ 
in 1D, in contrast with the situation for $d > 1$, where this operator can acquire a non-zero
expectation value (i.e. there can be off-diagonal long-range order). Indeed, 
in one dimension $\langle e^{i\phi(x)} \rangle = 0$ also in finite-size systems (see
appendices~\ref{appc} and ~\ref{appd}).

  We next consider long wave-length fluctuations of the density, which have been parametrized
as  $\rho_{0} + \Pi(x)$. $\Pi(x)$ describes {\it locally}  small fluctuations of wave-length $ \gg \rho^{-1}_{0}$. 
However, this does not suffice to describe all  possible {\it low-energy} density fluctuations.  It is thus necessary
to distinguish between ``low-energy'' and long ``wave-length''; There can be low-energy density fluctuations
with wave-length $\sim \rho^{-1}_0$ or shorter. To see this in a non-trivial case, it suffices to consider the case of the impenetrable-boson (Tonks) gas. It is known~\cite{Girardeau60,Lieb63b} that the excitations of such a system are those of a  free fermi gas of the same density. 
In a Fermi gas  the momentum of the fastest particle (the Fermi momentum) is related to the density by the formula $p_F = \pi \rho_0$. There are two Fermi points (see Fig.~\ref{fig1}),  corresponding to $p = \pm p_F$. The low-energy and long wave-length density fluctuations of this system are the small momentum particle-hole excitations around each one of the Fermi points. However, a 
fermion can also be excited with very low energy from one Fermi point to the other, thus producing a density fluctuation that oscillates as $\cos 2p_F x = \cos 2 \pi \rho_0 x$. The  long wave-length density fluctuations described by  $\Pi(x)$ lead to small changes in the local Fermi momentum: $p_{F}(x) = p_F + \pi \Pi(x)$. This affects  shorter wave-length density fluctuations, which now oscillate as $\cos 2 \int^{x} dx' p_F(x') = \cos 2 \Theta(x)$, where we have introduced the auxiliary field $\Theta(x)$, related to $\Pi(x)$ by means of the 
following expression~\cite{EfetovLarkin75,Haldane81b}:
\begin{equation}\label{eq2.5}
\frac{1}{\pi} \partial_x \Theta(x) = \rho_{0} + \Pi(x).
\end{equation}
Note that the integral on the left hand-side of (\ref{eq2.5})
from $x=0$ to $x = L$ must be equal to total 
particle number operator $N$, namely 
\begin{equation}\label{eq2.6}
\Theta(L) - \Theta(0) = \pi N.
\end{equation}
\begin{figure}[b]
\centerline {
\includegraphics[width=9cm]{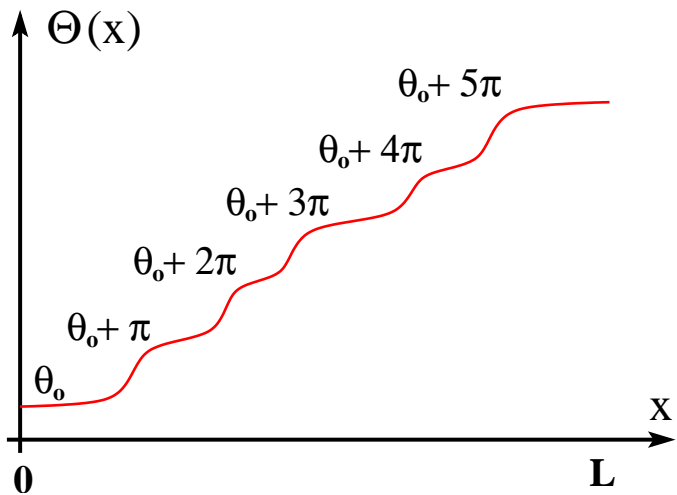}}
\caption{Typical configuration of $\Theta(x)$. Its value 
changes by $\pi$ every time $x$ surpasses the 
location of a particle. Note that $\Theta(x)$ is a slowly varying field.}
\label{fig2}
\end{figure}
This is an important {\it topological} property of $\Theta(x)$. It tells us
that  on a {\it global} scale the changes in $\Theta(x)$ are related to  changes in the total 
particle number. Hence, the configurations of $\Theta(x)$ can be regarded as 
functions that increase monotonically from $x = 0$ to $x = L$  (see Fig.~\ref{fig2}).  
Thus it is tempting to  associate the location of  particles  with the points 
where $\Theta(x)$ equals an integer multiple of $\pi$. 
Physically this amounts to trading 
discrete particles  by  {\it solitons} or {\it kinks} 
in $\Theta(x)$ (see Fig.~\ref{fig2}).  The construction of a low-energy projection of the
full density operator, $\rho(x)$, which reflects the discrete 
nature of the particles and therefore  can
describe the shorter wave-length low-energy fluctuations 
follows from this idea~\cite{Haldane81b}. First one sets:
\begin{equation}\label{eq2.7}
\rho(x) = \partial_x \Theta(x) \sum_{n=-\infty}^{+\infty} \delta(\Theta(x) - n\pi),
\end{equation}
which in the particle-soliton sense is equivalent 
to the first quantized form of the density operator, $\rho(x) = 
\sum_{i=1}^{N} \delta(x -x_{i})$ provided that $x_i$ is interpreted as the position operator of
the particle-soliton and one uses  $\delta[f(x)]=\delta(x-x_0)/|f^{\prime}(x_0)|$, where $f(x_0) = 0$.  
The previous expression can be rewritten in a more useful
way with the help of Poisson's summation formula:
\begin{equation}\label{eq2.8}
\sum_{n=-\infty}^{+\infty} f(n) = \sum_{m=-\infty}^{+\infty}\, \int_{-\infty}^{+\infty} dz \,  f(z)\:  e^{2m\pi i z},
\end{equation}
which yields:
\begin{equation}\label{eq2.9}
\rho(x) = \frac{1}{\pi} \partial_x \Theta(x)\: \sum_{m=-\infty}^{+\infty} e^{2m i \Theta(x)} = \left[\rho_{0} + \Pi(x) \right]
\: \sum_{m=-\infty}^{+\infty} e^{2 m i \Theta(x)}.
\end{equation}
This is the sought representation of the density operator. The $m = 0$ term  is precisely $\rho_0 + \Pi(x)$,
and describes the  long wave-length fluctuations (i.e. those with momenta $|q| \ll \rho_0$), 
the  $m= \pm 1$ terms describe fluctuations with $q \approx \pm 2\pi \rho_0$,  while 
$m = \pm 2$  those with $q \approx \pm 4 \pi \rho_0$, etc. (see Fig.~\ref{fig5}).

 We next take up the construction 
of the boson field operator $\Psi^{\dagger}(x)$ in 
terms of $\Theta(x)$ and  $\phi(x)$. According to Eq.~(\ref{eq2.2}), 
this requires finding a representation  for  the square root 
of the density operator, Eq.(\ref{eq2.7}). At this point it is useful to recall Fermi's trick:
$[\delta(\theta)]^2 = A \: \delta(\theta)$,  where the constant $A$  
depends on the particular way   the Dirac delta function is defined.  
Extracting the square root 
yields $\sqrt{\delta(\theta)} = A^{-1/2} \: \delta(\theta)$.  Thus,
using Poisson's formula (\ref{eq2.8}) and 
multiplying by $e^{-i \phi(x)}$ 
from the  right, one arrives at\cite{Haldane81b}:
\begin{equation}\label{eq2.10}
\Psi^{\dagger}(x) \sim 
\left[\rho_{0} + \Pi(x) \right]^{\frac{1}{2}} \sum_{m=-\infty}^{+\infty} e^{2mi\Theta(x)} e^{-i\phi(x)}.
\end{equation}
The symbol $\sim$ means that the field operator is 
given by the expression on the right up to a 
prefactor. This prefactor is not determined independently 
of the way we choose to exclude the high-energy fluctuations (i.e. the fast modes)
from the low-temperature description.  
This is usually done  using a  cut-off in real or momentum space, and
different cut-off schemes lead to different prefactors. 
\begin{figure}[hb]
\centerline {
\includegraphics[width=9cm]{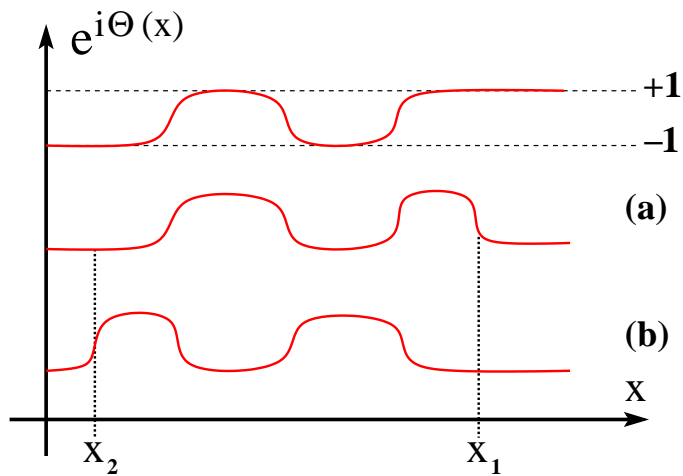}}
\caption{Effect on the string operator $e^{i\Theta(x)}$ of 
the creation of a particle at two
different points $x_1$ (a) and $x_2$ (b). When a particle is  created (or removed)
at $x_1$,  the string operator at $x_2 < x_1$ is not affected (a). However, when
the particle is created (or removed) at $x_2$ the value of the string at $x_1$ picks
up an extra factor of $-1$. Since the operators  $\Psi^{\dagger}_F(x)$  
contain only odd powers of $e^{i\Theta(x)}$, they are sensitive to these changes, 
which implies that they anti-commute.}
\label{fig3}
\end{figure}

  The construction of a fermionic (i.e. anti-commuting) field operator 
is also possible. One only needs to realize that, since
$\Theta(x)$ jumps by $\pi$ every time a particle 
is surpassed (Fig.~\ref{fig3}), the {\it string operator} $e^{i \Theta(x)}$
alternates  between $+1$ and $-1$ (see Fig.~\ref{fig3}). 
This implies that 
\begin{equation}\label{fermionfield}
\Psi^{\dagger}_{F}(x) =    \Psi^{\dagger}(x) \: e^{i \Theta(x)}  
\sim \left[\rho_{0} + \Pi(x)  \right]^{\frac{1}{2}}
 \sum_{m=-\infty}^{+\infty}
e^{(2m+1)i\Theta(x)}\: e^{-i\phi(x)}
\end{equation}
anti-commutes at different positions. 
To see this, consider the product 
$\Psi^{\dagger}_{F}(x_2) \Psi^{\dagger}_{F}(x_1)$
assuming that  $x_1 > x_2$, for instance.   Since
the operator $\Psi^{\dagger}_{F}(x_1)$ acts
first by creating a  particle  at $x_1$,
it does not  have any effect on the string
operator at $x_2 < x_1$ (see Fig.~\ref{fig3}, case (a)). 
On the other hand, if we instead consider
$\Psi^{\dagger}_{F}(x_1) \Psi^{\dagger}_{F}(x_2)$
 ($x_1 > x_2$), the 
operator $\Psi^{\dagger}_{F}(x_2)$ acts first, thus creating  
at $x_2 < x_1$.  Therefore, the
string $e^{i\Theta(x_1)}$ of the second field
operator  picks up an extra minus sign (see Fig.~\ref{fig3},
case (b)) relative to the  case where a particle  is created at $x_1$ first. 
Thus  the operator defined by~(\ref{fermionfield}) anti-commutes
at different locations,  and therefore 
describes fermions instead of bosons. 
The same conclusion can be reached
for a product of two annihilation (or one creation and one annihilation) operators
at different points~\footnote{Recovering the full  anti-commutation
relations requires a more careful construction of the field operator than the one presented here. 
The reader interested  in this {\it constructive} approach to bosonization 
should consult Ref.~\onlinecite{Haldane81a}.}.

\begin{figure}[b]
\centerline{
\includegraphics[width=8.5cm]{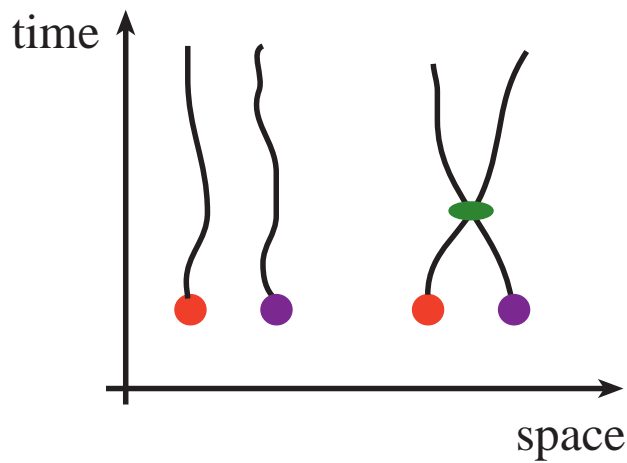}}
\caption{When two particles are exchanged in one-dimension 
to test their statistics, they cannot avoid collision. Thus the statistical
phase and the scattering phase shift cannot be separated.}
\label{fig4}
\end{figure}

\begin{figure}[t]
\centerline{
\includegraphics[width=14cm,height=5cm]{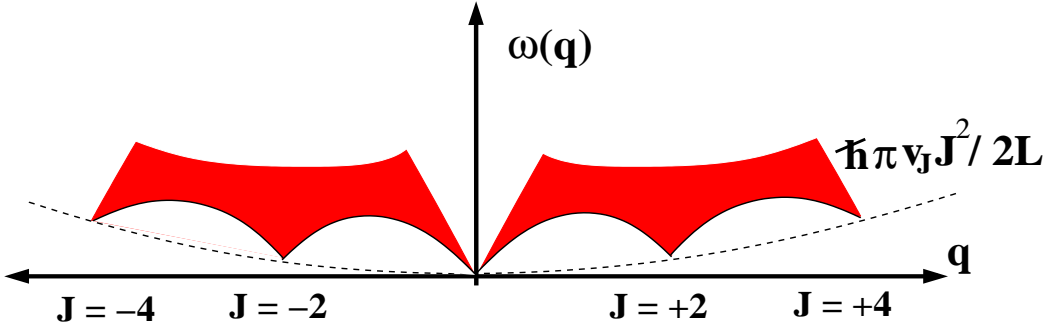}}
\caption{Typical low-energy  excitation spectrum  of a Luttinger liquid with periodic boundary
conditions.  The shaded area corresponds to the continuum 
of multi-particle-hole excitations. The Hamiltonian $H_{\rm eff}$, 
Eq.~(\ref{modhampbc}) describes the excitation spectrum for $q \approx \pi J \rho_0$,
with $J = 0, \pm 2, \pm 4, \ldots$.  Near $q \approx 0$ the excitations 
are linearly dispersing phonons: $\omega(q) \simeq \hbar v_s|q|$ 
(i.e. superpositions of particle-hole excitations about the same Fermi point,
see Fig.~\ref{fig1}). For $q \sim \pi J \rho_0$, where $J = \pm 2, \pm 4, \ldots$, 
the excitations involve the creation of $J$ current quanta in the ring.  For instance, 
$J=\pm 2$ correspond to  the low-energy transitions between Fermi points
shown in Fig.~\ref{fig2} (dashed arrow). For delta-interacting bosons, this picture 
holds in all regimes, and for the particle-hole spectrum agrees with the spectrum obtained 
from Bethe ansatz in Ref.~\cite{CastroNeto94}.}
\label{fig5}
\end{figure}

 The explicit construction of the low-energy representations
of commuting (i.e. bosonic) and anti-commuting (i.e. fermionic) fields
shows how this approach treats bosons and fermions on
equal footing. On physical grounds, the fact that  in 1D transforming 
bosons into fermions is possible  is not  a surprise. 
The reason~\cite{Haldane81b} is  that when one tries to exchange 
two interacting particles (or elementary excitations) in 1D  they 
must necessarily collide, and 
therefore the statistical phase cannot be separated from the 
phase shift associated with the collision (see Fig.~\ref{fig4}).

\subsection{Low-energy effective Hamiltonian and momentum operators}\label{ss1.2}
So far we have considered the 
kinematics of the low-energy 
description of a 1D quantum fluid. In other words,
we have introduced the variables that describe low-temperature 
states of the system and found their relationship
to  density and  field operators. The next step is to
consider the dynamics, that is, the Hamiltonian.  
The starting point for our considerations is the  
following Hamiltonian  for bosons interacting via a 
general two-body potential,
$v(x)$~\footnote{Three-body and higher-body interactions do not modify the form of the low-energy
effective Hamiltonian $H_{\rm eff}$, Eq.~(\ref{eq2.14}), only the precise dependence of the phase
and density stiffness on the microscopic parameters.}:
\begin{eqnarray}\label{eq2.1}
H  =  \frac{\hbar^2}{2M}  \int_{0}^{L} dx \: 
 \partial_x \Psi^{\dagger}(x) \partial_x \Psi(x)  
+ \frac{1}{2}  \int_{0}^{L} dx\; dx' \,  v(x-x') \rho(x) \rho(x').
\end{eqnarray}
Before proceeding  any further, a number of comments 
are in order: The above Hamiltonian is 
assumed to describe  the situation where all particles lie in the lowest level of a
transverse confining potential. This happens when the chemical potential, $\mu$, is smalller
than the transverse confinement energy, $\hbar \omega_{\perp}$.
Regarding the longitudinal confinement, 
we will assume in the following that it is either absent (case of
periodic boundary conditions) or that it can be approximated 
by two infinite barriers placed at $x=0$ and $x = L$ (case of open
boundary conditions). The effect
of a {\it  smooth}  potential in the longitudinal direction will 
be discussed in Sect.~\ref{trap}. By restricting ourselves
to the lowest transverse  level, we assume the system to
be effectively one-dimensional.  If $u_{0}^{\perp}(y,z)$ denotes
the lowest  transverse orbital,  the three dimensional
boson field operator can be written as 
$\Psi^{\dag}_{3d}({\bf r}) = \Psi^{\dag}(x) \:   u_{0}^{\perp}(y,z)+
 \tilde{\Psi}^{\dag}({\bf r})$, where $\tilde{\Psi}^{\dag}({\bf r})$ describes
bosons in the higher energy transverse levels. 
At low temperatures (i.e. $T <  \mu < \hbar \omega_{\perp}$), 
only  virtual transitions to higher transverse levels are permitted, 
which lead to a renormalization of the interaction 
potential $v(x)$ (see e.g. Ref.~\onlinecite{Olshanii98})

 To obtain the low-energy effective Hamiltonian we use the operator
identities derived in the previous section keeping only the leading terms,
which are quadratic in the gradients of the slowly varying fields $\Theta(x)$ 
and $\phi(x)$ (alternatively, one can linearize the equations of motion for the density and
the phase-gradient in terms of the gradients $\phi$ and $\Theta$).  The result 
can be generally written as
\begin{equation}\label{eq2.14}
H_{\rm eff} = \frac{\hbar}{2\pi} \int_{0}^{L} dx \: 
\left[ v_J \left(\partial_x \phi(x) \right)^2 +  v_N \: 
\left( \partial_x\Theta(x) - \pi \rho_{0} \right)^2  \right].
\end{equation}
The history of this Hamiltonian goes back to the pioneering work of Tomonaga~\cite{Tomonaga50}
on one-dimensional electron gases. He wrote it in a very different way, but as we shall see in the following
sections, both forms are essentially  equivalent and describe the same collection of harmonic oscillators,
whose quanta, the ``phonons'' correspond to  low-energy density and phase fluctuations. The first
application to a bosonic system was done by Efetov and Larkin~\cite{EfetovLarkin75}, who considered a 
system of tightly bound electron pairs (the ``BEC'' limit of a 1D superconductor) and wrote $H_{\rm eff}$ in a
form similar to Eq.~(\ref{eq2.14}). The form used here is 
due to Haldane~\cite{Haldane81a,Haldane81b} who, building
upon and extending the work of Tomonaga~\cite{Tomonaga50}, Lieb and Mattis~\cite{LiebMattis65}, Luther~\cite{Luther74},  Efetov and Larkin~\cite{EfetovLarkin75},...  introduced 
the concept of (Tomonaga-) Luttinger
liquid as a {\it universality  class} of 1D systems with gapless, linearly-dispersing,  excitations.

For an interaction whose range $R \gg \rho^{-1}_0$, one should 
replace the second term in $H_{\rm eff}$ by
\begin{equation}
\frac{1}{2\pi^2}  \int_{0}^{L} dx\; dx' \,  v(x-x') (\partial_{x}\Theta(x)-\pi \rho_0) (\partial_{x'}\Theta(x') - \pi \rho_0). 
\end{equation}
However, provided  that the Fourier transform of the interaction potential $v(q) = \int^L_0 dx\, e^{-iqx} v(x)$ is 
not singular as $q \to 0$,
Eq.(\ref{eq2.14}) will hold at sufficiently low energies, and below we shall assume that
this is indeed the case (a notable exception is the Coulomb interaction).
The term proportional to $(\partial_x\Theta-\pi\rho_0)^2$ in Eq.~(\ref{eq2.14}) thus 
corresponds to the interaction energy of the long wave-length density fluctuations. The
density stiffness $v_N$  has  dimensions of  velocity. In the following subsection, it is shown
that it is inversely proportional to the  compressibility of the fluid.  The phase stiffness $v_J$ has also velocity
units and it is found to be proportional to the superfluid fraction. Therefore, 
both parameters  must be regarded as phenomenological.  
They can be extracted from an exact solution of the microscopic model
(when available),   from numerical calculations or, ultimately, from 
experimental data, which still allows one to correlate the results from different experiments.  

 For many models on the continuum   (i.e. those which do not require  a lattice to be defined) the term 
proportional to $(\partial_x \phi)^2$ in Eq.~(\ref{eq2.14}) follows  
from  the kinetic energy operator after using Eq.~(\ref{eq2.10}), i.e.
\begin{equation}
\frac{\hbar^2}{2M}\: \int_{0}^{L} dx \: \partial_x \Psi^{\dagger}(x) 
\partial_x \Psi(x) \approx \frac{\hbar^2 \rho_{0}}{2 M} 
\int_{0}^{L} dx \: \left( \partial_x \phi(x) \right)^2.
\end{equation}
Hence $v_J  = \hbar \pi \rho_{0}/m \equiv v_F$, where $v_F$, 
is the Fermi velocity of a gas of  free spinless  fermions of 
density $\rho_{0}$.  As we show in the following subsection, this 
relationship between $v_J$ and $v_F$ is ensured by Galilean invariance,
and will not hold if this symmetry  is broken. This is generally the case of lattice models, such like 
a 1D system of bosons hopping in a sufficiently deep optical lattice 

  Another interesting operator is the total momentum, which can also be  expressed
in terms of $\Theta(x)$ and $\phi(x)$. The derivation is similar to that of the Hamiltonian, 
starting from the second quantized form,
\begin{equation}
P =  \frac{\hbar }{2i}
\int^{L}_{0} \left[ \Psi^{\dagger}(x) \partial_x \Psi(x)
- \partial_x \Psi^{\dagger}(x) \Psi(x) \right],
\end{equation}
we keep the leading terms in the gradients of $\Theta(x)$ and $\phi(x)$, and obtain
\begin{equation}\label{peff}
P =  \frac{\hbar}{\pi} \int^{L}_{0} \partial_x \Theta(x) \partial_x \phi(x). 
\end{equation}

 We close this subsection with a remark about notation. It is customary 
in the literature on bosonization to work with the field
\begin{equation}
\theta(x) = \Theta(x) - \pi \rho_{0} x,
\end{equation}
instead of $\Theta(x)$. Hence, $\Pi(x) = \partial_x \theta(x)/\pi$. It is also common 
to introduce the parameters $K = \sqrt{v_J/v_N}$ 
and $v_s = \sqrt{v_N v_J}$, such that
\begin{eqnarray}\label{ham}
H_{\rm eff} =  \frac{\hbar v_s}{2} \int_{0}^{L} dx \: 
\left[ \frac{K}{\pi} \left(\partial_x \phi(x) \right)^2 + \frac{\pi}{K}\: \left( \Pi(x) \right)^2  \right] 
= \frac{\hbar v_s}{2\pi} \int_{0}^{L} dx \: \left[ K \left(\partial_x 
\phi(x) \right)^2 + \frac{1}{K} \: \left( \partial_x \theta(x) \right)^2  \right]
\end{eqnarray}
As we shall see shortly, $v_s$ is the phase velocity of the 
low-energy excitations (sound waves). However, the dimensionless parameter 
$K$ is related to the strength of quantum fluctuations (see Appendix~\ref{appb}).

\subsection{Particles in a ring: periodic boundary conditions.}\label{modespbc}
 Our first task in this subsection will be to
find appropriate mode expansions for the fields $\Theta(x)$ and $\phi(x)$,  such that
the Hamiltonian, Eq.~(\ref{eq2.14}), is diagonalized. 
Before doing it, we need to find out how to implement the boundary conditions
in terms of these fields.  Therefore, consider a system obeying periodic boundary 
conditions.  As commonly 
introduced in the literature, this  seems a mere mathematical  
convenience that simplifies the calculations
before taking the thermodynamic limit. However, nowadays there exist 
experimental realizations of these  BC's. In the case of cold atoms one can think of a
quantum degenerate atomic gas in a tight toroidal trap~\cite{Sauer01}.  
Thus, if the boson field obeys   $\Psi^{\dagger}(x+L) = \Psi^{\dagger}(x)$, 
then equations~(\ref{eq2.6}) and~(\ref{eq2.9},\ref{eq2.10}) imply that:
\begin{eqnarray}
\Theta(x + L) &=& \Theta(x) + \pi N, \label{thetamod}\\
\phi(x+L) &=& \phi(x) + \pi J \label{phimod},
\end{eqnarray}
where $N$ is the particle-number operator, and $J$ is an operator 
whose eigenvalues are even integers so that $(-1)^J = +1$.
These are very important {\it topological} properties of the 
phase and density fields.  In particular, they imply that the system
has states that can be labeled by the eigenvalues of $J$ and $N$.
We show below that $N$ and $J$ are conserved quantities (i.e. they commute with $H_{\rm eff}$)
and can be used to label topologically excited states of the system. In 
the case of $N$ this is not surprising because the microscopic Hamiltonian, Eq.~(\ref{eq2.1}),
conserves the total  particle number. However, we will see that
$J$ is associated with the possibility of quantized persistent  currents.

  We next write down  the mode expansions for $\Theta(x)$ and $\phi(x)$, 
which obey~(\ref{thetamod}) and (\ref{phimod}), as well as the commutation relation: 
\begin{equation}\label{CRthetaphi}
\left[\partial_x  \Theta(x),\phi(x')  \right] = i\pi \delta(x-x').
\end{equation}
The appropriate expressions read:
\begin{eqnarray}
\Theta(x) &=& \theta_{0}  + \frac{\pi x}{L} N + \frac{1}{2} 
\sum_{q \neq 0} \Big|\frac{2\pi K}{qL} \Big|^{\frac{1}{2}} e^{-a 
|q|/2} 
\big[ e^{i q x} b (q) + e^{-i q x} b^{\dagger}(q)\big], \label{mpbc1}\\
\phi(x) &=& \phi_{0}  + \frac{\pi x}{L} J + \frac{1}{2} \sum_{q 
\neq 0} \Big|\frac{2\pi}{qL K} \Big|^{\frac{1}{2}} e^{-a |q|/2}  
{\rm sgn}(q)
\big[e^{i q x} b (q) + e^{-i q x} b^{\dagger}(q)  \big].\label{mpbc2}
\end{eqnarray}
The operators $b(q)$ and $b^{\dagger}(q)$ have the commutation relation:
\begin{equation}
[b(q), b^{\dagger}(q')] = \delta_{q,q'},
\end{equation}
commuting otherwise. Momenta are quantized as $q = 2 \pi m/L$, 
where $m=0,\pm 1, \pm 2, \pm 3, \ldots$ The pairs 
$(N, \phi_{0})$ and  $(J,\theta_{0})$ are 
conjugate action-angle variables which obey:
\begin{eqnarray}
\left[N, e^{-i\phi_{0}}\right] &=&  e^{-i \phi_{0}}, \\
\left[J, e^{-i \theta_{0} } \right] &=&  e^{-i \theta_{0}}\\
\left[J,N \right] &=& \left[\phi_{0}, \theta_{0} \right] = 0.
\end{eqnarray}

Introducing the expressions (\ref{mpbc1}) and (\ref{mpbc2}) into Eq.~(\ref{eq2.14}), one obtains:
\begin{equation}\label{modhampbc}
H_{\rm eff} =  \sum_{q \neq 0} \hbar \omega(q)\:  b^{\dagger}(q) b(q) 
+  \frac{ \hbar\pi v_s }{2 L K} (N-N_{0})^2 +  \frac{\hbar\pi v_s  
K}{2 L} J^2 + {\rm const.},
\end{equation}
where $\omega(q) = v_s |q|$ for $q \ll \rho^{-1}_0$.  It now becomes clear that this Hamiltonian 
describes collective phonon-like excitations (sometimes called Tomonaga bosons~\cite{Tomonaga50}),
which disperse linearly  in the long wave-length limit $|q|\to 0$. This result is perhaps not striking
if one deals with bosons because it is already obtained from Bogoliubov's theory (although in 1D this
theory is inconsistent in several respects, which can be  cured only in the weakly interacting limit~\cite{Popov,Andersen02,Mora02}). However, for interacting fermions it has striking consequences,
because it  means the absence of  individual (i.e. particle-like) excitations in the low-energy spectrum, 
and the break-down of Fermi-liquid theory in 1D.

The expansions (\ref{mpbc1})  and (\ref{mpbc2})   diagonalize 
the momentum operator as well:
\begin{equation}\label{modmompbc}
P = \frac{\hbar \pi}{L} N J + \sum_{q \neq 0} \hbar q \, 
b^{\dagger}(q) b(q).
\end{equation}
As $\Theta(x)$, $\phi(x)$,  $H_{\rm eff}$ and $P$  
describe long wave-length fluctuations, the momentum sums must be cut-off at a momentum  
$a^{-1}  \lesssim  \min\left\{R^{-1}, q_c \right\}$.  For short range interactions, 
$a^{-1} \lesssim q_c$;  $q_c$ is then fixed by demanding that $\hbar v_s q_c = \mu$, i.e. it is an estimate 
of the momentum where the excitation spectrum deviates from the linear behavior.

From the previous expression for the Hamiltonian the (inverse)  adiabatic   compressibility at zero temperature
can  be obtained:
\begin{equation}\label{compress}
\kappa^{-1}_S = \rho_0^{2} \: \left( \frac{\partial\mu}{\partial\rho}\right)_{\rho=\rho_0} =  
\rho^2_0 \: L \left[ \frac{\partial^2 E_0(N)}{\partial N^2} \right]_{N=N_0} 
= \hbar \pi v_N \rho^2_{0} = \frac{\hbar \pi v_s \rho^2_{0}}{K},
\end{equation}
where $E_0(N) = \langle H \rangle_{N}$ is expectation value of the Hamiltonian, Eq.~(\ref{modhampbc}),
taken over the ground state with  $ N$ particles; hence  $v_N \propto \kappa_S^{-1}$. 
Using the previous expressions, we can also write:
\begin{equation}\label{vn}
v_N = \frac{1}{\pi \hbar} \, \left( \frac{\partial \mu}{\partial \rho}\right)_{\rho = \rho_0} = \frac{L}{\pi\hbar} 
\left[ \frac{\partial^2 E_0(N)}{\partial N^2} \right]_{N=N_0},
\end{equation}
which will be  useful in extracting the density stiffness from the Bethe-ansatz solution of 
the delta-interacting bosons in Sect.~\ref{luttinger}.

  We next find the relationship between the phase stiffness, $v_J$,
and the superfluid fraction, which we denote as $\rho_s$ below. 
To this purpose, we  consider  {\it twisted} boundary conditions instead
of PBC's:
\begin{equation}\label{twistbc}
\Psi^{\dag}(x+L) = e^{-i\alpha} \Psi^{\dag}(x).
\end{equation}
Therefore, the  phase field obeys the modified boundary conditions: $\phi(x+L) = \phi(x) + 
\pi J + \alpha$, such that $(-1)^J = +1$. This  means that in Eq.~(\ref{mpbc2}) we
have to shift $J \to J + \alpha/\pi$. As a result, the ground state energy and momentum are also
shifted (recall that in the ground state $\langle N\rangle = N_{0}$ and 
$\langle J\rangle  = 0$):
\begin{eqnarray}
\label{changenrg1}
\langle H_{\rm eff} \rangle_{\alpha\neq 0}  - \langle H_{\rm eff} \rangle_{\alpha = 0} &=& \frac{1}{2} \left(\frac{\hbar v_J}{\pi}\right) \left(\frac{\alpha}{L}\right)^2 L,\\
\langle P \rangle_{\alpha\neq 0}  - \langle P  \rangle_{\alpha = 0}  &=& \hbar N_{0} \left(\frac{\alpha}{L}\right).
\end{eqnarray}
Thus we see that the system responds to the twist in the BC's by
drifting as whole with  constant velocity $v_{sf} = (\langle P \rangle_{\alpha\neq 0}  - 
\langle P  \rangle_{\alpha = 0})/(MN_{0}) = \hbar \alpha / M L$. The superfluid
fraction, $\rho_{s}$, can be obtained by regarding the shift in the ground state energy 
as the kinetic energy of the  superfluid  mass  $M (\rho_s L)$:
\begin{equation}
\label{changenrg2}
\langle H_{\rm eff} \rangle_{\alpha\neq 0}  - \langle H \rangle_{\alpha = 0} = \frac{1}{2}M (\rho_{s} L) v^2_{sf} 
= \frac{\hbar^2\rho_{s}}{2M}  \left(\frac{\alpha}{L}\right)^2 L.
\end{equation}
By comparing~(\ref{changenrg1}) and~(\ref{changenrg2}) we can identify 
\begin{equation}
\frac{\hbar v_J}{\pi} = \frac{\hbar^2 \rho_s}{M} .  
\end{equation}
Otherwise, in view of Eq.~(\ref{changenrg1}), we could have defined the phase stiffness in an analogous
way to the density stiffness:
\begin{equation}\label{vj}
v_J  =   \frac{\pi L}{\hbar} \left(\frac{\partial^2 E_0(\alpha)}{\partial\alpha^2}\right)_{\alpha=0},
\end{equation}
where $E_0(\alpha) = \langle H \rangle_{\alpha\neq 0}$. Hence, $v_J$ is  seen to be related to the 
response of the system to a phase twist just like $v_N$ is related to the response 
to a change in particle number. For charged particles 
the phase twist can be thought as  the result of a magnetic flux that threads
the ring~\cite{Giamarchi95}.

We next specialize to Galilean invariant systems.  As we have already 
anticipated in the previous section,  for these systems
$v_J = v_F = \hbar \pi \rho_0/M$, and therefore $\rho_s = \rho_{0}$. To prove this identity
we  perform a Galilean {\it boost} where the particle momenta are shifted: $p_i \to p'_i = p_i + \hbar Q$. 
The Hamiltonian and momentum operators transform as:
\begin{eqnarray}
\label{galileiham}
H' &=& H + \frac{\hbar Q}{M}  P + N \frac{(\hbar Q)^2}{2M},\\ 
P' &=& P + \hbar N Q. 
\end{eqnarray}
Therefore, the ground state  energy and momentum  are shifted: $E'_{0} - E_{0} = N (\hbar Q)^2/2M$,
and $P'_{0} = \hbar N Q$.  Upon comparing  the latter expression  for $P$ with  
Eq.~(\ref{modmompbc}), we see that, to leading other,  
$Q = \pi J/L$. Hence, $E'_{0} = E_{0} + (\hbar \pi)^2 \rho_{0} J^2/(2L)$.
However, from Eq.~(\ref{modhampbc}), the energy of the boosted state 
is $E'_{0} = E_{0} + \hbar \pi v_J J^2/(2 L)$. From  these last
two expressions it follows that $v_J = v_F$.

	The previous analysis also allows to understand why no term involving the product
$\partial_x\Theta(x)\partial_x\phi(x)$ appears in the low-energy Hamiltonian, Eq.~(\ref{eq2.14}). The reason is 
that this product is the momentum density (cf. Eq.~(\ref{peff})). Therefore, the presence of such a 
term in the Hamiltonian would indicate that we have not chosen the 
reference frame where the system is at rest. Thus one can get rid of it 
by a suitable Galilean transformation (cf. Eq.~(\ref{galileiham})).

 We end this subsection with a discussion of the selection rules for the
eigenvalues of $J$ and $N$. For bosons with PBC's  we have found the following selection
rule: $(-1)^J = +1$, i.e. $J$ has eigenvalues that are even integers. However, obtaining
the selection rule for fermions becomes somewhat messy using the above methods.
To derive the selection rules for both bosons and fermions we recall here a construction due to Mironov 
and Zabrodin~\cite{Mironov91}. The construction  works directly with the many-particle 
wave function and therefore does not rely on the previous formalism. Consider a {\it rigid} translation
of the system by a distance $a$, i.e.
\begin{equation}
e^{i \hat{P}a/\hbar} \Phi(x_1, x_2, \ldots, x_N) = \Phi(x_1 + a, x_2, \ldots, x_N + a)
\end{equation}
Let $\Phi$ be an eigenstate of the momentum operator $\hat{P}$
with eigenvalue $P$ and set $a = L/N = \rho^{-1}_{0}$,  $x_1 = x, x_2 = x + a, \ldots,
x_N = x + (N-1)a$. Thus,
\begin{equation}
e^{i Pa/\hbar} \Phi(x, x+a, \ldots, x+(N-1)a) =  
\Phi(x+a, x+2a, \ldots, x) = (\pm 1)^{N-1} \Phi(x,x+a,\ldots,x+(N-1)a),
\end{equation}
where the plus sign corresponds to bosons and the minus to fermions.
The last equation follows from permuting the particle coordinates in the
translated wave function to sort them in increasing order. 
The operation involves $N-1$ transpositions and therefore  for fermions 
the signature of the permutation is  $(-1)^{N-1}$. 
If we choose the wave function to be an eigenstate of $\hat{P}$ with
$P = \hbar \pi J N/L$, we arrive at the selection rule:
\begin{equation}\label{selectionrule}
(-1)^{J} = (\pm 1)^{N-1},
\end{equation}
which reduces to $(-1)^J = +1$ for bosons, and to: 
\begin{equation}
(-1)^{J} = -(-1)^{N}
\end{equation}
for fermions. In the thermodynamic limit these selection rules are not very important because
one can effectively treat $N$ and $J$ as having continuous eigenvalues. Furthermore, the 
terms in the Hamiltonian~(\ref{modhampbc}) proportional to $(N-N_0)^2$ 
and $J^2$ are of order $L^{-1}$ and disappear as $L\to \infty$. However, 
for the  mesoscopic systems which concern us here, 
the selection rules can be  important in certain situations where one needs to keep
track of finite-size effects and the discreteness of particles.

\subsection{Particles in a box: open  (Dirichlet) boundary conditions.}\label{modesobc}

We now turn our attention to systems with open  (Dirichlet) boundary conditions (OBC's). In cold atom 
systems these conditions have been already experimentally realized (in an approximate way)~\cite{Hansel01} 
using a microchip trap where the potential was shaped to a square-well with very high barriers,
which can be approximated by perfectly reflecting walls.  Less restrictively,  one can 
assume that there are two points, $x = 0$ and $x = L$, where current density vanishes, and this  corresponds to
the equilibrium state in the experiment  of Ref.~\onlinecite{Wada01}, where $^4$He was confined in a long nanopore. 
To implement the latter BC's, we first note that, from the continuity equation, 
\begin{equation}
\partial_t \rho(x,t) + \partial_x j(x,t) = 0,
\end{equation}
and using that $\rho(x,t) \approx \partial_x \Theta(x)/\pi$, it follows that the current
density $j(x,t) \approx - \partial_t \Theta(x,t)/\pi$.  Hence, demanding that $j(x= 0) = 0$ 
amounts to  $\partial_t \Theta(x = 0, t) = 0$, i.e. $\Theta(x = 0, t) = \theta_B =$ const. 
However, what is often understood by open boundary conditions is something more
restrictive than this: it is demanded that  $\rho(x = 0, t) = 0$. This is achieved by the further requirement 
that $\theta_B  \neq  0, \pm \pi, \pm 2\pi, \ldots$
In other words, $\Theta(x = 0)$ must be {\it pinned} at real number that is not a
multiple of $\pi$. To understand this, we need to go back to Eq.~(\ref{eq2.7}),
\begin{equation}
\rho(x) = \partial_x \Theta(x)\: \sum_{n=-\infty}^{+\infty} \delta(\Theta(x) - n\pi).
\end{equation}
From this expression, one can see that $\rho(x = 0) = 0$ provided that 
$\Theta(x= 0) = \theta_B \neq n\pi$, where $n$ is an integer. If $\rho(x = 0)$ vanishes so does 
$\Psi^{\dag}(x = 0) \propto \sqrt{\rho(x=0)}$. What about the other end of a finite system  (i.e. $x = L$)?  
The property 
\begin{equation}
 \left[ \Theta(L) - \Theta(0) \right] = N\pi
\end{equation}
also fixes  $\Theta(L) = \Theta(0) + N\pi = \theta_{B} +  N\pi$. Thus the boundary condition is
automatically satisfied at $x = L$.

   We now find the appropriate mode
expansions for $\Theta(x)$ and $\phi(x)$. The requirements are the same
as in the previous subsection, namely that the BC's  and the 
commutation relation of  $\partial_x \Theta(x)$ and $\phi(x)$, Eq.~(\ref{CRthetaphi}), must
be fulfilled for $0 < x, x' < L$. Thus we arrive at the  following expressions  (remember in
what follows that $\theta_B$  is  just a real number, not an operator):
\begin{eqnarray}
\label{modesobc1}
\Theta(x) &=& \theta_{B} + \frac{\pi x }{L}N  + i \sum_{q > 
0} \Big( \frac{\pi K}{q L} \Big)^{\frac{1}{2}}\: e^{-a q /2} 
\sin(qx) \big[b(q) - b^{\dagger}(q)\big],\\
\phi(x) &=& \phi_{0} + \sum_{q > 0}  \Big( \frac{\pi}{q L K} 
\Big)^{\frac{1}{2}}\: e^{-a q /2} \cos(qx)  \big[ b(q)+  
b^{\dagger}(q)\big] \label{modesobc2}
\end{eqnarray}
where $a^{-1} \lesssim \min\left\{R^{-1}, q_c = \mu/\hbar v_s\right\}$ and $q = m \pi/L$, with 
$m=1,2,3,\ldots$ Notice that 
in this case only one pair of action-angle operators is needed, 
$(N,\phi_{0})$. This is because in a box, as opposed to a ring, there
cannot be persistent currents, and therefore $J$  is not a good quantum number. 

  The above mode expansions diagonalize the 
Hamiltonian and render it as follows:
\begin{equation}
H_{\rm eff} = \sum_{q > 0} \hbar \omega(q) \: b^{\dagger}(q) b(q) + 
 \frac{\hbar v_s \pi}{2 LK} (N-N_{0})^2,
\end{equation}
with $\omega(q) = v_s q$ for $q \ll \rho_{0}$. The restriction $q > 0$ means
that $q$ cannot be interpreted as the momentum of the excitation  
but as its wave number. The sound waves in a system with OBC's are standing waves
and therefore do not carry momentum. Consistently, the momentum operator vanishes, 
i.e. $P = 0$~\footnote{Note that $P$ measures the momentum of the system relative to its 
center of mass (i.e. the momentum carried by the excitations). There is an extra term, which accounts
for the momentum of the center  of mass, and which would describe the motion of the system as a whole, including the trap (box). Here we have assume to be in a reference frame where the box is at rest.}.

\subsection{Effect of a slowly varying confining potential}\label{trap}

In most of current experimental setups  cold atoms  are confined
in harmonic traps. In this section we are going to discuss how the harmonic-fluid
approach must be modified when a smooth potential is applied in the longitudinal direction.
We shall distinguish two situations. In the first we  consider that the 
external potential is weak and show that much of what has been said above applies
with small modifications. However, the harmonic potential does not belong to this class.
In other words, it is not weak, but this does not mean that the harmonic-fluid approach
cannot be adapted to this case. However, calculations are no longer analytically
feasible (except in certain limits).   Some results for harmonically trapped 
gases of bosons~\cite{Gangardt03} and fermions~\cite{Recati03} are 
already available in the literature, but since in general no explicit expressions for the
correlation functions are available it
is  hard to extract much qualitative information from these results. 
In the Tonks limit, one can rely on the fermion-boson corresondence 
established by Girardeau~\cite{Girardeau60}, which also allows
to make a beautiful connection with the theory of random matrices in the gaussian 
unitary ensemble (GUE).   This  connection makes it possible to obtain the density profile~\cite{Mehta}, the 
form of the Friedel oscillations~\cite{Kalisch02} as well as asymptotic forms for the one-body density 
matrix~\cite{Forrester03}.  It is also worth mentioning that for fermionic systems one 
can use  constructive bosonization~\cite{Haldane81a},
and some analytical results can be thus obtained~\cite{Xianlong03}. However, the problem with this approach
is that the interactions between cold fermions do not have a simple form in the harmonic-oscillator basis. 
Therefore, it becomes hard to assess the effect of the different matrix elements on the low-energy physics.

Introducing an external potential amounts to adding to the Hamiltonian in Eq.~(\ref{eq2.1})
the following term 
\begin{equation}\label{eq3.1}
U_{\rm ext} = \int_{0}^{L} dx \: u_{\rm ext}(x) \rho(x).
\end{equation}
Provided  that $u_{\rm ext}(x)$ is weak (i.e. $|u_{\rm ext}(x)| \ll \mu$) 
and varies slowly over distances on the scale of $a$, it couples only to the slow part
of the density operator, namely to $\partial_x \Theta(x)/\pi$. Hence,
\begin{equation}\label{eq3.2}
H_{\rm eff} = \int_{0}^{L} dx \: \left\{   \frac{\hbar v_s}{2 \pi} \left[ K \left(\partial_x \phi(x) \right) +  \frac{1}{K} \left( \partial_x \Theta(x) - \pi \rho_{0} \right)^2 \right]  +  \frac{u_{\rm ext}(x)}{\pi}  \partial_x \Theta(x) \right\}.
\end{equation}
Up to a constant, $H_{\rm eff}$ can be  rewritten as:
\begin{equation}\label{eq3.3}\label{hamdens}
H_{\rm eff}  = \frac{\hbar v_s}{2 \pi} \int_{0}^{L} dx \: 
 \left[ K \left(\partial_x \phi(x) \right)^2 + 
\frac{1}{K} \left( \partial_x \Theta(x)  - \pi   \rho_{0}(x) \right)^2  \right] ,
\end{equation}
where $\rho_{0}(x) = \rho_{0} + \delta \rho_{0}(x)$, and
\begin{equation}\label{linearesponse}
\delta \rho_{0}(x) = - \frac{K}{\hbar \pi v_s} u_{\rm ext}(x) = - \left( \frac{d\rho}{d\mu} 
\right) \: u_{\rm ext}(x).
\end{equation}
The last expression follows from the relationship found in Eq.~(\ref{compress})
between $K/\hbar \pi v_s$ and the compressibility. The result from 
linear response theory is thus recovered, since $u_{\rm ext}(x)$ enters 
the Hamiltonian as (minus) the chemical potential.
In order to render the Hamiltonian to its form in the absence of $u_{\rm ext}$ it is convenient to shift
\begin{equation}\label{shift}
\tilde{\Theta}(x) = \Theta(x) - \pi \int_{0}^{x} dx' \: 
\left[ \rho_{0}(x') - \rho_{0} \right],
\end{equation}
such that the second term in Eq.~(\ref{hamdens}) becomes:
\begin{equation}
\frac{\hbar v_s}{2\pi K}\int^{L}_{0} dx  \left(\partial_x \tilde{\Theta}(x) - \pi \rho_{0} \right)^2.
\end{equation}
The expansions in modes given in previous sections can now be used to diagonalize $H_{\rm eff}$ 
with $\tilde{\Theta}(x)$ playing the role of $\Theta(x)$. This leads to the same form for the spectrum,
but the shift (\ref{shift})  must be taken into account  when computing correlation functions. Thus, 
the density and field operators become:
\begin{eqnarray}
\rho(x) &=& \left[ \rho_{0}(x) + \frac{1}{\pi}\partial_x \tilde{\theta}(x) \right] \sum_{m = -\infty}^{+\infty} 
e^{2m i \pi \int_{0}^{x} dx' \: \rho_{0}(x')} \: e^{2mi\tilde{\theta}(x)},\\
\Psi^{\dag}(x) &\sim& \left[\rho_0(x) + \frac{1}{\pi}\partial_x \tilde{\theta}(x) \right]^{1/2}
\sum_{m=-\infty}^{+\infty} e^{2mi\pi \int_{0}^{x} dx' \: \rho_{0}(x')}\: 
 e^{2mi\tilde{\theta}(x)}\: e^{-i\phi(x)},\\
\Psi^{\dag}_F(x) &\sim& \left[\rho_0(x) + \frac{1}{\pi}\partial_x \tilde{\theta}(x) \right]^{1/2}
\sum_{m=-\infty}^{+\infty} e^{(2m+1)i\pi \int_{0}^{x} dx' \: \rho_{0}(x')}\: 
 e^{(2m+1)i\tilde{\theta}(x)}\: e^{-i\phi(x)},
\end{eqnarray}
where $\tilde{\theta}(x) = \tilde{\Theta}(x) - \pi \rho_{0}x$.  

	The above treatment is essentially correct provided that
 the external potential represents weak perturbation, 
 which has been quantified by requiring that $|u_{\rm ext}(x)| \ll \mu$,
for  $0 < x < L$,  and that the external potential is a slowly varying function. In terms of its 
Fourier its transform the latter means that $u_{\rm ext}(q) \approx 0$  for $|q| \gtrsim a^{-1} \approx \mu/\hbar v_s$. 
Whereas the last condition does certainly hold for a shallow harmonic trapping potential, 
its effect cannot be considered weak, especially near the ends of the atomic cloud where typically $u_{\rm ext}(x) \approx \mu$. Therefore, away from the center of the trap this potential cannot be treated within
linear response theory and the above results do not apply. However, one can always redo the
harmonic-fluid approach by considering fluctuations around a smooth density profile $\rho_0(x)$, 
which can be  obtained, e.g.  from the equation of state 
in the local-density approximation (LDA)~\cite{Dunjko01}. Thus, it is not hard to see that 
the effective Hamiltonian becomes:
\begin{equation}\label{lda}
H_{\rm eff} = \frac{\hbar}{2\pi} \int^{L}_{0} dx \left[ v_J(x) (\partial_x \phi(x))^2 + 
v_N(x) \left( \partial_x \Theta(x)-\pi \rho_0(x) \right)^2\right],
\end{equation}
where in the LDA $v_J(x) = \hbar \pi \rho_0(x)/M$ (i.e. the local
Fermi velocity) and
\begin{equation}
v_N(x) = \frac{1}{\hbar \pi}\left(\frac{\partial\mu}{\partial\rho}\right)_{\rho=\rho_0(x)},
\end{equation}
The equation of state $\mu = \mu(\rho)$ is obtained from the  Bethe-ansatz 
solution~\cite{Lieb63a,Lieb63b}.  Thus, diagonalization of~(\ref{lda}) 
can only be accomplished numerically in the general case, and will not be pursued here. 
\begin{figure}[b]
\centerline {
\includegraphics[width=11cm]{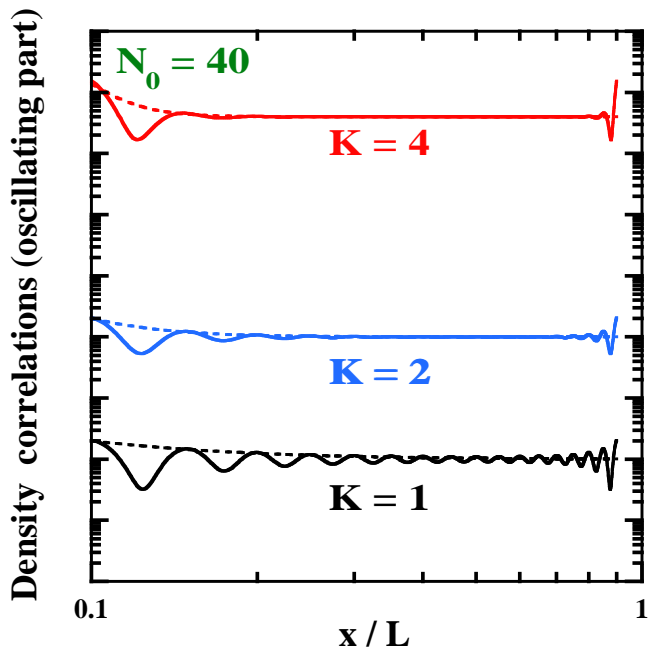}}
\caption{Oscillating part (i.e. $m=1$ term in~(\ref{densitypbc})) 
of the density correlation function as a function of distance, $x$ (normalized
to the system's size), for several values of $K$ (the different curves are shifted for clarity). As $K$ decreases 
the system becomes more ``density stiff'', which
means that density correlations between distant points are enhanced. However, for large $K$ density correlations
decay very rapidly with distance. This strong suppression of density fluctuations is characteristic of Bose-Einstein
condensates in higher dimensions.  The dashed line represents the power-law behavior expected for the envelope
in the thermodynamic limit $L\to \infty$.}
\label{fig6}
\end{figure}
\section{Correlation functions}\label{correlators}
\subsection{Periodic Boundary Conditions (PBC's)}
\begin{figure}[b]
\centerline {
\includegraphics[width=10cm]{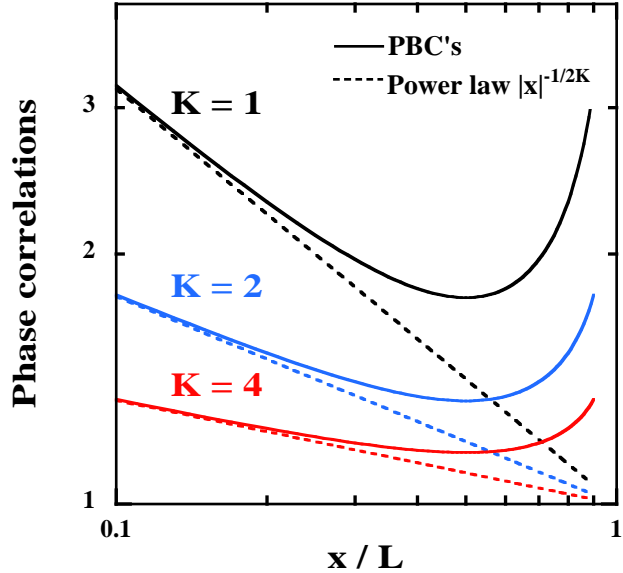}}
\caption{
Phase correlations in a ring as a function of distance, $x$ (normalized to the system's size), for several values of the
dimensionless parameter $K$. Note that as $K$ decreases the system becomes less ``phase stiff'', i.e. 
phase correlations decay faster. For large values of $K$ phase correlations decay very slowly, leading to a larger
degree of phase coherence between distant points.  The dashed lines  
correspond to the power-law behavior occurring in the thermodynamic limit $L \to +\infty$. The final upturn of the curves is due to the periodicity of the correlation functions in a ring.}\label{fig7} 
\end{figure}
 We begin by considering the static correlation
functions at $T = 0$ (see Sect.~\ref{temperature} for results at finite temperature) 
for particles in a ring, i.e. periodic boundary conditions.  
The time-dependent correlation functions can be  also obtained, but will
not be discussed in this paper (except for the dynamic density response function, which is discussed
in Sect.\ref{tips}). Here we only notice that it suffices to perform 
a careful analytical continuation to real time $\tau \to it$ in the  
expressions given in appendices~\ref{appc} and \ref{appd}. 
Time-dependent correlations can be important for future experiments such
like the two-photon Raman out-coupling experiment 
recently proposed by Luxat and Griffin~\cite{Luxat03}.

When computing correlation functions within the harmonic-fluid approach,
two strategies are possible.  One can directly work with the mode
expansions given in sections~\ref{modespbc} and \ref{modesobc} using the 
properties of the exponentials of  linear combinations  of the $b(q)$
and $b^\dagger(q)$ operators. Another possibility is to use the more sophisticated
techniques of Conformal Field Theory (CFT). The latter is our choice here
and it is explained in the appendices~\ref{appc} and~\ref{appd}.  Therefore,
the reader unfamiliar with these methods should consult  the appendices for
full details. In this subsection we present the results for the correlation functions 
for both fermions and bosons in a ring. 

 Let us first consider  the 
density correlation function,  which does not depend on statistics.  We first notice that
the structure of the density operator, Eq.~(\ref{eq2.9}),
implies that the density correlation function is a series of harmonics
of the Fermi momentum $p_{\rm F} = \pi \rho_{0}$. The origin of this  
has been already discussed in Sect.~\ref{ss1.1}.  In our calculations, we have  kept 
only the  leading (i.e. the slowest decaying) term of each harmonic. Thus,
\begin{eqnarray}
\label{densitypbc}
\langle \rho(x) \rho(0) \rangle &=& \left\{ \frac{1}{\pi^2}\langle 
\partial_{x}\Theta(x) \partial_{x}\Theta(0)  \rangle
+ \rho^2_{0} \sum_{m = -\infty}^{+\infty} e^{2  \pi i m \rho_{0} x}
\langle A_{2m,0}(x)  A_{-2m}(0) \rangle_{\rm pbc} \right\} \nonumber \\
&=& \rho^2_{0} \left\{ 1 - \frac{K}{2\pi^2} 
\left[\frac{1}{\rho_{0} d(x|L)} \right]^2 + 
\sum_{m > 0} a_{m} 
\left[\frac{1}{\rho_{0} d(x| L)} \right]^{2m^2K} 
\cos \left(2 \pi m  \rho_{0} x\right)  \right\}.
\end{eqnarray}
This result is valid in the {\it scaling} limit, i.e. for $|x| \gg a$, where $a$ is the short-distance
cut-off introduced in previous sections. The {\it vertex} operators $A_{m,n}(x,\tau)$ 
are defined in  Appendix~\ref{appc}. The coefficients $a_m$ are non-universal, in other words,
they depend on the  microscopic details of the model  and in general cannot be fixed 
by  the harmonic-fluid  approach. The function $d(x|L) = L |\sin(\pi x/L)|/\pi$ 
is called {\it cord} function: it measures the length of a cord between two
points separated by an arc $x$ in a ring of circumference $L$. 

 We next take up the boson and fermion one-particle density matrices. Upon using
Eqs.~(\ref{eq2.10}) and (\ref{fermionfield}), we obtain:
\begin{eqnarray}
\langle \Psi^{\dagger}(x) \Psi(0) \rangle & = & \rho_{0}
\sum_{m=-\infty}^{+\infty} e^{2m i \pi \rho_{0} x} \langle 
A_{2m,-1}(x) A_{-2m,+1}(0)  \rangle_{\rm pbc} \nonumber \\ 
&=& \rho_{0} \left[  \frac{1}{\rho_{0} d(x|L)} \right]^{\frac{1}{2K}}
 \left\{  b_0 + \sum_{m= 1}^{+\infty} b_m 
\left[  \frac{1}{\rho_{0} d(x|L)} \right]^{2m^2 K} \cos \left(2\pi m \rho_{0} 
x \right) \right\}, \label{g1pbc}\\
\langle \Psi^{\dagger}_{F}(x) \Psi_{F}(0) \rangle &=& \rho_{0}
\sum_{m=-\infty}^{+\infty} e^{2\pi i (m + \frac{1}{2}) \rho_{0} x} \langle 
A_{2m+1,-1}(x) A_{-2m-1,+1}(0)  \rangle_{\rm pbc} \nonumber \\
&=& \rho_{0}
 \sum_{m= 0}^{+\infty} f_m 
\left[ \frac{1}{\rho_{0} d(x|L)} \right]^{2(m+ \frac{1}{2})^2 K+
\frac{1}{2K}} \sin \left|2\pi (m+\frac{1}{2}) \rho_{0} 
x \right|.
\end{eqnarray}
Also in these expressions the {\it dimensionless} coefficients $b_m$ and $f_m$
are non-universal. The results of  
Ref.~\onlinecite{Haldane81b} are recovered in the thermodynamic 
limit $L \to + \infty$, which effectively amounts to performing the replacement 
$d(x|L) \to |x|$ in the above expressions~\cite{notation}. In the Tonks limit $K = 1$ and
the leading term of (\ref{g1pbc}) agrees with the exact  
asymptotic results obtained in Refs.~\onlinecite{Lenard72,Forrester03}:
\begin{equation}
\langle \Psi^{\dag}(x)\Psi(0) \rangle_{\rm pbc} = \rho_0\,  b^{\rm Tonks}_0 \left[\frac{1}{\rho_0 d(x|L)}\right]^{1/2}, 
\end{equation}
where $b^{\rm Tonks}_0 =2^{-1/3}  \sqrt{\pi e}\,  A^{-6}\simeq 0.5214$ 
($A \simeq   1.2824271$ is Glaisher's constant).

 In Fig.~\ref{fig6} we have plotted the $m = 1$ term of $\langle \rho(x) \rho(0) \rangle$ to illustrate
how the density correlations are enhanced as $K$ is decreased. Thus for $K=1$ (free fermions or 
the Tonks limit of delta-interacting bosons) the distant points are more correlated in density than
for large $K$ (very attractive fermions or  bosons with  weakly repulsive interactions). 
This means that the oscillatory terms become more important as $K$ decreases, whereas for large
$K$ they can be safely neglected, as it is  done in the Bogoliubov approximation. On the other hand, as shown
in Fig.~\ref{fig7}, distant points become less phase correlated for small $K$ 
as the result of density and phase being conjugated
fields. The system is then said to be less ``phase stiff''. However,
for larger $K$ phase correlations are enhanced, and  only in this sense one can 
speak of ``Bose-Einstein condensate'' even though, strictly speaking, there is not 
such a thing as a condensate (i.e. long range-order) in one dimension  (for instance $\langle e^{i\phi(x)} \rangle_{\rm pbc} = 0$). At most, all that exists in the thermodynamic limit is an {\it algebraic} decay of phase correlations, which  is often called {\it quasi long-range} order (QLRO).

\subsection{Open Boundary Conditions (OBC's)}

  In this subsection we consider the ground state correlations for particles in 
a box. The expressions for the correlation functions
in this case are somewhat more  complicated than for periodic boundary conditions. 
The reason is that in a box translational invariance is lost
and  two-point correlation functions  depend on both arguments separately and 
not only on their difference. Furthermore,  the ground state expectation value of some operators
becomes non-trivial, that is, there are non-trivial one-point correlation functions.
This is the case of the density operator (see below).
However, the operator $e^{i\phi(x)}$ continues to have zero-expectation value,
implying the absence of any continuous symmetry-breaking  even though the system is finite and bounded.

\begin{figure}[hb]
\centerline {
\includegraphics[width=18cm,height=9cm]{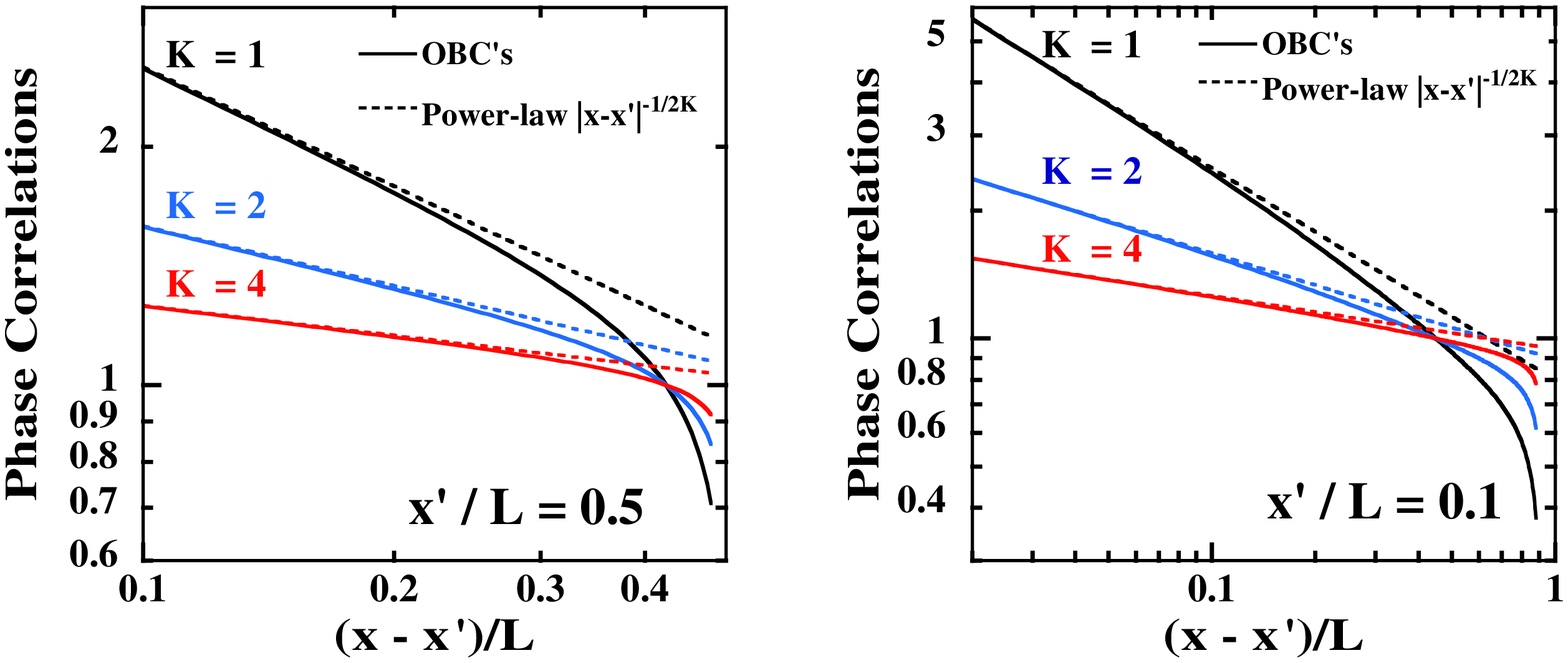}}
\caption{Phase correlations for a system of bosons in a box. In the left plot, $x'$ has been
fixed to the center of the box, while phase correlations are sample as $x$ is shifted towards the
right end. In the right plot, $x'$ is close to the left end and as  $x$ increases we move towards
the right end. As remarked in the case of a ring, the system becomes less phase stiff as $K$ is decreased.
This means that in this case the boundary conditions have a large effect on the phase correlations in
the Tonks limit ($K=1$) than in the weakly interacting limit ($K\gg 1$).}
\label{fig8}
\end{figure}

  We first compute the ground state expectation value of the density:
\begin{eqnarray}
\langle \rho(x) \rangle = \rho_{\rm o } \sum_{m=-\infty}^{+\infty} e^{2i m\pi \rho_{0} x}
 \langle A_{2m,0}(x) \rangle_{\rm obc} 
= \rho_{0} \left\{ 1 +  \sum_{m > 0}
 c_m \left[ \frac{1}{\rho_{0} d(2x|2L)} \right]^{m^2K}
\cos (2 m \pi \rho_{0} x + \delta_m) \right\},
\end{eqnarray}
where the coefficients $c_m$ and $\delta_m$ are model dependent~\cite{coeffs}.
The above expression is valid in the scaling limit, which for the above 
expression means that $\min\{x,L-x\} \gg  a$, i.e. sufficiently far from 
the boundaries.  It is noticeable that $\langle \rho(x)\rangle$ exhibits Friedel
oscillations, independently of the statistics of the constituent particles. 
This is not surprising in view of the many common features exhibited
by interacting bosons and fermions  in one-dimension,  and whose origin
was already discussed in Sect.~\ref{sect1}. In the thermodynamic limit,  
the replacement $d(2x|2L) \to |2x|$ shows that the different
oscillating terms in the expression for $\langle \rho(x)\rangle$ decay from the boundary 
as power laws with increasingly large exponents: $m^2 K$ for $m = 1, 2, \ldots$
indicating that only the leading two terms are important.

  Next we consider the more complicated two-point correlation functions.
We begin with the density-density correlation function,
\begin{eqnarray}
\langle \rho(x) \rho(x') \rangle &=&
\left\{ \frac{1}{\pi^2} \langle \partial_{x} \Theta(x) 
\partial_{x'} \Theta(x') \rangle_{\rm obc}  + 
\sum_{m,m' = -\infty}^{+\infty}  e^{2i \pi \rho_{0}(m x - m' x')}
\langle A_{2m,0}(x) A_{-2m',0}(x')\rangle_{\rm obc} \right\} \nonumber \\
&=& \rho^{2}_{0}\: \Bigg\{ 1  - \frac{K}{2\pi^2} 
\left[ \frac{1}{\rho_{0}d(x-x'|2L)} \right]^2  - \frac{K}{2\pi^2}
 \left[ \frac{1}{\rho_{0}d(x+ x'|2L)} \right]^2 \nonumber \\
&&+ \sum_{m,m' = -\infty}^{+\infty} a_{m,m'} 
 \left[ \frac{d(x+x'|2L)}{d(x-x'|2L)}\right]^{2m m' K}
 \frac{  e^{2 i \pi \rho_{0}  (m x - m' x')}  } 
{\left[\rho_{0} d(2 x|2L) \right]^{m^2 K} 
\left[\rho_{0} d(2 x'|2L) \right]^{m'^2 K}}
\Bigg\}.\label{densobc}
\end{eqnarray}

 Using the results of Appendix \ref{appd} one can also obtain
the one-particle density matrices of bosons and fermions.
The corresponding expressions read:
\begin{eqnarray}
\label{g1}
\langle \Psi^{\dagger}(x) \Psi(x') \rangle &=& 
\rho_{0} \sum_{m,m' = -\infty}^{+\infty}  e^{2i \pi \rho_{0}(m x - m' x')}
\langle A_{2m,-1}(x) A_{-2m',1}(x')\rangle_{\rm obc} \nonumber \\
&=& \rho_{0}\:
\left[ \frac{\rho^{-1}_{0}
\sqrt{d(2x|2L) d(2x'|2L)}}{d(x+x'|2L) d(x-x'|2L)}\right]^{\frac{1}{2K}}
 \sum_{m,m'=-\infty}^{+\infty} b_{m,m'}\: 
e^{-i (m+m')\pi {\rm sgn}(x-x')/2} \nonumber \\
&& \times \left[ \frac{d(x+x'|2L)}{d(x-x'|2L)}\right]^{2mm'K}   
 \frac{  e^{2 i \pi \rho_{0}  (m x - m' x')}  } 
{\left[\rho_{0} d(2 x|2L) \right]^{m^2 K} 
\left[\rho_{0} d(2 x'|2L) \right]^{m'^2 K}} \label{g1obc} \\
\langle \Psi^{\dagger}_{F}(x) \Psi_{F}(x') \rangle &=& 
\rho_{0} \sum_{m,m' = -\infty}^{+\infty}   
e^{2 i \pi \rho_{0}  \left[ (m + \frac{1}{2}) x -  (m' + \frac{1}{2}) x' \right]}
\langle A_{2m+1,-1}(x) A_{-2m'-1,1}(x')\rangle_{\rm obc} \nonumber \\
&=& \rho_{0}\:
\left[ \frac{\rho^{-1}_{0}
\sqrt{d(2x|2L) d(2x'|2L)}}{d(x+x'|2L) d(x-x'|2L)}\right]^{\frac{1}{2K}}
 \sum_{m,m'=-\infty}^{+\infty} f_{m,m'}\: 
e^{-i (m+m' + 1)\pi {\rm sgn}(x-x')/2} \nonumber \\
&\times& \left[ \frac{d(x+x'|2L)}{d(x-x'|2L)}\right]^{2(m + \frac{1}{2})
(m' + \frac{1}{2})K}   
 \frac{  e^{2 i \pi \rho_{0}  \left[ (m + \frac{1}{2}) x -  (m' + \frac{1}{2}) x' \right]}  } 
{\left[\rho_{0} d(2 x|2L) \right]^{(m + \frac{1}{2})^2 K} 
\left[\rho_{0} d(2 x'|2L) \right]^{(m' + \frac{1}{2})^2 K}}. 
\end{eqnarray}
The coefficients $a_{m,m'}, b_{m,m'}$ and $c_{m,m'}$ are model-dependent complex numbers.
In contrast with the case of periodic boundary conditions, where these prefactors can shown 
to be real, this cannot be established in the present case~\cite{cft}.
The above expressions are accurate in the scaling limit, which for two-point
correlation functions means that   $|x-x'|\gg a$ as well as  $\min\{x,x',L-x,L-x'\} \gg  a$.
It is also worth mentioning that  the bulk behavior is recovered for $|x-x'| \ll  \min\{x,x',L-x,L-x'\}$, i.e. 
mostly near the center of the box (see Fig.~\ref{fig8}). In this limit, the less oscillating terms are those with
$m = m'$, which exhibit, in the thermodynamic limit, the same algebraic decay as the correlations for 
the ring geometry. It is also worth pointing out that in the Tonks limit (i.e. for $K=1$), the leading term
of Eq.~(\ref{g1obc}) agrees with the asymptotic result obtained in Ref.~\onlinecite{Forrester03},
\begin{equation}
\langle \Psi^{\dag}(x)  \Psi(x')  \rangle_{\rm obc} = \rho_0 \, b^{\rm Tonks}_{00} \left[ \frac{\rho^{-1}_{0}
\sqrt{d(2x|2L) d(2x'|2L)}}{d(x+x'|2L) d(x-x'|2L)}\right]^{\frac{1}{2}},
\end{equation}
where $b^{\rm Tonks}_{00} = 2^{-1/3}  \sqrt{\pi e}\,  A^{-6}\simeq 0.5214$ 
($A \simeq   1.2824271$ is Glaisher's constant). Note that  this is the same 
prefactor as the one found for PBC's, which is not suprising given that both correlation 
functions have the same bulk limit.

 It is interesting to point out that for particles in a box there are two types of 
exponents that characterize the decay of correlation functions.  Consider for instance the 
density correlation function, Eq.~(\ref{densobc}). Its bulk behavior  (i.e. near the center of the box)
is governed by  the exponent $2m^2 K$ for  a harmonic that oscillates 
as $e^{\pm 2m \pi\rho_0(x-x')}$. However, the asymptotic behavior when one of the coordinates 
is taken near the boundary is governed by a different exponent. Thus if we set $x' \approx a$ and $x\gg x'$, 
a term that oscillates  as $e^{\pm 2i m\pi\rho_0 x}$in  the correlation function falls off  with the exponent 
equal to $m^2 K$, i.e. the same exponent that we have already encountered 
for the Friedel oscillations of the density.  
For the boson density matrix the corresponding exponents
are $m^2 K + 3/4K$ whereas for the fermions they are $(m+1/2)^2 K + 3/4 K$. It is important to
stress that the presence of the boundary breaks  the (Lorentz) invariance of $H_{\rm eff}$  making
space and time directions non-equivalent. To see this,  consider a dynamic correlation function such 
like the Green's function  $\langle \Psi^{\dag}(x,t)\Psi(x',0)\rangle$, for bosons, or 
$\langle \Psi^{\dag}_F(x,t ) \Psi(x',0)\rangle$, for fermions ($t> 0$). Using the expressions of Appendix~\ref{appd},
and performing the analytical continuation $\tau \to i t$ after setting $x=x' \approx a$, 
one obtains that the leading term decays as $t^{-1/K}$ for $t \gg a/v_s$ (for both bosons and fermions).
This in contrast with the  behavior of the same functions in the bulk, where space and time
are equivalent (i.e. Lorentz invariance is restored) and one expects a behavior like 
$t^{-1/2K}$ for bosons and $t^{-(1/K + K)/2}$ for fermions. \footnote{There is a simple way
to understand these exponents. The exponent that governs the decay of a correlation
function in the bulk is twice the scaling dimension of the operator $d = h+\bar{h}$,
$h$ and $\bar{h}$ being its conformal dimensions (see appendix~\ref{appc}
for details). Thus, the different terms in the bosonic field have $d = m^2 K + 1/4K$, whereas for the
fermionic field $d = (m+1)^2 K + 1/4K$. In the presence of a boundary, which breaks
Lorentz invariance, one has to introduce an additional scaling dimension, termed {\it boundary}
dimension: $d_{B} = 1/2K$  for both bosons and fermions (i.e. half 
the exponent of  $\sim t^{-1/K}$, the asymptotic behavior 
near the boundary). Thus, for a two-point correlation function of the same operator,
if one of the arguments lies at the boundary, the spatial decay of  the correlation
function is given by the exponents $d + d_B = m^2 K + 3/4K$ for bosons
and $d + d_B = (m+1/2)^2 K + 3/4K$ for fermions.}
 
   To illustrate both finite-size and confinement effects we show in Fig.~\ref{fig8} 
the behavior of leading term in Eq.~(\ref{g1}), which corresponds 
to the phase fluctuations $\langle e^{-i\phi(x)} e^{i\phi(0)}\rangle_{\rm obc}$ (i..e. $m=m'=0$).
As mentioned above, the correlation function is now a function of $x$ and $x'$
separately.  Thus, in Fig.~\ref{fig8} we have considered two situations. In the plot
on the left, we take $x'$ to be the center of the box. Thus it can be seen that the
power-law (bulk) behavior is a good approximation for $x$ around $L/2$, whereas the
phase correlations  deviate from it  as $x$ approaches the right end of the box 
(i.e. for $x-x' \gtrsim 0.3 L$. These deviations become more important as $K$ 
is decreased and the system approaches the Tonks limit ($K = 1$). This is because,
as remarked in the previous subsection, the system becomes less phase stiff as $K$ 
decreases. Thus, for smaller $K$ the boundary conditions have a larger effect in decreasing
the phase correlations near the ends of the box. However, for larger $K$ the system exhibits
a larger phase stiffness, and phase correlations decay very slowly, even near the boundaries.
Furthermore, the deviations from the power-law behavior are smaller. However, these deviations
become much larger with distance if now $x' = 0.1 L$, i.e. when $x'$ is fixed near the left end of the
box. Then, the bulk power-law is no longer a good approximation and we can observe
deviations from it for a decade: from $x -x' \approx 0.1 L$ to $x-x' \approx L$.  The reasons for this
deviations is that initially the correlations decay according to the bulk power-law, but they rapidly
cross-over to a decay governed by the exponent $3/4K$ (recall that $m=m'=0$). Near the
right boundary, the correlation function must vanish. This is because the pinning of the density field $\Theta(x)$
at $x=L$ makes fluctuate the phase at that point  wildly.

\subsection{Correlations at finite temperature}\label{temperature}
 
   As  described in Appendix~\ref{appc} the same methods that have allowed us to obtain the correlation functions of a finite ring at zero temperature also allow to obtain expressions for the correlation functions of an infinite system at finite temperature (see below).  
 
 	At finite temperature one finds (see Appendix~\ref{appc}) that correlations decay exponentially with 
distance (as opposed to the algebraic decay found at $T=0$ for $L\to +\infty$). Therefore, 
in the expressions below, we shall keep only the leading terms, which effectively means that we cut-off
the series of harmonics at $m=0$ (for the boson density matrix, and at $|m| = 1$ for the density correlation function and the fermion density matrix. Thus the following expressions are obtained:
\begin{eqnarray}
\langle \rho(x) \rho(0) \rangle_T &=& \rho^2_0\left\{1 - 
\frac{K}{2\pi^2} \left[ \frac{\pi/L_T}{ \rho_0 \sinh\left( \pi x/L_T\right) } \right]^2 + B \cos(2\pi \rho_0 x)
\left[ \frac{\pi/L_T}{\rho_0\sinh\left(\pi x/L_T\right)}\right]^{2 K}
   + \ldots \right\},\\
\langle \Psi^{\dag}(x)\Psi(0)\rangle &=&   \rho_0\: A \left[ \frac{\pi/L_T}
{\rho_0\sinh\left(\pi x/L_T\right)}\right]^{\frac{1}{2K} } + \ldots,\label{bosonT}\\
\langle \Psi^{\dag}_F(x) \Psi_F(0) \rangle_{T} &=&  \rho_0\:  C \left[  \frac{\pi/L_T}
{\rho_0\sinh\left(\pi x/L_T\right)}\right]^{\frac{K}{2} +\frac{1}{2K} } \sin|\pi \rho_0 x| + \ldots,
\end{eqnarray}
the dimensionless coefficients $A, B$, and  $C$ are model dependent. In Sect.~\ref{cutoffs} we shall discuss how to fix the prefactor for the bosonic density matrix, Eq.~(\ref{bosonT}).

 From the previous expressions  one can see 
that  the behavior of  the correlations ``crosses over'' from  algebraic
decay for  $a \ll |x| \lesssim L_T$ to exponential  $\sim \exp\left[ - |x|/L_c(T) \right]$  for $|x| \gg L_c(T)$.
The characteristic decay length $L_c(T)$ depends on the correlation function. Thus,  for  the 
non-oscillating  part of $\langle \rho(x)\rho(0)\rangle_T$, $L_c(T) \sim L_T$. However, 
for the oscillating part  $L_c(T) \sim L_{\rho}(T) = \hbar v_s /\pi KT = \hbar v_N/\pi T$.  
For the phase correlations, however,   the characteristic distance is of the order of
$L_{\phi}(T) = \hbar v_s K/\pi T = \hbar v_J/\pi T  = \hbar^2\rho_0/MT$, the latter expression being
valid only in systems with Galilean invariance.  One can use the previous expressions also in
finite-size systems provided  the temperature is high enough. The value of $T$ for which these 
characteristic lengths become of the order of the system size, $L$, give us a 
rough estimate of the temperature scale below which finite-size and confinement effects are important. 
Since, in general, $L_T \neq L_{\phi} \neq  L_{\rho}$, this temperature 
scale  depends on the correlation function. Thus, for the phase 
correlations,  $L_{\phi}(T) \approx L$ for $T_{\phi} \approx \hbar 
v_J/\pi L = \hbar^2 \rho_0/ML$,  the latter expression applying to 
systems with Galilean invariance. However,  for the density correlations,  
$L_T/\pi \approx L$ or $L_{\rho}(T) \approx L$, whichever gives the highest value of $T$.  
Hence,  $T_{\rho} = \max\{\hbar v_s/\pi L, \hbar v_s/\pi K L \}$.

\section{Delta-Interacting bosons as a Luttinger liquid}\label{delta}

\subsection{Extracting the Luttinger liquid parameters from the Bethe-ansatz solution}\label{luttinger}
\begin{figure}[ht]
\centerline {
\includegraphics[width=10cm]{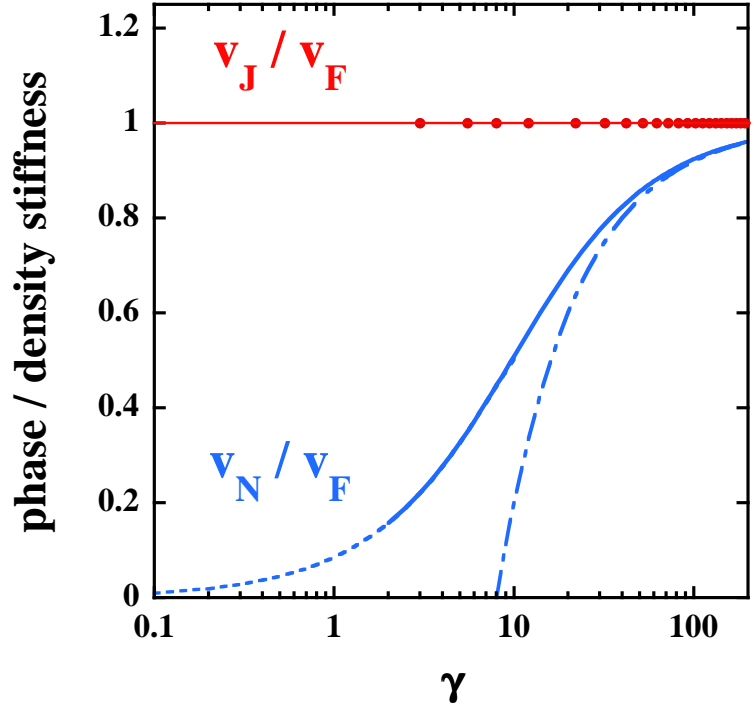}}
\caption{Density and phase stiffness as a function of $\gamma = M g/\hbar^2\rho_0$. The
dots are the result of the numerical calculation of $v_J$ from the Bethe-ansatz equations. 
The dashed line corresponds to the Bogoliubov result for $v_N$, Eq.~(\ref{bogol1}).
The dotted-dashed line is the large-$\gamma$ asymptotic result, Eq.~(\ref{largegamma1}).}\label{fig9}
\end{figure}

  In what follows, we shall illustrate many of the concepts introduced above by considering a model of bosons
interacting with a zero-range potential~\cite{Lieb63a,Lieb63b}. This is relevant to 
cold atomic vapors,  where atoms mainly interact through the s-wave scattering channel, and their
interaction is parametrized by the s-wave scattering length, $a_s$. Olshanii~\cite{Olshanii98} has 
shown that the interaction between atoms  confined in 1D wave guide with transverse harmonic confinement 
is well described by  a zero-range interaction potential  $v(x-x') = g \: \delta(x-x')$, where
 \begin{equation}
 g = \frac{4\hbar^2}{M \ell_{\perp}} \frac{a_s/\ell_{\perp}}{1 - C (a_s / \ell_{\perp})}, 
 \end{equation}
where $a_s$ is the three-dimensional scattering length, $C = 1.4603\ldots$,
 and $\ell_{\perp} = \sqrt{ \hbar/M\omega_{\perp}}$ is the transverse
oscillator length. There is a single dimensionless parameter, $\gamma = M  g/\hbar^2\rho_0$, 
which characterizes the different regimes (some authors prefer to use the 1D gas parameter 
$\rho_0 |a_{1D}|$, which is related to $\gamma$ by $\gamma  = 2/\rho_0|a_{1D}|$). 
\begin{figure}[b]
\centerline {
\includegraphics[width=10cm]{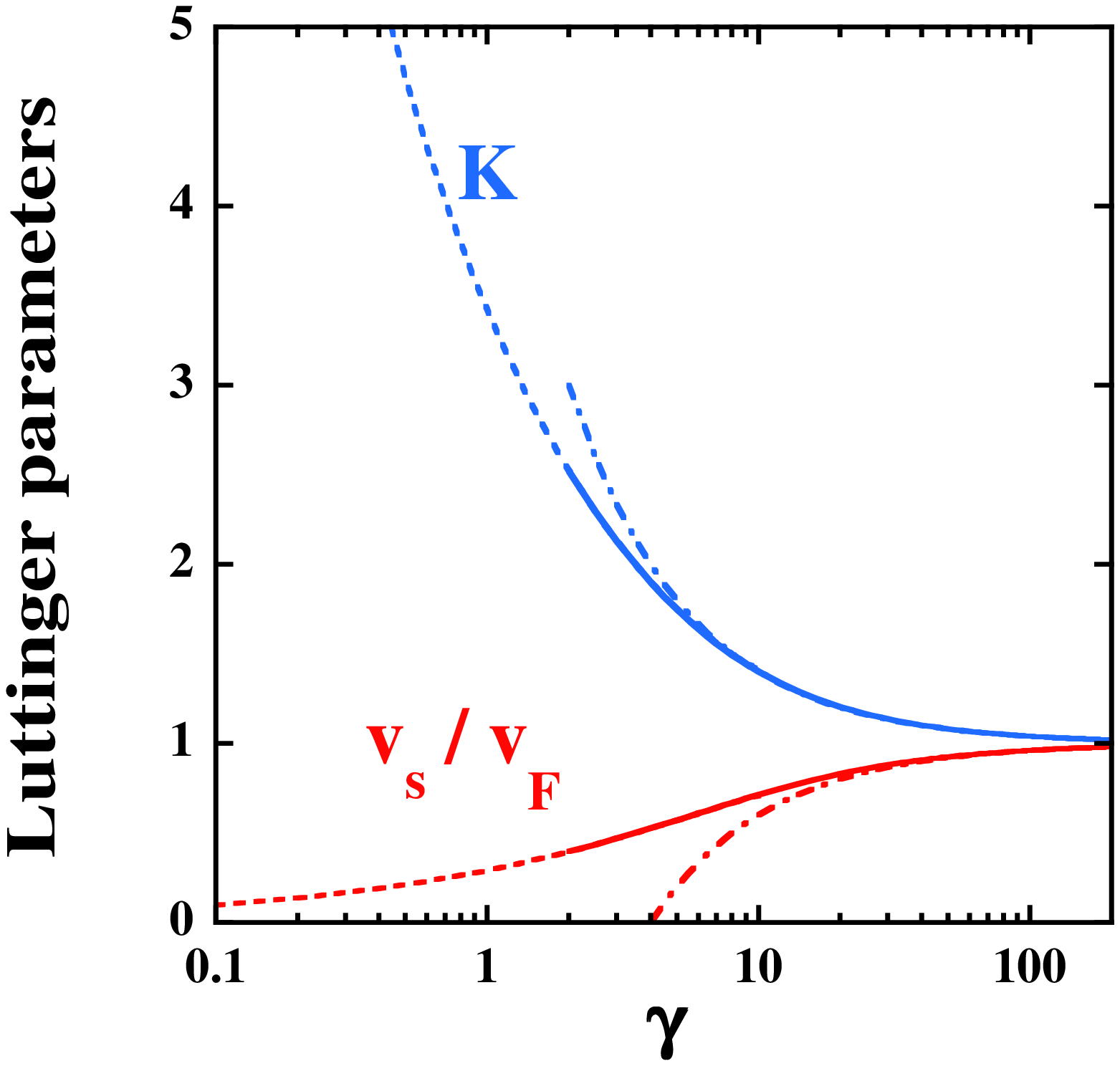}}
\caption{Luttinger-liquid parameters vs. $\gamma = M g/\hbar^2\rho_0$. The dashed lines are the 
small $\gamma$ approximations obtained from Bogoliubov theory whereas the dotted-dashed lines
correspond to the asymptotic expressions for large $\gamma$.}\label{fig10}
\end{figure}
  The model thus defined is exactly solvable. Its solution for periodic boundary 
conditions was found by Lieb and Liniger~\cite{Lieb63a,Lieb63b}, whereas the case with open boundary
conditions was solved by Gaudin~\cite{Gaudin71}. Lieb and Liniger were also able 
to compute the ground state energy per particle as well as the chemical potential and the 
sound velocity by numerically solving a set of integral equations.   One can thus obtain the phase
and density stiffness, $v_J$  and $v_N$,  by writing integral equations
which allow one to obtain these parameters. These equations 
were written down by Haldane~\cite{Haldane81c}. However, we
shall not follow this path here but instead, we shall adopt a different approach~\cite{Schulz90,Mironov91}.  
We first numerically find  the solution of the Bethe-ansatz equations~\cite{Lieb63a,Gaudin80},
\begin{equation}\label{BA}
k_{i} L = 2\pi I_i + \alpha - 2 \sum_{j=1}^{N_0} \tan^{-1}\left( \frac{k_i-k_j}{\gamma\rho_0}\right). 
\end{equation}
When deriving these equations we have assumed that the boundary conditions are not periodic but  
twisted as in (\ref{twistbc}), where $\alpha$ is the twist in the phase. In the ground state the set of  integers $\left\{I_i\right\}_{i=1,\ldots,N_0} =\left\{ - (N_0-1)/2, -(N_0-3)/2,  \ldots, (N_0-3)/2, (N_0-1)/2\right\}$~\cite{Lieb63a,Gaudin80}.  
The ground state energy for $N_0$  particles and a phase twist $\alpha$  can 
be computed from the solution to the above equations,
\begin{equation}
E_0(N_0,\alpha) = \sum_{i=1}^{N_0}\frac{\hbar^2 k^2_i}{2M}
\end{equation}
and using Eq.~(\ref{vn}) and (\ref{vj}) $v_N$ and $v_J$ can be numerically computed~\footnote{We have
checked the convergence of  the results as a function of $N_0$.}. The results of these calculations are shown in figures~\ref{fig9} and \ref{fig10}. Fig.~\ref{fig9} shows the behavior of the phase ($v_J$) and density stiffness ($v_N$) as a function of the dimensionless parameter $\gamma = M g/\hbar^2 \rho_0$. 
We have  numerically obtained $v_J$ to show explicitly that, because of the Galilean invariance of the
model, $v_J = v_F$. 

  As already pointed out by Lieb and Liniger~\cite{Lieb63a,Lieb63b}, 
the Bogoliubov approximation for the ground state energy 
yields a very accurate expression of the sound velocity for $\gamma \lesssim 10$. Hence,
\begin{equation}\label{bogol1}
v_N = v_F \left(\frac{\gamma}{\pi^2} \right) \left(1 - \frac{\sqrt{\gamma}}{2\pi}  \right).
\end{equation}
However, for $\gamma \gg 1$ one can extract an asymptotic expression for $v_N$ from 
either the expressions for $\mu$  obtained for large $\gamma$ by Lieb and Liniger~\cite{Lieb63a}
or from a strong coupling expansion of energy~\cite{Haldane81a,Cazalilla03}, which yields:
\begin{equation}\label{largegamma1}
v_N = v_F \left(1 - \frac{8}{\gamma}\right)
\end{equation}

 In Fig.~\ref{fig10} we have plotted the   parameters $v_s$ and $K$ as function of $\gamma$. They also
show good agreement with the asymptotic results at small $\gamma$,
\begin{eqnarray}\label{bogol2a}
v_s &=& v_F \: \frac{\sqrt{\gamma}}{\pi}\left( 1 - \frac{\sqrt{\gamma}}{2\pi} \right)^{1/2},\\
K &=& \frac{\pi}{\sqrt{\gamma}}\left( 1 - \frac{\sqrt{\gamma}}{2\pi} \right)^{-1/2},
\end{eqnarray}
and large $\gamma$,
\begin{eqnarray}
\label{largegamma2a}
v_s &=& v_F \left( 1 - \frac{4}{\gamma} \right),\\
\label{largegamma2b}
K &=& \left(1 + \frac{4}{\gamma}\right).
\end{eqnarray}
 From Fig.~\ref{fig10} one can see that as $\gamma$ varies from zero to infinity, $K$ varies from 
infinity to $1$. Thus, $K = 1$ corresponds to the Tonks limit, which  is also the value of $K$ for 
non-interacting fermions. Finally, in table~\ref{table1} we provide a ``translation table'' between $K$ and $\gamma$ for the  values of $K$ used in plots of  this article.
\begin{table}
\caption{\label{table1} 
Translation table $\gamma$ to $K$ for some  values of $K$ used in this paper.}
\begin{ruledtabular}
\begin{tabular}{cc}
$\gamma$ & $K$ \\ 
\hline
$+\infty$ &  $1$\\
$3.517$  & $2$\\
$0.7126$ & $4$
\end{tabular}
\end{ruledtabular}
\end{table}

\subsection{Remarks on cut-offs and prefactors}\label{cutoffs}
 
   One of the disadvantages of the harmonic-fluid approach is that it cannot provide explicit expressions
for the prefactors of the correlation functions. As obtained within this approach, they are cut-off dependent
quantities and their values depend on the particular regularization scheme used in the calculation This is a signature of {\it non-universality}, i.e. the prefactors depend on the microscopic details of the model. Thus, 
they must be fixed by other means, either by comparing with numerical results or by adopting a physically 
sensible regularization scheme (this is possible in certain models) and using rather sophisticated 
form-factor methods (see e.g.~\cite{GNT98} and references therein). Sometimes 
expressions for the prefactors can be obtained in certain limits by
working directly with the microscopic model. This is the case for the Tonks limit of delta-interacting 
bosons~\cite{Lenard72,VaidyaTracy79} or for a number of lattice models, such like the quantum-Ising 
model  and or the XY model~\cite{Lieb61,McCoy68}. 
\begin{figure}[b]
\centerline {
\includegraphics[width=10cm]{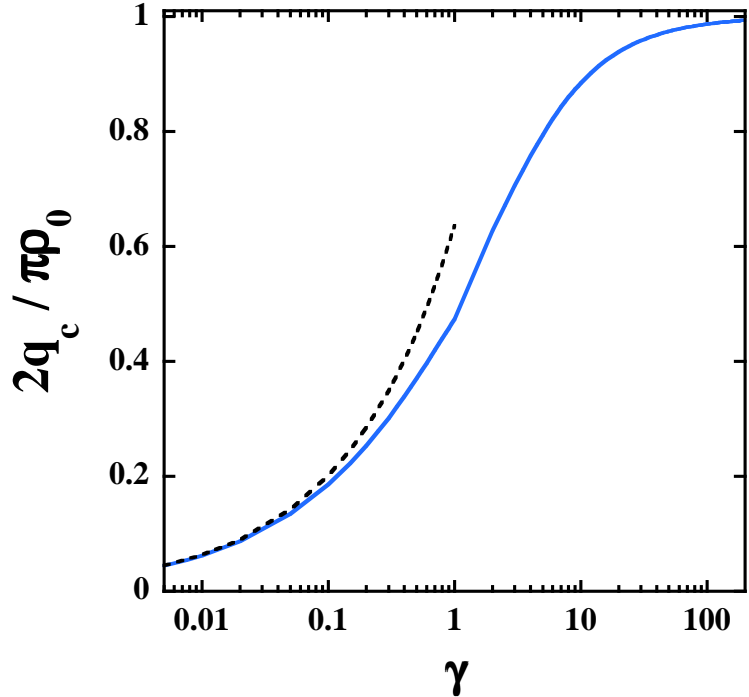}}
\caption{Large momentum cut-off vs. $\gamma$. At large $\gamma$, $q_c \to \pi \rho_0/2$, whereas
for small $\gamma$, $q_c \approx \xi^{-1} = \sqrt{\mu \rho_0}/\hbar$, i.e. the inverse healing length.
}\label{fig11}
\end{figure}

 Before jumping into the discussion of  the prefactors for the model of delta-interacting bosons, 
let us pause for a moment to consider a question that will be relevant for the fixing of the 
prefactors, namely the dependence of the wave-number cut-off $q_c = \mu/\hbar v_s$ on $\gamma$.  
As mentioned in previous sections, for delta-interacting bosons
 the different regimes are characterized by a single  dimensionless 
parameter,  $\gamma = Mg/\hbar^2 \rho_0$, where $g$ is the strength of the interaction.
For weakly interacting bosons (i.e. $\gamma \to 0$)  $q_c \to \xi^{-1} = \gamma^{1/2}\rho_0$, where 
$\xi = \hbar/\sqrt{M\mu}$ is the healing length. On the other hand, for strongly interacting bosons
($\gamma \to +\infty$), $q_c \to \pi \rho_0/2$.  The full dependence can be obtained
from the numerical solution of the Bethe-ansatz equations~(\ref{BA}).
The result for $q_c$ has been plotted in Fig.~\ref{fig11}. 

	As we have mentioned at the beginning of this section, within the harmonic-fluid approach the
prefactors depend of the cut-off, $a \lesssim q^{-1}_c$ (see Appendix~\ref{appc}).  In the weakly  interacting regime, 
Popov~\cite{Popov80,Mora02} was able to obtain the prefactor of the one-body 
density matrix . In our notation, his result reads:
\begin{equation}
g_1(x) = \langle \Psi^{\dag}(x) \Psi(0) \rangle = \rho_0 \left( \frac{e^{2-C} \xi}{4|x|} \right)^{\sqrt{\gamma}/2\pi},
\end{equation}
for $\gamma \ll 1$. In the previous expression $C = 0.57721\ldots$ is Euler's constant and $\xi$ the
healing length, $\xi = \rho^{-1}_0 \gamma^{-1/2}$. Taking into account that the numerical 
factor $e^{2-C}/4 = 1.037 \simeq 1$, we notice that the above result can be obtained within the
harmonic-fluid approach by setting the short-distance cut-off $a = \xi$. Indeed, this makes a lot
of sense since, as discussed above, 
in the weakly interacting limit one expects that $a = q^{-1}_{c} = \xi$.  
Things become more interesting when one expresses both the exponent and the prefactor
in terms of the Luttinger-liquid parameter $K$, so that one can make contact with the way $g_1(x)$ was
written in previous sections,
\begin{equation}
\langle \Psi^{\dag}(x) \Psi(0) \rangle = \rho_0 \frac{A_{\rm Popov}(K)}{|\rho_0 x|^{1/2K}}, 
\end{equation}
where we have used that, in the weakly interacting limit, 
$K = \pi \gamma^{-1/2}\gg 1$. The dimensionless 
prefactor (neglecting the difference  $e^{2-C}/4 -1 \simeq 0.037$) is
\begin{equation}\label{apopov}
A_{\rm Popov}(K) = \left( \frac{K}{\pi} \right)^{1/2K}.
\end{equation}
In view of this result, one  may be very tempted to use 
this expression for any value of $K$ relevant to this model (i.e. $1\le K < +\infty$). 
The result, when compared with
the diffusion Monte Carlo  data reported in 
Ref.~\onlinecite{Giorgini02} and exact results~\cite{Lenard72,VaidyaTracy79,Forrester03}, 
is very surprising (see Table~\ref{table2}).
It is found that Eq.~(\ref{apopov})  is accurate within less than $10\%$  over the whole range of $K$ values,
becoming essentially exact for $K\gg 1$, as expected. This fact is quite 
remarkable taking into account that from 
$K \approx 10$ to $K=1$ the prefactor changes by a almost a factor of two.

\begin{table}
\caption{Numerical prefactor of the one-body density matrix as obtained from Eq.~(\ref{apopov}) vs. 
exact and diffusion Monte Carlo (DMC) results.}\label{table2}
\begin{ruledtabular}
\begin{tabular}{ccccc}
$\gamma = Mg/\hbar^2\rho_0$ & $K$ & $K$ from the exponent (Exact/DMC) 
& $A_{\rm Popov}(K)$ & $A_{\rm Exact/DMC}$ \\
\hline\hline 
$+\infty$ & $ 1$ & 1 &$0.5642$ & $0.5214$\\
$2\times 10^3$ &  $1.002$ & $1.0$ & $0.565$ & $0.530$ \\
$6.6667$ & $1.584$ & $1.582 $& $0.806 $ & $0.760$\\
$2.0$ & $2.523$ & $2.525$ & $0.957$ & $0.951$\\   
$0.0667$ & $12.425$ & $12.469$ & $1.06$ & $1.06$\\
$ 2\times 10^{-3}$ & $70.50$ & $70.42$ & $1.02$ & $1.02$ 
\end{tabular}
\end{ruledtabular}
\end{table}
\subsection{Tips to compare with  the experiments in cold gases}\label{tips}
  
    In the previous sections we have equipped ourselves with the tools to characterize the behavior
of one-dimensional cold-atom systems.  In this characterization the Luttinger-liquid parameters $v_s$ 
and $K$ play an prominent role. The latter is especially important because it enters the exponents of the
correlation functions, and governs their algebraic decay at zero temperature. Furthermore, the value
of $K$ also tells us to which kind of instabilities the system can become unstable when perturbed. A well-known
example~\cite{Haldane81b,Buchler03} is the quantum-phase transition to a Mott-insulator in the presence
of a commensurate periodic potential, which takes place (strictly speaking at $T=0$ and $L\to \infty$) for
arbitrarily weak potentials at $K<2$. At finite temperature,
the  behavior of the system is characterized by the correlation lengths $L_T$, $L_{\phi}(T)$ and $L_{\rho}(T)$,
which are combinations of the parameters $v_s, K$ and the temperature, $T$.  As we are interested
in the bulk properties of 1D systems, we shall focus in this section on the 
implications of the correlation functions previously derived  for the experimental observations.

\begin{figure}[t]
\centerline {
\includegraphics[width=10cm]{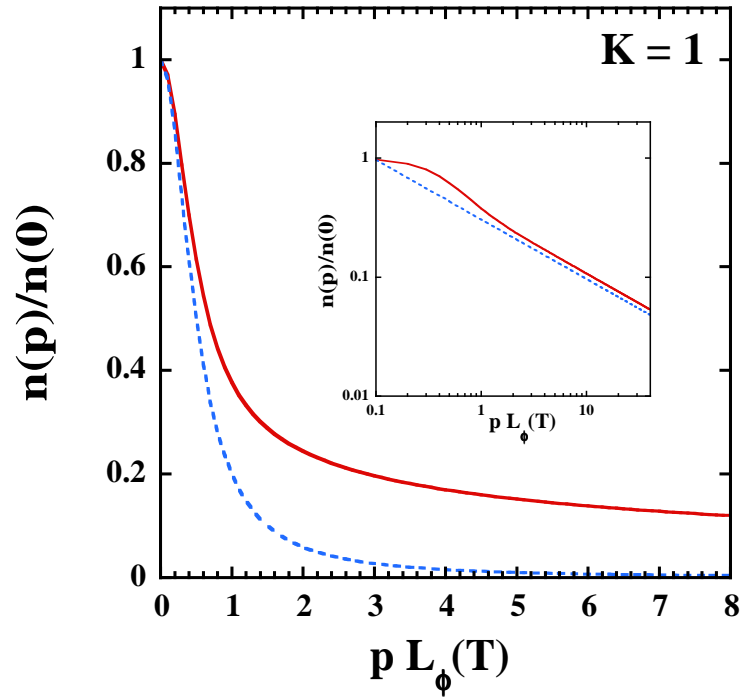}}
\caption{Momentum distribution at finite temperature for the Tonks gas ($K=1$). The inset
shows that the asymptotic behavior is not as $p^{-2}$, as expected for a lorentzian
but rather as $p^{1/2K-1} = p^{-1/2}$.}
\label{fig12}
\end{figure}

	We begin our discussion by considering the finite temperature momentum distribution. This quantity can be accessed by performing Bragg scattering measurements at large momentum transfer~\cite{Richard03,Grebier03,Luxat03}. At temperatures
$T \gg T_{\phi} = \hbar^2 \rho_0/M L$, one can safely neglect finite-size and boundary effects. In this temperature range the effects of the inhomogeneity in harmonically trapped systems can be treated using the local density approximation as described in Refs.~\onlinecite{Richard03,Grebier03}.  For a uniform system, the momentum distribution is defined as  the Fourier transform of $g_{1}(x,T)$, i.e. 
\begin{equation}
n(p,T) \simeq \int_{-\infty}^{+\infty} dx\, e^{-ipx}\: g_1(x,T)
\end{equation}
After introducing Eq.~(\ref{bosonT}) into this expression and approximating the prefactor by 
$A_{\rm Popov}(K) = (K/\pi)^{1/2K}$,  we obtain the following expression:
\begin{equation}\label{form1}
n(p,T) \simeq \left( \frac{2K}{\pi}\right)^{1/2K} \left( \frac{\rho_0 L_{\phi}(T)}{K}\right)^{1-\frac{1}{2K}}
\Gamma\left(1-\frac{1}{2K}\right) {\rm Re}\left[ 
\frac{\Gamma\left(1/4K  + i  p L_{\phi}(T)/2K\right)}
         {\Gamma\left(1-1/4K  + i  p L_{\phi}(T)/2K \right)}
 \right]
\end{equation}
where $L_{\phi}(T) = \hbar^2 \rho_0/M T = \rho_0 \Lambda^2_T$ ($\Lambda_T = \hbar/\sqrt{MT}$ is the 
de Broglie thermal wave-length) is the phase correlation length. This expression for $n(p,T)$ is different
from the lorentzian form commonly used throughout the literature. 
The lorentzian  results from taking the Fourier transform of
\begin{equation}\label{g1Texp}
g_1(x,T) \simeq \rho_0 \left(\frac{K^2}{\pi\rho_0 L_{\phi}(T)}\right)^{1/2K} e^{-|x|/2 L_{\phi}(T)},
\end{equation}
which is the asymptotic form of $g_1(x,T)$ for $|x|\gg L_T$. Hence,
\begin{figure}[t]
\centerline {
\includegraphics[width=10cm]{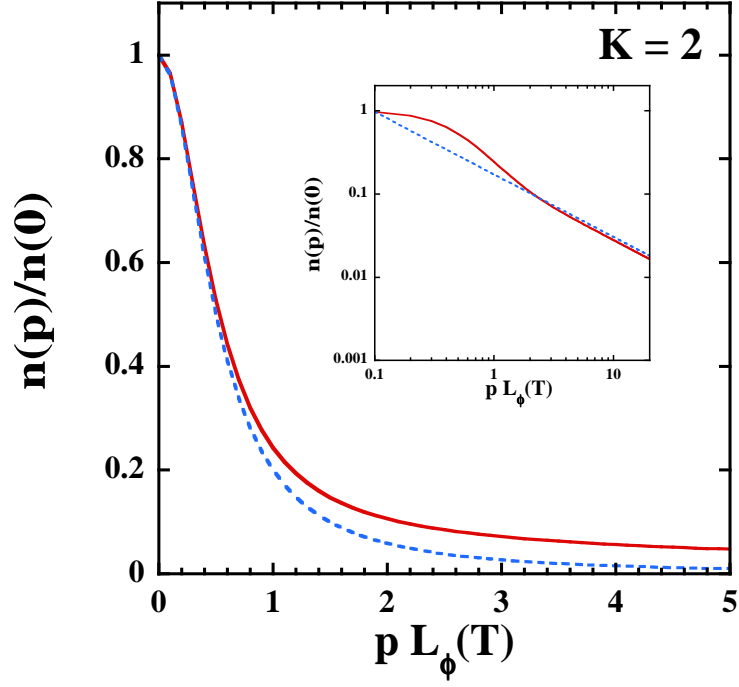}}
\caption{Momentum distribution at finite temperature for a system with $K=2$. The inset
shows that the asymptotic behavior is not as $p^{-2}$, but rather as $p^{1/2K-1} = p^{-3/4}$.}
\label{fig13}
\end{figure}
\begin{equation}\label{form2}
n'(p,T) \simeq 4 \left( \frac{K^2}{\pi} \right)^{\frac{1}{2K}} \left(\rho_0 L_{\phi}(T)  
\right)^{1-1/2K} \frac{1}{1+\left(2 p L_{\phi}(T)\right)^2}.
\end{equation}
In Figs.~\ref{fig12}, \ref{fig13}, and \ref{fig14} we compare both forms, Eq.(\ref{form1}) and (\ref{form2});
$n(p,T)/n(0,T)$ and $n'(p,T)/n'(0,T)$ are plotted against $p L_{\phi}(T) $. 
Our calculations assume  the scaling limit, which in these plots 
means that $|p| \lesssim q_c$. For $|p| \gg q_c$, Olshanii and Dunjko 
have  shown~\cite{Olshanii02} that $n(p,T) \sim p^{-4}$. From figures~\ref{fig12}, \ref{fig13}, and \ref{fig14} one can draw the conclusion that the lorentzian is a good approximation for large $K$, i.e. for weakly interacting bosons. The width  of the lorentzian is 
proportional to $L_{\phi}(T) \propto \rho_0/T$, which thus provides a measure of this ratio. However, 
in the Tonks limit the momentum distribution at  $T > T_{\phi}$  looks more like a stretched lorentzian.
The reason is that, in the large momentum limit, $n(p,T)$ as given by Eq.~(\ref{g1Texp}), behaves
as a power-law: $p^{-(1-1/2K)}\sim p^{-1/2}$, for $K=1$. Experimentally, extracting the Luttinger-liquid parameter from $n(p,T)$ may be hard because the true $n(p,T)$ may not display this algebraic 
behavior in a sufficiently wide range of $p$ (set aside the complications that averaging $g_1$ using the local
density approximation may introduce). However, there may be some chance to observe  the more qualitative behavior shown by these plots: as the system's parameters are tuned 
(e.g. by using a Feshbach or the confinement-induced resonance~\cite{Olshanii98}) 
towards the Tonks limit, the momentum distribution should become more ``stretched'' and less Lorentzian-like.
\begin{figure}[t]
\centerline {
\includegraphics[width=10cm]{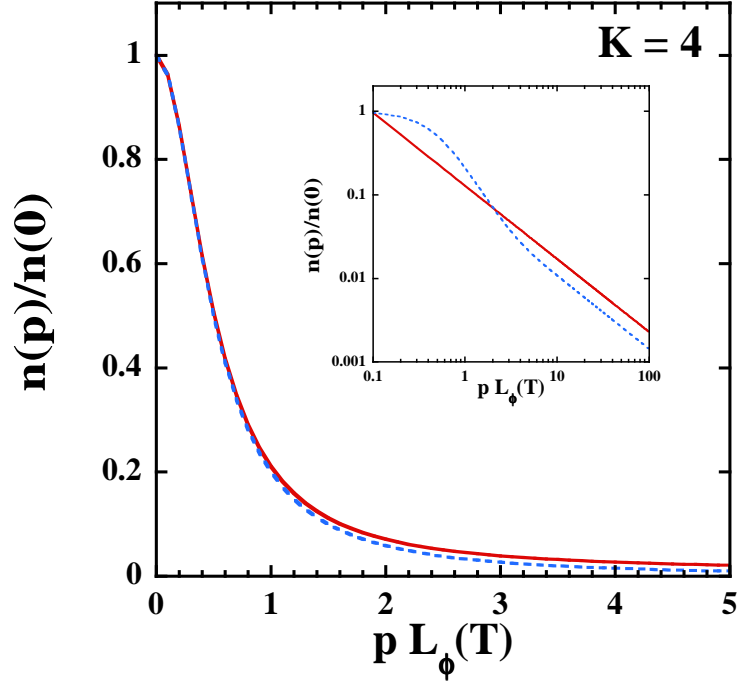}}
\caption{Momentum distribution at finite temperature for a system with $K=4$. The inset
shows that the asymptotic behavior is not as $p^{-2}$,  but rather as $p^{1/2K-1} = p^{-7/8}$.}
\label{fig14}
\end{figure}

  A more quantitative estimate of the Luttinger-liquid parameters near the center of the trap
can be obtained from the dynamic density response function, 
\begin{eqnarray}
\chi(q,\omega) &=& \int dx \: dt\, e^{i\omega t - i q x}\: \chi(x,t),\\ 
\chi(x,t) &=&-\frac{i}{\hbar} \theta(t) \: \langle \left[\rho(x,t),\rho(0,0)\right] \rangle_T,
\end{eqnarray}
whose imaginary part is measured in low-momentum Bragg scattering experiments~\cite{Zambelli00}. 
In the low momentum ($=$ long wave-length)
limit, we can replace $\rho(x,t)$ by $\Pi(x,t) = \partial_x\theta(x,t)/\pi$. Thus, using the
equations of motion for $\Pi(x,t)$, one can obtain the result ($\eta \to 0^{+}$):
\begin{equation}
\chi(q,\omega) = -\frac{q^2}{\pi \hbar} \frac{v_J}{(\omega + i\eta)^2  - \omega^2(q)},
\end{equation}
where $\omega(q) = v_s |q|$ for $|q| \ll q_c$. Hence,
\begin{equation}
{\rm Im}\: \chi(q,\omega) = \frac{v_J q^2}{2 \hbar \omega(q)} 
\left[\delta\left(\omega - \omega(q)\right)   - \delta\left(\omega + \omega(q)\right) \right].
\end{equation}
Since  $v_J = v_s K$, the weight becomes $K q/2$. Thus by measuring the dispersion $\omega(q)$ 
one can obtain $v_s$, whereas from the weight of ${\rm Im} \chi(q,\omega)$,
$K$ can be measured. Inhomogeneity effects can be treated
also in this case within the local density 
approximation, as discussed in Ref.~\onlinecite{Zambelli00}. In this respect,
we notice that ${\rm Im} \: \chi(q,\omega)$ has the same structure as  the result from 
Bogoliubov approximation in $d>1$. The only difference 
is that it holds, as well as the expression for $n(p,T)$ derived above, for {\it all values}  
of the interaction parameter $\gamma$, and not only 
in the weakly interacting limit.

\section{Jastrow-Bijl wave functions and Luttinger liquids}\label{WFS}

  In Sect.~\ref{sect1} it has been shown that the low-energy spectrum of a Luttinger
liquid is completely exhausted by phonons with linear dispersion. This phonons are nothing
but collective oscillations of the density and the phase. In view of  this fact, one may wonder what
kind of correlations are built into the ground state wave  function by these collective excitations. 
The question has been considered in different contexts 
by several authors~\cite{ReattoChester67,Pham9901}. Here we restrict ourselves to the
case of particles in  a box (OBC's).  Taking inspiration from the 
work of Reatto and Chester~\cite{ReattoChester67}
we write the ground state wave function for $N_0$ bosons in a box as  follows:
\begin{equation}
\Phi_0(x_1,\ldots,x_N) = \Phi_c(x_1,\ldots,x_N) \tilde{\Phi}_0(x_1,\ldots,x_N).
\end{equation}
where $\Phi_c(x_1,\ldots,x_N)$ describes the correlations between the particles in the ground
state whereas $\tilde{\Phi}_0(x_1,\ldots,x_N)$ describes their independent motion, which for $N_0$
bosons in a box is given by
\begin{equation}
\tilde{\Phi}_0(x_1,\ldots,x_{N_0}) = \prod_{i=1}^{N_0} \sin\left(\frac{\pi x_i}{L}  \right),
\end{equation}
that is, all the $N_0$ lie in the lowest-energy orbital $\sin(\pi x/L)$. As for $\Phi_c(x_1,\ldots,x_{N_0})$ we
will obtain an asymptotic expression valid for $|x_i-x_j| \gg a$, for $i,j = 1,\ldots, N_0$. In this limit the correlations
between particles are dominated by the zero-point motion of the collective excitations.

 In order to obtain the asymptotic behavior of $\Phi_c$,  we 
consider the ground-state wave function of the 
low-energy Hamiltonian (in this section $\hat{A}$ denotes the operator
of the corresponding classical variable $A$), 
\begin{equation}
\hat{H}_{\rm eff} = \frac{\hbar v_s}{2} \int_{0}^{L} dx\:
\left[ \frac{\pi}{K} \hat{\Pi}^2(x) + \frac{K}{\pi} \left( \partial_x \hat{\phi}(x) 
\right)^2  \right].
\end{equation}
For OBC's, it is convenient to use the following expansions:
\begin{eqnarray}
\hat{\Pi}(x) &=& \hat{\Pi}_{0} + \sqrt{\frac{2}{L}} \sum_{q > 0} \hat{\Pi}_q\, \cos(q x) , \\ 
\hat{\phi}(x) &=& \hat{\phi}_{0} + \sqrt{\frac{2}{L}} \sum_{q > 0}  \hat{\phi}_q\, \cos(q x),
\end{eqnarray} 
where $q = n\pi/L$,  with $n$ a positive integer and $\hat{\Pi}_0 = \hat{N}-N_0$ .
The canonical commutation relations now read
\begin{eqnarray}
\left[ \hat{\Pi}_{0}, \hat{\phi}_{0}\right] &=& i,\\
\left[\hat{\Pi}_q, \hat{\phi}_{q'} \right] &=& i \, \delta_{q, q'}.
\end{eqnarray}
 We  shall work in first quantization, which means that 
the wave function $\Phi_c$ is regarded  as functional of $\Pi(x)$.  
From the second of the above commutation rules
\begin{equation}
\hat{\phi}_{q} = -i \frac{\partial}{\partial \Pi_q}.
\end{equation}
Therefore, when expressed in terms of  Fourier modes the
Hamiltonian takes the form
\begin{equation}
\hat{H}_{\rm eff} = \frac{\hbar\pi v_s}{2 K} 
\sum_{q > 0} \left[ \hat{\Pi}^2_{q} + \left(\frac{qK}{\pi} \right)^2 \hat{\phi}^2_q \right]
= \frac{\hbar\pi v_s}{2 K} 
\sum_{q > 0} \left[ \Pi^2_{q} - \left(\frac{qK}{\pi}\right)^2 
\frac{\partial}{\partial \Pi^2_q} \right],
\end{equation}
i.e. it represents a collection of  decoupled harmonic oscillators, and 
therefore its ground state is just a product of gaussians:
\begin{equation}
\Phi_{c} = \exp \left\{-\frac{\pi}{2K} 
\sum_{q>0}{\frac{\Pi^2_q}{q}}\right\}. 
\end{equation}
When expressed in terms of $\Pi(x)$,  this function becomes:
\begin{equation}\label{fuctional}
\Phi_{c} = \exp \left\{\frac{1}{2} \int_{0}^{L}
 dx \int_{0}^{L} dx' \: \Pi(x)  {\cal K}(x,x') \Pi(x') \right\},
\end{equation}
where 
\begin{equation}
{\cal K}(x,x') =  -\frac{2\pi}{LK} \sum_{q>0}  
\frac{e^{-a q}}{q}\cos(qx) \cos(q x')
\simeq \frac{1}{K} \ln\left| 4 \sin\left[\frac{\pi(x+x')}{2L} \right]
   \sin\left[\frac{\pi(x-x')}{2L} \right] \right|.
\end{equation}
Notice that $K(x,x')$ depends on the short-distance cut-off  $a$
and therefore, it cannot correctly describe short-distance correlations where
$x\approx x'$.  The second expression above is thus only asymptotically
correct. 

  As we discussed in Sect.~\ref{sect1}, the configurations of $\Pi(x)$ describe 
long wave-length density fluctuations, i.e. the ``slow'' part of the density. Thus, we are
quite tempted to make the replacement $\Pi(x) \to \rho(x) = \sum_{i=1}^{N_0}\delta(x_i-x_j)$.
However, in doing this we must be careful enough to remove the terms that involve the function
$K(x,x')$ evaluated at $x=x'$ since $K(x,x')$ cannot describe short-distance correlations
(In Ref.~\cite{Cazalilla02} we failed to notice this, and as a result the wave function obtained
there is not completely correct. See below for the correct expression). If we do so, the following
function is obtained in terms of  particle coordinates:
\begin{equation}\label{phiobc}  
\Phi_{c}(x_1,\ldots, x_{N_{0}}) =  \prod_{i<j} \left| \sin\left[\frac{\pi(x_i+x_j)}{2L}\right] 
\sin\left[\frac{\pi(x_i-x_j)}{ 2L}\right] \right|^{\frac{1}{K}} = \frac{1}{2^{N_0(N_0-1)}}
\prod_{i<j} \left|\cos\left(\frac{\pi x_i}{L} \right) - \cos\left( \frac{\pi x_j}{L} \right)  \right|^{1/K},
 \end{equation}
 Hence, the complete ground-state wave function reads
 \begin{equation}\label{jastrow}
 \Phi_0(x_1,\ldots,x_{N_0}) = {\cal N}\prod_{i<j} \left|\cos\left(\frac{\pi x_i}{L} \right) - \cos\left( \frac{\pi x_j}{L} \right)  \right|^{1/K} \prod_{i=1}^{N_0} \sin\left( \frac{\pi x_i}{L}  \right)
\end{equation}
 where $\cal N$ is the normalization constant. 
 
 There are several reasons to believe that the wave function~(\ref{jastrow}) is, at least
asymptotically, correct. First, in the non-interacting limit $K\to +\infty$ (i.e. $\gamma \to 0$ for delta-interacting
bosons), $\Phi_c  \to 1$, and one recovers the independent-particle ground state $\Phi_0=\tilde{\Phi}_0$. On the 
other hand, in the Tonks limit $K = 1$, $\Phi_0$ is the {\it  exact} ground state of a system of hard-core
bosons in a box~\cite{Forrester03}. Furthermore, if one repeats the above calculation for bosons in a ring, the
result is the  well-know Jastrow-Bijl function:
\begin{equation}\label{phipbc}
\Phi^{\rm ring}_{0}(x,\ldots, x_{N_{0}}) = {\cal N}\,
\prod_{i<j} \left|\sin \left[  \frac{\pi(x_i-x_j)}{L} \right] \right|^{\frac{1}{K}}
\end{equation}
It turns out that this is the exact ground state wave function of the Calogero-Sutherland
model~\cite{Sutherland98}, which is a model of hard-core bosons with long-range interactions, and
whose Hamiltonian reads\footnote{The Calogero-Sutherland model is also known to be a Luttinger
liquid. See N. Kawakami and S.-K. Yang, Phys.~Rev.~Lett.~{\bf 67}, 2493 (1991).}:
\begin{equation}
H_{CS} = -\frac{\hbar^2}{2M} \sum_{i=1}^{N_{0}} 
\frac{\partial^2}{\partial x^2_i} + \frac{\hbar^2}{M} \sum_{i<j} \frac{K^{-1}  
(K^{-1}-1)}{\left[d(x_i-x_j|L)\right]^2}.
\end{equation}
It is also known that this model also has a solution under harmonic confinement~\cite{Sutherland98}, i.e. for
\begin{equation}
H^{\prime}_{\rm CS} =  -\frac{\hbar^2}{2M} \sum_{i=1}^{N_{0}} 
\frac{\partial^2}{\partial x^2_i} + \frac{M\omega^2}{2}
\sum_{i=1}^{N_{0}} x^2_{i} + \frac{\hbar^2}{M}
\sum_{i<j} \frac{ K^{-1} (K^{-1}-1)}{(x_i-x_j)^2}.
\end{equation}
The exact ground-state wave function  has a similar structure to Eq.~(\ref{jastrow}):
\begin{equation}\label{hcswf}
\Phi_{0}(x,\ldots,x_{N_{0}}) = 
\prod_{i<j} |x_i-x_j|^{1/K} \, \prod_{i=1}^{N_{0}} 
e^{-M \omega x^2_i/2\hbar}.
\end{equation}
There is also a good chance that the latter wave function may be asymptotically correct for
other models of interacting bosons in a harmonic trap.

\section{Conclusions}

   In this paper we have addressed the properties of one-dimensional systems of cold atoms using the
harmonic-fluid approach (also know as ``bosonization''). Besides reviewing the method in pedagogical 
detail, we have argued that it allows to treat boson and fermion systems both in strongly and weakly
interacting limits. We have also shown how concepts and results obtained using the Bogoliubov-Popov approach~\cite{Popov,Popov80} and its modifications~\cite{Andersen02,Mora02} can be naturally 
recovered using bosonization, which in our opinion is much simpler conceptually.  When combined with the conformal field theory methods explained in the appendices, it becomes a very powerful tool to obtain  the functional forms of the correlation functions for various geometries (ring, box with Dirichlet BC's), 
and also  finite-temperature correlations.  The method has 
some limitations, however, as it cannot provide explicit expressions for the prefactors of the correlation functions, which turn out to be model-dependent. 
Moreover, for strongly interacting systems it becomes difficult, without further input, to relate the phenomenological parameters $v_J$ and  $v_N$ to the microscopic parameters of the model at hand. Nevertheless, for a relevant model of bosonic cold atoms, we have shown
that one can successfully extract those parameters from the exact (Bethe-ansatz) solution. Furthermore, the prefactor of the one-body density matrix could be fixed with less than $10\%$ error 
by expressing a result obtained in the weakly interacting limit~\cite{Popov80} in terms of $K =Ê\sqrt{v_J/v_N}$.
These results should allow for a quantitative comparison with the experiment. In this respect, we have 
discussed how to extract the Luttinger-liquid parameters $K$ and $v_s = \sqrt{v_Nv_J}$ (i.e. the sound
velocity) from  Bragg-scattering measurements of the density response function and finite-temperature
momentum distribution.  For the latter quantity, we have argued that, as a system is tuned into the 
Tonks regime, the finite-$T$ momentum distribution should
become more stretched, as compared to the lorentzian form exhibited by  $n(p,T)$ in the 
weakly interacting limit. Looking forward to experiments where these predictions can be tested, 
we hope that the present work will foster the use of the harmonic-fluid approach in the 
study of 1D cold-atom systems.

\acknowledgments

   This work was started while the author was a postdoctoral fellow in the condensed matter group of the 
Abdus Salam International Centre for Theoretical Physics (ICTP), in Trieste (Italy). I would like to acknowledge
the hospitality  of the center as well as 
the stimulating atmosphere that I found there during the initial stages of this work. 
I am particularly indebted to A. Nersesyan for many valuable 
discussions and for sharing many of his deep insights  into bosonization with me. I am also grateful to 
M. Fabrizio, A. Ho,  V. Kravtsov, Yu Lu, and E. Tosatti for useful conversations and remarks. Part of this work was
stimulated by discussions with T. Giamarchi during a visit to the   
\'Ecole de Physique in Geneva (Switzerland). I thank T. Giamarchi for his 
hospitality   as well as for many illuminating conversations and explanations.   
Fruitful discussions with I. Carusotto,  T. Esslinger, D. Gangardt, C. Kollath, E. Orignac, M. Olshanii,
A. Recati, L. Santos,  G. Shlyapnikov,  H. Stoof,  S. Stringari, D.S. Weiss, and W. Zwerger   
are also  acknowledged. I am also grateful to 
S. Giorgini for useful discussions and for kindly providing me with the fits to his Monte Carlo data. 
This paper was  brought to its present form during a stay at the Aspen 
Center for Physics (Aspen, USA), whose hospitality is also gratefully acknowledged. This research
has been supported through a  Gipuzkoa fellowship granted by  Gipuzkoako Foru Aldundia (Basque Country).

\appendix
\section{Commutation relations of $\Pi(x)$ and $\phi(x)$}\label{appa}

 In the main text we have often used the  commutation relation:
\begin{equation}\label{ap1}
\left[ \Pi(x), \phi(x') \right] = i \delta(x - x'),
\end{equation}
which states that $\Pi(x)$ and the phase $\phi(x)$ are canonically conjugated
in a low energy subspace. We now proceed to give a heuristic derivation of it (a different 
argumentation can be found in Appendix~\ref{appb} in terms of path integrals).
Our starting point will be the following commutation relation between the 
density and {\it momentum} density operators:
\begin{equation}\label{dj}
\left[ \rho(x), j_p(x') \right] = i\hbar  \partial_{x'} \delta(x-x') \rho(x'). 
\end{equation}
This  result can be most easily derived by working in first quantization, where
\begin{eqnarray}
\rho(x) &=& \sum_{i=1}^{N} \delta(x-x_i), \\
j_p(x) &=& \frac{1}{2}\sum_{i=1}^{N} \big[ p_i\delta(x-x_i) + \delta(x-x_i) p_i  \big].   
\end{eqnarray}
In the previous expressions $x_i$  and $p_i = -i\hbar \partial_{x_i}$ stand for  
the position and momentum operators, 
respectively. Next, we shall consider long  wave-length  fluctuations of the density and the current. 
Thus we set $\rho(x) \approx \rho_{0} + \Pi(x)$ and, upon linearizing,
\begin{equation}
j_p(x) = \frac{\hbar}{i} \left[ \Psi^{\dag}(x) \partial_x \Psi(x) - \partial_x \Psi^{\dag}(x) \Psi(x)\right] 
\approx \hbar \rho_{0} \partial_x \phi(x),
\end{equation}
we arrive at
\begin{equation}\label{dcr}
\left[ \Pi(x), \partial_{x'} \phi(x') \right] = i \partial_{x'}\delta(x-x').
\end{equation}
By integrating this expression over $x'$, it reduces to
\begin{equation}
\left[ \Pi(x), \phi(x') \right] = i \delta(x-x') + C.
\end{equation}
The integration constant $C$ must be zero since from this
commutation relation  it must follow that
\begin{equation}
[N, e^{-i\phi(x)} ] = e^{-i\phi(x)}, 
\end{equation}
which  follows from the fact that the field operator $\Psi^{\dag}(x)$ 
adds particles to the system. Thus we have provided 
a justification for Eq.~(\ref{ap1}). Alternatively, one could have written 
Eq.~(\ref{dcr}) as
\begin{equation}
\left[\partial_x \Theta(x), \partial_{x'}\phi(x') \right] 
= i\pi \partial_{x'} \delta(x-x') = -i\pi \partial_x \delta(x-x'),
\end{equation}
which upon integration over $x$ becomes:
\begin{equation}
\left[ \Theta(x), \partial_{x'} \phi(x') \right] = -i\pi \delta(x-x') + C'
\end{equation}
To show that the constant $C' = 0$ one has to work a little bit more in
this case. For open boundary conditions one has to work out this 
result for $0 < x, x' <L$  by using the mode expansions, Eqs.~(\ref{modesobc1},\ref{modesobc2}), 
and then formally taking the limit $a \to 0^+$.  For periodic boundary 
conditions, however, this is required for the  expression:
\begin{equation}
\left[J,\theta_{0}  \right] = i
\end{equation}
to hold. Thus,  after setting $C' = 0$ and defining the canonical
momentum $\Pi_{\phi}(x) = \partial_x \phi(x)/\pi$, we can write the above
commutation relation as:
\begin{equation}
\left[\Pi_{\phi}(x), \Theta(x')  \right] = i\delta(x-x').
\end{equation}
This is just a different representation of the {\it duality} of the fields $\Theta(x)$ and $\phi(x)$,
which provide two complementary descriptions of the low-energy
physics.  Finally, we urge those readers unhappy with 
the rather non-rigorous treatment of operators in this appendix and 
Sect.~\ref{sect1}  to consult appendix~\ref{appb} to be reassured of the
correctness of the results.

\section{Path Integral Formulation}\label{appb}

 Our goal in this appendix is to  present a somewhat different, but at the same time
complementary derivation of some aspects of the harmonic-fluid  approach. To this purpose, we will
employ the path integral formalism. We also compute the low-temperature limit of the
partition function to show that the spectrum described by this formulation has the same structure
and degeneracies as the one obtained  from the Hamiltonian given in Eq.~(\ref{modhampbc})
of Sect.~\ref{modespbc}.

 As explained in  e.g.  Ref.~\onlinecite{Negele},  the partition function  of a bosonic
system  in the grand canonical ensemble, $Z$  can be written as a coherent-state path integral:
\begin{equation}
Z = \int\: [d\psi^* d\psi] \: e^{-S[\psi^*,\psi]}.
\end{equation}
The functional $S[\psi^*,\psi]$ is the (euclidean) action, and for
the  the Hamiltonian in Eq.~(\ref{eq2.1}) has the following form
\begin{eqnarray}\label{appa12}
S[\psi^*,\psi] =  \int_{0}^{\hbar \beta} \frac{d\tau}{\hbar} \int_{0}^{L} dx \: 
\Big[ \hbar \psi^*(x,\tau) \partial_{\tau} \psi(x,\tau) - \mu \: 
\psi^{*}(x,\tau) \psi(x,\tau) + \frac{\hbar^2}{2m} |\partial_x \psi(x,\tau)|^2\nonumber \\
+ \frac{1}{2} \int_{0}^{L} dx' \: v(x-x«) \psi^{*}(x,\tau) \psi^{*}(x',\tau) \psi(x',\tau) \psi(x,\tau) \Big].
\end{eqnarray}
Using ``polar coordinates", $\psi(x,\tau) = \sqrt{\rho(x,\tau)} e^{i\phi(x,\tau)}$ and 
$\psi^{*}(x,\tau) = \sqrt{\rho(x,\tau)} e^{-i\phi(x,\tau)}$, where 
$\rho(x,\tau)$ and $\phi(x,\tau)$ are  {\it real functions} (i.e. not operators), the action 
becomes:
\begin{eqnarray}\label{appa13}
S &=&  \int_{0}^{\hbar \beta} \frac{d\tau}{\hbar}
 \int_{0}^{L} dx \: \Big[ i \hbar \rho(x,\tau) \partial_{\tau} \phi(x,\tau)  
+ \frac{\hbar^2}{2m} \rho(x,\tau) \left( \partial_x \phi(x,\tau) \right)^2 
+ \frac{\hbar}{2}\partial_{\tau}\rho(x,\tau)  \nonumber \\ 
& & \,\,\,\,\,\  + \frac{\hbar^2}{8m} \left(\frac{\partial_x \rho(x,\tau)}{\rho(x,\tau)}\right)^2
 - \mu \: \rho(x,\tau)+    \frac{1}{2}
\int_{0}^{L} dx' \:  \rho(x,\tau)   v(x-x')\rho(x',\tau) \Big].
\end{eqnarray}
Next we proceed to give a more explicit meaning to the coarse-graining
procedure used in Sect.~\ref{sect1}.  As explained there, we first split:
\begin{eqnarray}
\rho(x,\tau) &=& \rho_<(x,\tau) + \rho_>(x,\tau),\\
\phi(x,\tau) &=& \phi_<(x,\tau) + \phi_>(x,\tau),
\end{eqnarray}
where $\rho_>(x,\tau)$ and $\phi_>(x,\tau)$ describe the fast modes, i.e. those with momenta
higher than $a^{-1} = \min\{R^{-1}, q_c\}$, and frequencies higher than $\omega_c \sim \mu$. 
The fields $\rho_<(x,\tau), \phi_<(x,\tau)$ describe the slow modes. To
give a more mathematically precise definition, we can define the slow part 
of a given field, $h(x,\tau) = \rho(x,\tau), \phi(x,\tau)$  as
\begin{equation}
h_<(x,\tau) = \int dx' d\tau' \: f(x-x',\tau-\tau') h(x',\tau'), 
\end{equation}
where $f(x,\tau)$ is a slowly varying function over distances $\sim a$ and imaginary times
 $\sim \hbar/\omega_c$. The low-temperature description of the system is obtained 
by coarse-graining the action,  i.e. by integrating out the fast modes. Thus 
we define the effective-low energy action by
\begin{equation}\label{coarse-grain}
e^{-S_{\rm eff}[\Theta,\phi]} = \int [d\rho_> d\phi_>] \, e^{-S[\rho_>,  \phi_>, \Theta, \phi]},
\end{equation}
where, just as we did in Sect.~\ref{sect1}, we have parametrized the slow modes by 
$\phi \equiv \phi_<(x,\tau)$ and $\Theta(x,\tau)$, such that $\rho_<(x,\tau) = \partial_x\Theta(x,\tau)/\pi$.
In the general case, performing the functional integral involved in Eq.~(\ref{coarse-grain}) it is not feasible.
However, on physical grounds and from the structure of the microscopic action, Eq.~(\ref{appa13}), one can
guess the following form~\footnote{Ultimately, the choice of the terms in $S_{\rm eff}$ can only be 
justifed on the basis of a renormalization-group analysis of the problem. For a single-component
fluid on the continuum, quadratic terms in the derivatives of $\Theta$ and $\phi$ suffice
to describe the low-energy spectrum (excluding damping phenomena).}:
\begin{eqnarray}\label{appa14}
 S_{\rm eff}[\Theta,\phi] =   \int_{0}^{\hbar \beta} d\tau \int_{0}^{L} dx \: 
\Big[  \frac{i}{\pi} \partial_x \Theta(x,\tau) \: \partial_{\tau}  \phi(x,\tau) 
+  \frac{v_{s} K}{2\pi} \left( \partial_x \phi(x,\tau) \right)^2  + \frac{v_s}{2\pi K}  
\left( \partial_x \Theta(x,\tau)  - \pi \rho_{0} \right)^2  \Big].
\end{eqnarray}
For weakly interacting systems one can perform the coarse-graining perturbatively , and to the lowest
order this amounts to  keeping only the
quadratic terms in the gradients of $\Theta$ and $\phi$. Thus, for instance, for
delta-interacting bosons one has $v_J = v_s K = \hbar \pi \rho_0/M = v_F$, as discussed in Sect.~\ref{sect1},
and $v_N = v_s/K = g/\hbar \pi + {\rm O}(g^2)$. For this model, one can also show, using 
the effective fermionic Hamiltonian reported in Ref.~\cite{Sen03,Cazalilla03}, that the action has the form 
(\ref{appa14}), with $v_J = v_F$ and $v_N = v_F\left(1- 8 \gamma^{-1} + {\rm O}(\gamma^{-2})\right)$.
Such a derivation will be given elsewhere~\cite{unpub} (the result for $v_N$ can be also obtained from the
expressions given by Lieb and Liniger for the chemical potential at large $\gamma$).

 Note that the  imaginary  term (i.e. the Berry phase) $i \hbar \partial_x\Theta \: \partial_{\tau}\phi/\pi$ 
indicates that $\partial_x\Theta/\pi$ (and hence $\Pi(x,\tau)$) is canonically conjugated to $\phi$, 
just as in the original action the Berry phase reflected that $\psi$ and $\psi^{*}$
are canonically conjugated fields (see e.g. Ref.~\onlinecite{Negele}).

  For bosons, the path integral must be performed over configurations that obey 
$\psi(x,\tau+\hbar \beta) = \psi(x,\tau)$ besides the periodic boundary conditions,
 $\psi(x+L,\tau) = \psi(x,\tau)$.  Hence, from the expressions for the field operators in
 terms of $\Theta$ and $\phi$, these fields must obey the following boundary conditions:
\begin{eqnarray}
\Theta(x + m L, \tau + n\hbar \beta) &=& \Theta(x,\tau) + m \pi N + n \pi P,\\
\phi(x+ m L,\tau + n \hbar \beta) &=& \phi(x,\tau) + m \pi J +  n \pi Q,
\end{eqnarray}
where $m,n$ are arbitrary integers, and $(-1)^{J} = (-1)^Q = +1$.

After obtaining the low-energy effective action, $S_{\rm eff}$, we notice that it describes
a quadratic theory. Therefore, we  have two choices: 
to  integrate out $\Theta$ or  to integrate  out $\phi$.  This yields two apparently different
representations of the theory.  If we integrate out $\phi$ and introduce $\theta = \Theta - \pi\rho_0 x$
we obtain 
\begin{equation}\label{appa15}
S_{\rm eff}[\theta] = \frac{1}{2\pi K} \int_{0}^{\hbar \beta} d\tau \int_{0}^{L} dx \: 
\Big[ \frac{1}{v_s} \left(\partial_{\tau}\theta\right)^2 +  v_s \left(\partial_x \theta\right)^2\Big].
\end{equation}
Had  we integrated out $\Theta$ instead,  
\begin{equation}\label{appa16}
S_{\rm eff}[\phi] =  \frac{K}{2\pi} \int_{0}^{\hbar \beta} d\tau \int_{0}^{L} dx \: 
\Big[ \frac{1}{v_s} \left( \partial_{\tau}\phi \right)^2 + 
 v_s \left(\partial_x \phi \right)^2\Big] + S_B.
\end{equation}
where $S_{B} = i \rho_{0} \int^{L}_{0} dx \: \left[\phi(x,\hbar \beta) - \phi(x,0) \right]$. This 
is a ``surface'' term that can be in principle dropped. However, we shall retain it in the computation of the
partition function that  follows because we are working with a finite-size system.  Notice  that
both representations, Eqs.~(\ref{appa15}) and~(\ref{appa16}), are {\it dual} to each other, in the sense
that one can be obtained from the other by  means of the replacements $\theta \leftrightarrow \phi$
and $K^{-1} \leftrightarrow K$. This property has important consequences, in particular the
correlation functions of  one field alone can be obtained from the corresponding correlation function 
of the other field alone by means of the replacement $ K \to K^{-1}$. Another important feature of these 
representations  is that they show that the role played by $K$ is that of an effective ''temperature''  or 
Planck's constant: When  $K$ is large, the density field, $\theta$,  fluctuates wildly 
whereas the phase field, $\phi$,  behaves almost classically. For instance, this is what happens in a 
system of weakly interacting bosons (a ``quasi-condensate'' at finite $T$, which resembles more
and more a BEC as the temperature is lowered). On the other hand, 
when $K$ is small it is the phase that fluctuates more violently whereas the density 
behaves more classically. This corresponds to a system whose behavior approaches crystallization.
However, in both cases no symmetry breaking can take place in one-dimension, as both phase 
transitions (BEC and  crystallization) break continuous symmetries. At most, as it is emphasized
in the main text,  one gets quasi-long range order.

    We finally compute the partition function using the representation of the action in terms of the
phase field, Eq.~(\ref{appa16}). Upon writing $\phi(x,\tau) = \tilde{\phi}_{0}(\tau) + 
\pi x J/L + \pi \tau Q/(\hbar \beta) + \tilde{\phi}(x,\tau)$,
where $(-1)^J = (-1)^Q= 1$, and 
$\tilde{\phi}_{0}(\tau + \hbar \beta) = \tilde{\phi}_{0}(\tau)$ is the spatially homogeneous part of
$\phi(x,\tau)$, whereas 
\begin{equation}\label{appa17}
\tilde{\phi}(x,\tau) =  \frac{1}{\hbar \beta L} \sum_{\omega, q\neq 0} e^{i(q x - w \tau)}\: \tilde{\phi}(q,\omega) 
\end{equation}
with $q = 2\pi m/L$ and $\omega = 2\pi n/(\hbar \beta)$, and $m \neq 0,n$  integers. At this point it must be stressed
that these Fourier expansions must be cut-off at a momentum $q \sim a^{-1}$. Effectively, 
this can achieved by restricting ourselves to temperatures much lower than $\hbar v_s/a \lesssim \mu$. 

 Using the above decomposition,  the partition function can be written as the product:
\begin{equation}\label{appa18}
Z = Z_{0} Z_Q Z_J \tilde{Z}.
\end{equation}
where 
\begin{eqnarray}
\label{appa20}
Z_{0} &=&  \int_{0}^{2\pi} d\tilde{\phi}_{0}(0) \: \int_{\tilde{\phi}_{0}(0) = \tilde{\phi}_{0}(\hbar \beta)}
 \left[d\tilde{\phi}_{0} \right] \,\, \exp \left\{-\frac{K L }{2\pi v_s} \int_{0}^{\hbar \beta} d\tau\:
 \left( \frac{ d \tilde{\phi}_{0}}{d\tau} \right)^2 \right\}, \\
\label{appa21}
Z_Q &=& \sum_{{\rm even}\: Q} \exp\left\{ - \frac{\pi  K L Q^2}{2 \beta \hbar v_s} \right\}\, e^{i\pi N_{0} Q}, \\
\label{appa22}
Z_J &=& \sum_{{\rm even } J} \exp \left\{-\beta \frac{ \hbar \pi  v_s K J^2 }{2 L} \right\}, \\
\label{app23}
\tilde{Z} &=& \int  [d \tilde{\phi}] 
\: \exp \left\{{\frac{K}{2\pi v_s L \hbar \beta} \sum_{q \neq 0, \omega} 
\left[ \omega^2 + (v_s q)^2 \right] |\tilde{\phi}(q,\omega|^2 } \right\}.
\end{eqnarray}
Notice that $Z_{0}$ is formally equal to the imaginary-time 
propagator of a particle of mass $M_{0} =  K L/(\pi v_s)$ 
moving  on a line, subjected to the boundary condition that after a time 
$\tau  = \hbar \beta$ it should return to the starting point: $\tilde{\phi}_{\rm}(\hbar \beta) =
\tilde{\phi}_{0}(0)$.  From this observation and the 
form of the imaginary-time propagator  in one dimension: $G(\tau) = 
\sqrt{M_{0}/2\pi \tau}  \exp\{-M_{0} ( \tilde{\phi}(\tau) - \tilde{\phi}_{0}(0))^2/2\tau\}$, 
we obtain that
\begin{equation}
Z_{0} =   \int_{0}^{2\pi} d\tilde{\phi}_{0}(0)\:  G(\hbar \beta) = \sqrt{\frac{2 K L}{ \hbar \beta v_s}}.
\end{equation}
The integral over the initial position $0 <\tilde{\phi}_{0}(0) \leq 2\pi$ stems 
from the fact that the partition function is a trace.

 Next consider the other terms of the product. 
Using Poisson summation formula we can rewrite $Z_Q$ as follows (we 
set $Q = 2q$): 
\begin{eqnarray}
Z_Q &=&   \sum_{q=-\infty}^{+\infty} \exp\left\{ - \frac{2 \pi  K L q^2}{\beta \hbar v_s} \right\}\, e^{2 i\pi N_{0} q},
=  \sum_{N = -\infty}^{+\infty} \int_{-\infty}^{+\infty} dz \: \exp
\left\{- \frac{\pi  L z^2}{2 \beta \hbar v_s K } + 2 \pi i (N-N_{0}) z \right\}  \\
&=& \sqrt{\frac{\hbar \beta  v_s}{2 K L}} \sum_{N = -\infty}^{+\infty} \: 
\exp\left\{-\beta \frac{\hbar \pi v_s}{2 K L} (N-N_{0})^2\right\},
\end{eqnarray}
Thus we can define
\begin{equation}
Z_{N,J} = Z_{0} Z_N Z_J =  \sum_{N=-\infty}^{+\infty} \exp\left\{-\beta \frac{\pi \hbar v_s K}{2 L} (N-N_{0})^2\right\}
\sum_{{\rm even}\,\, J}  \exp\left\{-\beta \frac{\pi \hbar v_s K}{2 L} J^2\right\}.
\end{equation}
The fact that $N$  runs over all integers follows from the requirement that  $Q$ is even, which is related to the
bosonic statistics of the particles, as it has been discussed above.

  Finally, we  recognize in $\tilde{Z}$  the partition function of a system of non-interacting
bosons~\cite{Negele} with linear dispersion $\omega(q) = \hbar v_s |q| > 0$,
\begin{equation}
\tilde{Z} = \prod_{q \neq 0} \left( 1 - e^{-\beta \hbar v_s |q|} \right)^{-1}.
\end{equation}
Hence, 
\begin{equation}
Z = \sum_{N=-\infty}^{+\infty} \exp\left\{-\beta \frac{\pi \hbar v_s K}{2 L} (N-N_{0})^2\right\}
\sum_{{\rm even}\,\, J}  \exp\left\{-\beta \frac{\pi \hbar v_s K}{2 L} J^2\right\}
 \prod_{q \neq 0} \left( 1 - e^{-\beta \hbar v_s |q|} \right)^{-1}.
\end{equation}
This result is the same that can be obtained from the Hamiltonian 
in Eq.~(\ref{modhampbc}). This shows that the path integral and
Hamiltonian formulations describe the same spectrum with the same degeneracies.

\section{Finite size/temperature correlation functions}\label{appc}

 In this appendix we shall compute two important correlations functions,
from which any two-point correlation function can be derived. Before we do it
we will introduce a lot of new technology. These methods are
related to a symmetry of the phonon model defined by $H_{\rm eff}$ 
known as {\it conformal} invariance, which has deep consequences.
This is a too vast subject to be covered here, and we refer the interested
reader to the literature~\cite{cft,Cardy87,GNT98}. However, in the discussion below we shall 
try to be as self-contained as possible. The introduction of this
technology  will be lengthy at the
beginning, but it really makes life much easier when complicated correlation
functions with different boundary conditions need to be computed. 
This means that part of the results of this appendix will be also used
in the following one. Here we begin by considering 
a system with periodic boundary conditions.
Let us first define the {\it vertex operators}:
\begin{equation}
A_{m,n}(x, \tau)  =  e^{i m \theta(x,\tau)} e^{i n \phi(x,\tau)}. 
\end{equation}
One of the correlation functions in which we are
interested is the following
\begin{equation}
\langle A_{m,n}(x,\tau) A_{-m,-n}(x,\tau) \rangle_{\rm pbc}.
\end{equation}
The other one is 
\begin{equation}
\langle \partial_x\theta(x,\tau) \partial_{x'} \theta(x',\tau') \rangle_{\rm pbc}.
\end{equation}
We have explicitly written that these correlation functions obey PBC's to
emphasize this aspect with respect to other correlation functions that will
show up below. In the above expressions $\tau$ stands for the 
imaginary time (see below).

Before we proceed any further, it is useful to introduce a new set of fields,
$\phi_L(x)$ and $\phi_L(x)$, implicitly defined by the following expressions:
\begin{eqnarray}
\phi(x) &=& \frac{1}{2\sqrt{K}} \left[\phi_R(x) - 
\phi_L(x) \right] \label{phi},\\
\theta(x) &=& \frac{\sqrt{K}}{2}
 \left[ \phi_R(x) + \phi_L(x) \right]. \label{theta}
\end{eqnarray}

	We next employ the mode expansions  presented in 
Sect.~\ref{modespbc}. It is convenient to work with operators that 
depend on the imaginary time $\tau$. To find the dependence on $\tau$
of  the fields $\phi_R$ and $\phi_L$, one  solves 
the equations of motion  for the normal modes. For instance,
\begin{equation}
\hbar \frac{d}{d\tau} b(q,\tau) = \left[H_{\rm eff}, b(q,\tau) \right] = -\hbar v_s |q| b(q,\tau),
\end{equation}
which yields $b(q,\tau) = b(q) e^{-v_s|q| \tau}$, etc.  When this is also done 
for $N, J$ as well as for $\phi_0$ and $\theta_0$, one finds
\begin{eqnarray}
\phi_L(x,\tau) &=& -\phi_{0L} - \frac{2\pi i}{L}  N_L (v_s \tau + i x)
+ \sum_{q > 0} \left(\frac{2\pi}{qL} \right)^{1\over 2} e^{+a q/2} 
\left[ b(-q) e^{-q(v_s \tau + i x)} +  b^{\dagger}(-q) e^{q(v_s \tau + i x)} \right], 
\label{hol}\\
\phi_R(x,\tau) &=&  +\phi_{0R} + \frac{2\pi i}{L} 
 N_R (v_s \tau - i x)
+ \sum_{q > 0} \left(\frac{2\pi}{qL} \right)^{1\over 2} e^{-a q/2} 
\left[ b(q) e^{-q(v_s \tau - i x)} +  b^{\dagger}(q) e^{q(v_s \tau - i x)} \right], 
\label{ahol}
\end{eqnarray}
where  $q = 2\pi m/L$ ($m = 0,\pm 1, \pm 2, \ldots$) and the {\it zero modes}:
\begin{eqnarray}
\phi_{0L} &=&     \sqrt{K} \phi_{0} - \frac{\theta_{0}}{\sqrt{K}},\\
\phi_{0R} &=& \sqrt{K} \phi_{0} +  \frac{\theta_{0}}{\sqrt{K}},\\
N_L &=& \frac{(N-N_{0})}{2\sqrt{K}} - \frac{\sqrt{K} J}{2}, \label{nl}\\
N_R &=& \frac{(N-N_{0})}{2\sqrt{K}} + \frac{\sqrt{K} J}{2} \label{nr}.\\
\end{eqnarray}
Note that $[N_L, \phi_{0L} ] = [  N_R, \phi_{0R}] = i$
but $[N_L, \phi_{0R}] = [ N_R, \phi_{0L} ] = 0$, as 
follows from $[N,\phi_{0}]  = [J, \theta_{0}] = i$. Thus, $\phi_R(x,\tau)$ and $\phi_L(x,\tau)$ commute,
which implies that the operator
\begin{equation}\label{amn}
A_{m,n}(x,\tau) = e^{i\beta(m,-n) \phi_L(x,\tau)} e^{i\beta(m,n) \phi_R(x,\tau)},
\end{equation}
where $\beta(m,n) = m\sqrt{K}/2 + n/2\sqrt{K}$. Furthermore,
\begin{equation}
H_{\rm eff} = \frac{\hbar v_s}{4\pi} \int_{0}^{L} dx \: 
\left[ (\partial_x \phi_L(x))^2 +  (\partial_x \phi_R(x))^2 \right]. \label{chiralham}
\end{equation}
Thus, the Hamiltonian splits into two independent parts. The
reason for this is that $\phi_L(x,\tau)$ and $\phi_R(x,\tau)$ represent modes propagating
in opposite directions. This can be more clearly seen 
by making an analytic continuation to real
time: $\tau \to i t$. Thus we see that $\phi_L(x,t) = \phi_L(x + v_s t)$ and therefore
describes the modes propagating to the left (hence the subindex $L$), 
whereas $\phi_R(x,t) = \phi_R(x - v_s t)$ represents the modes propagating to the right (hence the subindex $R$).
This  property  is called {\it chirality}.

To make contact with conformal field theory it is useful to introduce complex coordinates
$w = v_s \tau + i x$ and $\barw = v_s \tau- ix$, and to denote 
$z = e^{2\pi w/L}$ and $\barz = e^{2\pi\barw/L}$. Thus,
\begin{eqnarray}
\phi_L(x,\tau) &=& \phi_L(z) = -\phi_{0L} -  i N_L \ln z
+ \sum_{m=1}^{+\infty} \frac{1}{\sqrt{m}} 
\left[z^{-m} \: b\left(-\frac{2\pi m}{L}\right)  +  z^m\: b^{\dagger}\left(-\frac{2\pi m}{L}\right) \right], \label{hol2}\\
\phi_R(x,\tau) &=&  \phi_R(\barz) = +\phi_{0R} + i
 N_R \ln \barz
+ \sum_{m=1}^{+\infty} \frac{1}{\sqrt{m}} 
\left[  \barz^{-m}\:  b\left(\frac{2\pi m}{L}\right) +   \barz^{m} \: b^{\dagger}\left(\frac{2\pi m}{L}\right) \right].
\label{ahol2}
\end{eqnarray}
We also introduce the {\it chiral} vertex operators:
\begin{eqnarray}
V_{\beta}(z) &=& \, : e^{i\beta \phi_L(z)} : \, = e^{-i\beta\phi_{0L}} e^{+\beta N_L\ln z} \exp\left[ i\beta \sum_{m=1}^{+\infty} \frac{z^m}{\sqrt{m}}\: b^{\dagger}\left(-\frac{2\pi m}{L}\right) \right]
 \exp\left[ i\beta \sum_{m=1}^{+\infty} \frac{z^{-m}}{\sqrt{m}}\: b\left(-\frac{2\pi m}{L}\right) \right],\\
\bar{V}_{\beta}(\barz) &=& \, : e^{i\beta \phi_R(\barz)} : \,
 = e^{i\beta\phi_{0R}} e^{- \beta  N_R\ln \barz} 
\exp\left[ i\beta \sum_{m=1}^{+\infty} \frac{\barz^m}{\sqrt{m}}\:
 b^{\dagger}\left(\frac{2\pi m}{L}\right) \right]
 \exp\left[ i\beta \sum_{m=1}^{+\infty} 
\frac{\barz^{-m}}{\sqrt{m}}\: b\left(\frac{2\pi m}{L}\right) \right],
\end{eqnarray}
where $:\ldots:$  means the operators are {\it normal ordered} as indicated above. However, 
the vertex operator $A_{n,m}(x,\tau)$, as given by Eq.~(\ref{amn}), is {\it not} normal ordered.
To write it in normal order form one must take into account that:
\begin{eqnarray}
\label{pref1}
e^{i\beta \phi_L(z)} &=& a^{\beta^2/2}\: \left(\frac{2\pi z}{L}\right)^{\beta^2/2}\,  V_{\beta}(z),\\ 
e^{i\beta \phi_R(\barz)} &=& a^{\beta^2/2}\: \left(\frac{2\pi \barz}{L}\right)^{\beta^2/2}\,  \bar{V}_{\beta}(\barz),
\label{pref2}
\end{eqnarray}
where $a$ is the short-distance cut-off introduced in Sect.\ref{modespbc}.
However, in what follows we shall not keep track of those factors and will simply use the replacements
$e^{i\beta \phi_L(z)} \to V_{\beta}(z)$ and $e^{i\beta\phi_R(\barz)} \to \bar{V}_{\beta}(\barz)$ in Eq.~(\ref{amn}).
As described below, these factors can be restored at a latter time by performing a conformal transformation.

The vertex operators just introduced have an  interesting property 
that makes easier the computation of $n-$point correlation functions. This property is:  
\begin{eqnarray}
\langle V_{\beta_1}(z_1) \cdots  
V_{\beta_p}(z_p) \rangle &=& \prod_{j < k} (z_j - z_k)^{\beta_j\beta_k},
\label{vertexhol}\\
\langle \bar{V}_{\beta_1}(\barz_1) \cdots 
\bar{V}_{\beta_p}(\barz_p) \rangle &=& \prod_{j < k} (\barz_j - \barz_k)^{\beta_j\beta_k}
\label{vertexahol}
\end{eqnarray}
provided that $\sum_{i=1}^{p} \beta_j = 0$ ($\langle \cdots \rangle$ stands for the expectation over
the ground state of $H_{\rm eff}$, see below). This is some times called
the {\it neutrality condition}. It can be regarded as a consequence that  
the Hamiltonian, Eq.~(\ref{chiralham}) is invariant under the 
infinitesimal shifts $\phi_L \to \phi_L + \epsilon$ 
and $\phi_R \to \phi_R + \bar{\epsilon}$. Another way of proving this 
condition is to notice that
\begin{equation}\label{nc2}
\Bigl\langle \exp \Big\{i\sum_{j=1}^{p} \Big[\beta(m_j,-n_j)\phi_{0L} +
\beta(m_j,n_j)\phi_{0R}\Big] \Big\} \Bigr \rangle
= \Bigl\langle e^{i\sum_{j=1}^{p} m_j \theta_{0}} 
e^{i\sum_{j=1}^{p} n_j \phi_{0}}
\Bigr  \rangle
\end{equation}
vanishes unless $\sum_j m_j = \sum_j n_j = 0$ because the ground 
state is an eigenstate of $J$ and $N$. At this point it is important to remark that
in the calculations that follow we assume that in the ground state $\langle J \rangle = 0$,
$\langle N \rangle = N_0$ ($N_0$ being integer, which implies that we work in the 
canonical ensemble). This is the case for bosons, but for fermions the selection rule $(-1)^J = -(-1)^N$
requires  $N_0$  to be odd. However, if $N_0$ is even, then the ground state is degenerate 
and  $\langle J\rangle =\pm 1$.  The implications of this degeneracy are discussed below.
The neutrality condition as stated in Eq.~(\ref{nc2})
implies that the {\it only} allowed vertex operators
in this  field theory are those for which $\beta = \beta(m,-n)$
for the left-moving field and $\beta = \beta(m,n)$ for the right-moving field,
$m$ and $n$ being integers. Other  choices  lead to non-integral
exponentials of $\theta_{0}$ and $\phi_{0}$, which in general
are not physical.

Next we shall prove Eq.~(\ref{vertexhol}). We shall first
do it for two vertex operators (i.e. $p=2$) and later  explain how to extend the proof to 
$p > 2$ (we consider left-moving fields only, the proof for right-moving fields is identical).
Thus,
\begin{equation}
\langle V_{\beta}(z_1) V_{-\beta}(z_2) \rangle =
\langle e^{-i\beta \phi_{0L}} e^{\beta N_L \ln z_1}
e^{i\beta \phi_{0L}} e^{-\beta  N_L \ln z_2} e^{i \beta \Phi_L^{\dagger}(z_1)} e^{i\beta \Phi_L(z_1)}
 e^{-i \beta \Phi_L^{\dagger}(z_2)} e^{-i\beta \Phi_L(z_2)} \rangle.
\end{equation}
Using  the  identity $e^{A} e^{B} = e^{[A,B]} e^{B} e^{A}$
and  $[\Phi_L(z_1), \Phi_L^{\dagger}(z_2)] = -\ln(1-z_2/z_1)$,
where $\Phi_L(z) = \sum_{m=1}^{+\infty} z^{-m} 
b\left( -2\pi m/L\right)/\sqrt{m}$, one finds that
\begin{equation}\label{twopointvertex}
\langle V_{\beta}(z_1) V_{-\beta}(z_2) \rangle = e^{-\beta^2 \ln z_1}
e^{-\beta^2 \ln (1 - z_2/z_1)} = (z_1 - z_2)^{-\beta^2}.
\end{equation}
Thus the rule to compute the expectation value of an arbitrary number of
vertex operators is to first separate the zero modes from  $\Phi_L$ and $\Phi_L^{\dagger}$.
Next, commute all the terms involving $\Phi_L^{\dagger}$ and  $\phi_{0L}$ 
to the left of the terms  involving $\Phi_L$ and $N_L$. Every time one does so with a pair of
operators $i$ and $j$,
one gets factor $\exp\left[\beta_i \beta_j \ln(z_i - z_j)\right] = (z_i - z_j)^{\beta_i\beta_j}$. 
The final expression 
is thus rendered  into normal order, that is, it  has all the operators $\Phi_L,  N_L$ 
to the right of  $\Phi_L^{\dagger}, \phi_{0L}$. The expectation
value of the normal ordered product  equals unity provided  
the neutrality condition is obeyed (and the ground state is not degenerate). 

Using the above results,  
\begin{eqnarray}
\langle A_{m,n}(z_1,\barz_1) A_{-m,-n}(z_2,\barz_2) \rangle &=&
\langle V_{\beta(m,-n)}(z_1) V_{-\beta(m,-n)} (z_2) \rangle \,
\langle \bar{V}_{\beta(m,n)}(\barz_1) \bar{V}_{-\beta(m,n)} (\barz_2) \rangle 
\nonumber \\
&=&   (z_1 - z_2)^{-\beta^2(m,-n)}\, (\barz_1 - \barz_2)^{-\beta^2(m,n)}\label{amnzbz}
\end{eqnarray}
which follows upon using Eq.~(\ref{amn}) and
equations~(\ref{vertexhol},\ref{vertexahol}) . However, this result does not yet produce the 
correct correlation function for PBC's by simply replacing
$z \to  e^{2\pi(v_s \tau + ix)/L}$ and $\barz \to e^{2\pi(v_s \tau - ix)/L}$, because of the factors thrown 
away after normal ordering $A_{mn}(x,\tau)$. The  way to recover those 
factors is to regard the replacement 
\begin{eqnarray}\label{conftrans}
z = e^{2\pi w/L} \,\,\,\,\,\,\,\,\,\, \barz = e^{2\pi\barw/L}
\end{eqnarray}
as a {\it conformal transformation}
that maps the infinite complex plane onto an infinite cylinder of circumference 
$L$. In conformal field theory~\cite{cft,Cardy87},  it is known that  such transformations lead
to multiplicative renormalization of the correlation functions of  (primary) operators. 
Thus if $O_i(z,\barz)$  is a set of (primary) operators with the following two-point correlation functions 
in the coordinates $z,\barz$:
\begin{equation}\label{dimensions}
\langle O_i(z,\barz) O_i(0,0)\rangle = \frac{1}{z^{2h_i}} \: \frac{1}{\barz^{2\bar{h}_i}},
\end{equation}
where $h_i$ and  $\bar{h}_i$ are called {\it conformal dimensions}, then 
under an arbitrary conformal transformation $z = z(w)$ and $\barz = \barz(\barw)$,
their $n-$point correlation function transforms as~\cite{cft} 
\begin{equation}\label{transflaw}
\langle O_{i_1}(w_1, \barw_1) \ldots O_{i_p}(w_p, \barw_p) \rangle =
\langle O_{i_1}(z_1(w_1), \barz_1(\barw_1)) 
\ldots O_{i_p}(z_p(w_p), \barz_p(\barw_p)) \rangle \,\,
\prod_{i=1}^{p} \left( \frac{dw}{dz}\right)^{-h_i}_{w=w_i} 
\left(\frac{d\barw}{d\barz} \right)^{-\bar{h}_i}_{\barw = \barw_i} 
\end{equation}
It is not hard the see that when the transformation is given by Eq.~(\ref{conftrans}), the
terms $\left( dw/dz \right)^{-h_j}$, etc. give us back the factors in Eqs.~(\ref{pref1},\ref{pref2}).
Therefore, by applying the above formula to the chiral vertex operators, one obtains
the following expression:
\begin{eqnarray}
\langle A_{m,n}(w_1,\barw_1) A_{-m,-n}(w_2,\barw_2) \rangle_{\rm pbc} =
c_{m,n} \left( \frac{2 \pi L^{-1} e^{\pi (w_1+w_2)/L} }{e^{2\pi w_1/L} - e^{2\pi w_2/L}}\right)^{\beta^2(m,-n)}\: 
 \left( \frac{2 \pi L^{-1} e^{\pi (\barw_1+\barw_2)/L} }{e^{2\pi \barw_1/L} 
- e^{2\pi \barw_2/L}}\right)^{\beta^2(m,n)} \nonumber \\
= c_{m,n} \left[ \frac{\pi/L}{\sinh\left(\pi (w_1-w_2)/L\right)}\right]^{{m^2 K \over 4}+{n^2\over{4K}}}
 \left[ \frac{\pi/L}{\sinh\left(\pi (\barw_1-\barw_2)/L\right)}\right]^{{m^2 K \over 4}+{n^2\over{4K}}}
\left[ \frac{\sinh\left(\pi (w_1-w_2)/L\right)}{\sinh\left(\pi (\barw_1-\barw_2)/L\right)}  \right]^{m n \over 2}. 
\end{eqnarray}
The constants $c_{mn}$  are cut-off  and model dependent. 
Hence, upon setting $w_1 = i x$ and $w_2 = i x'$, 
\begin{equation}
\langle A_{m,n}(x) A_{-m,-n}(x') \rangle_{\rm pbc} =
\tilde{c}_{m,n} 
\left[ \frac{1}{\rho_{0} d(x-x'|L)}\right]^{{m^2 K \over 2}+{n^2\over{2K}}}
e^{imn \pi {\rm sgn}(x-x')/2},
\end{equation}
where we have introduced the mean density $\rho_{0}$ to get
a dimensionless prefactor $\tilde{c}_{m,n}$; the cord function
$d(x|L) = L |\sin(\pi x/L)|/\pi$. 

	As we have pointed out above, for fermions the ground state is degenerate if $\langle N\rangle = N_0$
is even. In this case $\langle J \rangle = \pm 1$, which using~(\ref{nl}) and (\ref{nl}) implies that
$\langle N_R \rangle = - \langle  N_L \rangle = \sqrt{K}/2 \langle J \rangle$. Therefore, 
the expectation value of fully normal ordered products of vertex operators no 
longer equals unity. Instead, for a two-point correlation function one is left with the factor 
\begin{equation}
\langle  e^{\beta(m,-n)  N_L \ln\frac{z_1}{z_2}} \rangle \langle 
e^{-\beta(m,n)  N_R  \ln \frac{\barz_1}{\barz_2}} \rangle   = \exp\left[ \beta(m,-n) 
\langle N_L\rangle \ln \frac{z_1}{z_2}   - \beta(m,n) \langle N_R\rangle\ln \frac{\barz_1}{\barz_2}\right]
\end{equation}
Upon making the replacements $z_{1,2} = e^{2\pi (v_s \tau_{1,2} + i x_{1,2})/L}$ and
$\barz_{1,2} = e^{2\pi (v_s \tau_{1,2} - i x_{1,2})/L}$, one obtains the following phase factor:
\begin{equation}
F_{m,n}(x_1-x_2,\tau_1-\tau_2) = e^{-\pi \langle J \rangle \left(m v_J (\tau_1-\tau_2) - i n (x_1-x_2)\right)/L},
\end{equation}
which after analytical continuation to real time ($\tau \to it$) becomes $F_{m,n}(x,t) = 
e^{- i \pi \langle J\rangle (mv_J t - n x) )/L }$. Thus, the left-moving part of the field operator 
$\Psi_F^{\dag}(x)$, $\psi_L \sim A_{+1,-1}$ must multiplied by the factor $F_{+1,-1}(x,t) =
e^{-i \pi \langle J\rangle (v_J t + x)/L}$, whereas the right-moving part $\psi_R \sim A_{-1,-1}$  by
$F_{-1,-1}(x,t) = e^{i \pi \langle J\rangle (v_J t - x)/L}$. When one works in the grand canonical emsemble 
similar phases will appear for both fermions and bosons because $\langle N \rangle = N_0(\mu)$ is not an
integer. In both cases the phases are of ${\rm O}(1/L)$ and simply disappear as $L\to +\infty$. However,
in finite systems they must be taken into account if detailed comparison with, e.g. numerics, is 
being sought. In what follows, however, we shall assume that $\langle J \rangle = 0$ and $N_0$ is 
a positive integer (odd in the case of fermions).

 We next turn to the computation of the second correlation function
mentioned above, namely $\langle \partial_x\theta(x\tau) \partial_{x'} \theta(x',\tau')
\rangle_{\rm pbc}$.  We first notice that $\partial_x = i\left(\partial - \bar{\partial}\right)$, 
where $\partial = (v^{-1}_s \partial_\tau - i \partial_x)/2$ and  
$\barpart = (v^{-1}_s \partial_\tau + i \partial_x)/2$.  Therefore, we first
compute:
\begin{eqnarray}
 \langle \partial \phi_L(z_1) \partial \phi_L(z_2) \rangle 
&=& -\frac{1}{\left(z_1 - z_2\right)^2},\\
 \langle \barpart \phi_R(\barz_1) \barpart \phi_R(\barz_2) \rangle &=& -\frac{1}{\left(\barz_1 - \barz_2\right)^2},
\end{eqnarray}
where the derivatives are taken with respect to $z$ and $\barz$. 
Using Eq.(~\ref{theta}) we write $\theta$ in terms of $\phi_L$ and $\phi_L$. Hence, 
\begin{eqnarray}
\langle \partial \theta(z_1,\barz_1) \partial \theta(z_2,\barz_2)\rangle
&=&  \frac{K}{4}  \langle \partial \phi_L(z_1) \partial \phi_L(z_2) \rangle
=  -\frac{K}{4} \frac{1}{\left(z_1 -  z_2\right)^2}, \label{pthetazbz1} \\
\langle \barpart \theta(z_1,\barz_1) \barpart \theta(z_2,\barz_2)\rangle
&=&  \frac{K}{4}   \langle \barpart \phi_R(\barz_1) \barpart \phi_R(\barz_2)
 = -\frac{K}{4} \frac{1}{\left(\barz_1 - \barz_2\right)^2}\label{pthetazbz2}
\end{eqnarray}
Applying the conformal transformation~(\ref{conftrans}) to these 
correlation functions, one obtains: 
\begin{eqnarray}
\langle \partial \theta(w_1,\barw_1) \partial \theta(w_2,\barw_2)\rangle_{\rm pbc} = -\frac{K}{4} \left[ \frac{\pi/L}{\sinh\left( \pi (w_1 - w_2)/L\right) } \right]^2,\\
\langle \barpart \theta(w_1,\barw_1) \barpart \theta(w_2,\barw_2)\rangle_{\rm pbc} = -\frac{K}{4} \left[ \frac{\pi/L}{\sinh\left( \pi (\barw_1 - \barw_2)/L\right) } \right]^2.
\end{eqnarray}
Hence,
\begin{eqnarray}
\langle \partial_x \theta(w_1,\barw_1) \partial_x \theta(w_2,\barw_2)\rangle_{\rm pbc} &=& 
-\langle (\partial - \barpart) \theta(w_1,\barw_1) 
(\partial - \barpart) \theta(w_2,\barw_2) \rangle_{\rm pbc}  \nonumber \\
&=& - \langle \partial \theta(w_1,\barw_1) \partial \theta(w_2,\barw_2)\rangle_{\rm pbc} 
-  \langle \barpart \theta(w_1,\barw_1) \barpart \theta(w_2,\barw_2)
\rangle_{\rm pbc} \\
&=& \frac{K}{4} \left\{   \left[ \frac{\pi/L}{ \sinh\left( \pi (w_1 - w_2)/L\right) } \right]^2 +
\left[ \frac{\pi/L}{ \sinh\left( \pi (\barw_1 - \barw_2)/L\right) } \right]^2
\right\} 
\end{eqnarray}
In the static case, one sets $w_1 = i x$ and $w_2 = i x'$, and the above expression
reduces to: 
\begin{equation}
\langle \partial_x \theta(x) \partial_x \theta(x') \rangle
= -\frac{K \rho^2_{0}}{2}  \left[\frac{1}{\rho_{0}d(x-x'|L)} \right]^2
\end{equation}

    Finally, we shall describe how to obtain correlation functions at finite temperature in an infinite system.
In the path integral formulation discussed in Appendix~\ref{appa}, finite temperature means that one has to sum over configurations of the bosonic fields satisfying $\psi(x,\tau+\hbar\beta) = \psi(x,\tau)$ (for fermions the boundary conditions are anti-periodic~\cite{Negele}). In terms of the complex variables $w = v_s \tau + i x$ and 
$\barw = v_s \tau + i x$, this means that $\psi(w +  L_T, \barw + L_T) = \psi(w,\barw)$, where $L_T = \hbar v_s 
\beta = \hbar v_s/T$ is the thermal length. If we compare this with the requirement of PBC's: 
$\psi(w + iL, \barw - i L) = \psi(w,\barw)$, we see that the replacements $iL \to L_T$, for the left-moving
part, and $-i L \to L_T$, for the right moving parts, lead to the sought correlation functions at finite temperature.
Otherwise,  performing the following conformal transformation:
\begin{eqnarray}\label{conftransT}
z = e^{2i\pi w/L_T} \,\,\,\,\,\,\,\,\,\, \barz = e^{-2i\pi\barw/L_T}
\end{eqnarray}
on Eqs.~(\ref{amnzbz}) and (\ref{pthetazbz1},\ref{pthetazbz2}) leads to the same results ($w = v_s \tau + i x$ and
$\barw = v_s \tau - i x$):
\begin{eqnarray}
\langle A_{m,n}(w,\barw) A_{-m,-n}(0,0) \rangle_{T} &=& 
c_{m,n} \left[ \frac{\pi/L_T}{\sin\left(\pi w/L_T\right)}\right]^{{m^2 K \over 4}+{n^2\over{4K}}}
 \left[ \frac{\pi/L_T}{\sin\left(\pi \barw/L_T\right)}\right]^{{m^2 K \over 4}+{n^2\over{4K}}}
\left[ \frac{\sin\left(\pi w/L_T\right)}{\sin\left(\pi \barw/L_T\right)}  \right]^{m n \over 2},\\
\langle \partial_x \theta(w,\barw)\partial_x \theta(0,0) \rangle_T &=& 
\frac{K}{4} \left\{   \left[ \frac{\pi/L_T}{ \sin\left( \pi w/L_T\right) } \right]^2 +
\left[ \frac{\pi/L_T}{ \sin\left( \pi \barw/L_T\right) } \right]^2
\right\} 
\end{eqnarray}
Upon setting $w = ix$ and $\barw = -ix$, one obtains:
\begin{eqnarray}
\langle A_{m,n}(x) A_{-m,-n}(0) \rangle_{T} &=& 
\tilde{c}_{m,n} e^{imn\pi{\rm sgn}(x)/2} \left[ \frac{\pi/L_T}{\rho_0|\sinh\left(\pi x/L_T\right)|}\right]^{{m^2 K \over 2}+{n^2\over{2K}}},\\
\langle \partial_x \theta(x)\partial_x \theta(0) \rangle_T &=& 
-\frac{K\rho^2_0}{2} \left[ \frac{\pi/L_T}{ \rho_0 \sinh\left( \pi x/L_T\right) } \right]^2. 
\end{eqnarray}
Thus we see that, at finite temperature, the correlation functions fall off exponentially with distance.

\section{Correlation functions with open (Dirichlet) boundary conditions.}
\label{appd}

 In this Appendix we  heavily rely on the results of
the previous Appendix. We first make the important observation that for open boundary
conditions   the density and phase fields can be expressed in terms of a {\it single} chiral boson field,
\begin{eqnarray}
\phi(x,\tau) &=& \frac{1}{2\sqrt{K}} \left[\phi_R(x,\tau) + \phi_R(-x,\tau) \right]\\
\theta(x,\tau) &=& \theta_{B} + \frac{\sqrt{K}}{2}  
\left[ \phi_R(x,\tau) - \phi_R(-x,\tau) \right].\label{thetaobc}
\end{eqnarray}
The main difference with the chiral (right-moving) field of the previous section is that 
the field $\phi_R(x)$ must be now defined for $-L<  x \leq +L$. 
Therefore, it will be a function of  $\barz = \exp\left[2\pi (v_s \tau - ix)/ 2L \right]$. In terms of this
complex coordinate, 
\begin{equation}
\phi_R(z) = \phi_{0R} + i N_R \ln \barz + \sum_{m = 1}^{+\infty}
\frac{1}{\sqrt{m}} \left[\barz^{-m} b\left(\frac{m\pi}{L} \right) +
 \barz^{m} b^{\dagger} \left( \frac{m\pi }{L}\right) \right].  
\end{equation}
The zero mode $[N_R, \phi_{0R}] = i$ since
$N_R = (N-N_{0})/\sqrt{K}$ and $\phi_{0R} = \sqrt{K}\: \phi_{0}$.
The reason why only one chiral field is need is because with OBC's the
wave number $q$ of the excitations can be only positive.
Furthermore, the presence of boundaries makes impossible the existence of 
persistent currents, and thus only $N$ can appear in the mode expansions. 
Therefore, only one chiral field operator can be constructed
out of $\phi_0, N$ and $b(q)$ and $b^{\dagger}(q)$. Such a field, however, must be
defined for $-L < x \leq  L$, such  that $q$ is quantized as $2\pi m/2L =
m \pi /L$, as required by OBC's. We chose the  interval $0<x\leq L$ to correspond
to the actual system, and below we shall assume that $x > 0$ always.
As in the previous Appendix, we can write:
\begin{equation}
A_{m,n}(x, \tau) = e^{im\theta(x,\tau)} e^{in\phi(x,\tau)} = 
e^{im\theta_{B}} \: e^{i\beta(m,n) \phi_R(x,\tau)} e^{i\beta(-m,n)\phi_R(-x,\tau)},
\end{equation}
where $\beta(m,n) = (m \sqrt{K} + n/\sqrt{K})/2$.
Notice as well  that since  $\phi_R(-x,\tau) = \phi_R(\barz^{*})$, 
the fields $\phi$ and $\theta$ are linear  
combinations of $\phi_R(\barz)$ and $\phi_R(\barz^*)$. At this point, 
it is worth pointing out that  in the present context 
$\barz^* \neq z$ because $z$ and $\barz$ are to be
treated as independent variables, whereas here $\barz^*$
denotes the complex conjugate of $\barz$. 
As we did above, we shall replace $A_{m,n}(z)$ by its
normal-ordered form:
\begin{equation}
A_{m,n}(\barz,\barz^{*})  = \bar{V}_{\beta(m,n)}(\barz)\,  \bar{V}_{\beta(-m,n)}(\barz^*).
\end{equation}
As a consequence,
\begin{equation}
\langle A_{m_1,n_1}(\barz_1,\barz^*_1) \ldots A_{m_p, n_p}(\barz_p,\barz^*_p) \rangle 
= {\rm const.} \times \langle V_{\beta(m_1,n_1)}(\barz_1) V_{\beta(-m_1,n_1)}(\barz^*_1)
\ldots   V_{\beta(m_p,n_p)}(\barz_p) V_{\beta(-m_p,n_p)}(\barz^*_p) \rangle.
\end{equation}
That is, for OBC's any $n$-point correlation function
of the fields $\theta$ and $\phi$ becomes a $2n$-point correlation function
of the chiral field $\phi_R$. This a particular version of a more general
result in boundary conformal field theory obtained by  Cardy~\cite{Cardy87}. In the last
expression the neutrality condition still holds. This fact can be again regarded
as a consequence of the invariance of the Hamiltonian
\begin{equation}
H_{\rm eff} = \frac{\hbar v_s}{4\pi} \int^{+L}_{-L} dx  \left[ \partial_x \phi_R(x) \right]^2
\end{equation}
under the infinitesimal shift $\phi_R \to \phi_R + \bar{\epsilon}$. 
It is worth to pause to examine the consequences of the neutrality condition for
OBC's. The condition requires that
\begin{equation}
\sum_{j = 1}^{p} \left[ \beta(m_j,n_j) + \beta(-m_j,n_j) \right] = 
\frac{1}{2\sqrt{K}} \sum_{j = 1}^{p} n_j = 0.
\end{equation}
Notice that now the condition does not involve $m_1, m_2, \ldots, m_p$, 
and therefore does not forbid the existence of non-vanishing 
$A_{m,0}(\barz)$, but implies that $\langle A_{n,m}(x,\tau) \rangle_{\rm obc} = 0$, for
any $n \neq 0$. Mathematically, this is
expected since $\theta$  is not affected  by the shift $\phi_R \to \phi_R + \bar\epsilon$
or, in other words, it does not contain $\phi_{0}$.  Therefore,
\begin{equation}
\langle A_{m,0}(\barz,\barz^*) \rangle
= c_m \, \langle \bar{V}_{\beta(m,0)}(\barz) \bar{V}_{\beta(-m,0)}(\barz^*) \rangle
= c_m \, (\barz - \barz^*)^{-\frac{m^2 K}{4}}.  
\end{equation}

 Next, to obtain  correlation functions at finite size, we perform the conformal
transformation
\begin{equation}\label{opentrans}
\barz = e^{2\pi \barw/(2L)} = e^{\pi \barw/L},
\end{equation}
which maps one half-plane onto a strip of width $L$. The transformation
law for chiral (primary) operators is:
\begin{equation}\label{conformalinv}
\langle O_{i_1}(\barw_1) \ldots O_{i_p}(\barw_p) \rangle  =
\langle O_{i_1}(\barz_1(\barw_1)) \ldots O_{i_p}(\barz_p(\barw_p))
 \prod_{j=1}^{p} \left(\frac{d\barw}{d\barz} \right)^{-\bar{h}_j}_{\barw = \barw_j}, 
\end{equation}
where the operators have the two-point correlation function $\langle O_{i}(\barz) O_{i}(0) \rangle = \barz^{-2\bar{h}_i}$. Using the above expression (recalling that $\barw = v_s \tau - i x$) 
\begin{equation}
\langle A_{m,0}(\barw,\barw^*)\rangle_{\rm obc} = \langle A_{m,0}(x) \rangle_{\rm obc}  =
c_m \left[\frac{ \pi e^{\pi(\barw + \barw^*)/L}/L }{e^{\pi w/L} - e^{\pi w^*/L} } \right]^{\frac{m^2K}{4}} = \tilde{c}_m\, \left[\frac{1}{\rho_{0}d(2x|2L)}\right]^{{m^{2}K \over 4}},
\end{equation}
where  the mean density $\rho_{0}$ has been introduced to make the pre-factor $\tilde{c}_m$ dimensionless.

Next  we  compute the two-point correlation function of the vertex operators:
\begin{eqnarray}
\langle A_{m,n}(\barz_1,\barz^*_1) A_{-m',-n}(\barz_2,\barz^*_2) \rangle &=& c^{n}_{m,m'}\,
(\barz_1-\barz^*_1)^{\beta(m,n)\beta(-m,n)} (\barz_1-\barz_2)^{\beta(m,n)\beta(-m',-n)}  \nonumber \\
&& \times (\barz_1-\barz^*_2)^{\beta(m,n)\beta(m',-n)}  (\barz^*_1-\barz_2)^{\beta(-m,n)\beta(-m',-n)} \nonumber \\
&& \times  (\barz^*_1-\barz^*_2)^{\beta(-m,n)\beta(m',-n)} (\barz_2-\barz^*_2)^{\beta(-m,-n)\beta(m',-n)}\,\, .
\end{eqnarray}
After performing the conformal transformation~(\ref{opentrans}),
it takes the form 
\begin{eqnarray}\label{bigeq}
\langle A_{mn}(\barw_1) A_{-m',-n}(\barw_2) \rangle_{\rm obc} = 
\tilde{c}^{n}_{m,m'}  
 \left[ \frac{1}{\rho_{0} d(\barw_1 - \barw^*_1 | 2L)} \right]^{\frac{m^2 K}{4} - 
\frac{n^2}{4K}}  \left[   \frac{1}{\rho_{0} d(\barw_2 - \barw^*_2|2L)}
\right]^{\frac{m'^2 K}{4} - \frac{n^2}{4K}} \nonumber \\
\times  \frac{\left[ \rho_{0} d(\barw_1 - \barw^*_2|2L) 
\right]^{(mm'K - {n^2 \over K })/2}}
{\left[  \rho_{0} d(\barw_1 - \barw_2|2L)  \right]^{(mm'K + {n^2 \over K })/2}}
  \left[ \frac{\sinh[\pi (\barw_1^*-\barw_2)/2L]}
{\sinh[\pi (\barw_1-\barw^*_2)/2L]} \right]^{n(m-m')/4}
   \left[ \frac{\sinh[\pi (\barw^*_1-\barw^*_2)/2L]}
{\sinh[\pi (\barw_1-\barw_2)/2L]} \right]^{n(m+m')/4}.
\end{eqnarray}
In the main text, we set $\barw_1 = -i x$ and $\barw_2 = -i x'$ so
that 
\begin{eqnarray}
\langle A_{mn}(x) A_{-m',-n}(x') \rangle_{\rm obc} = 
\tilde{c}^{n}_{m,m'}  
 \left[ \frac{1}{\rho_{0} d(2x | 2L)} \right]
^{\frac{m^2 K}{4} - \frac{n^2}{4K}}  
\left[   \frac{1}{\rho_{0} d(x' |2L)}
\right]^{\frac{m'^2 K}{4} - \frac{n^2}{4K}} \nonumber \\
\times  \frac{\left[ \rho_{0} d(x + x'|2L) 
\right]^{(mm'K - {n^2 \over K })/2}}
{\left[ \rho_{0} d(x-x'|2L)\right]^{(mm'K + {n^2 \over K })/2}}\, \, 
e^{i n(m+m') {\rm sgn}(x-x')/4}.
\end{eqnarray}
In the above expressions we have assumed that $ 0 < x = -{\rm Im}\: \barw_1,
x' = - {\rm Im} \: \barw_2 < L$.

Finally we need compute the following correlation function
\begin{equation}
 \langle \partial_{x} \theta(x) \partial_{x'} \theta(x') 
\rangle_{\rm obc}.
\end{equation}
To this end, we first notice that  Eq.~(\ref{thetaobc}) implies
\begin{equation}
\partial_x \theta(x,\tau) = \frac{\sqrt{K}}{2}
\left[ \partial_x \phi_R(x,\tau) + \partial_x \phi_R(-x,\tau) \right] =
\frac{\sqrt{K}}{2i} \left[ \barpart \phi_R(\barw) 
+ \barpart \phi_R(\barw^*) \right].
\end{equation}
Using
\begin{equation}
\langle \barpart \phi_R(\barz_1) \barpart \phi_R(\barz_2) \rangle
 = -\frac{1}{(\barz_1 - \barz_2)^2}
\end{equation}
we obtain
\begin{eqnarray}
\frac{K}{(2i)^2} \langle \left[ \barpart \phi_R(\barz_1)  +  
\barpart \phi_R(\barz^*_1) \right] \left[ \barpart \phi_R(\barz_2) + 
\barpart \phi_R(\barz^*_2) \right] \rangle = 
\frac{K}{4} \left[ \frac{1}{(\barz_1 - \barz_2)^2} 
+ \frac{1}{(\barz^*_1 - \barz^*_2)} +  
\frac{1}{(\barz^*_1 - \barz_2)} +
\frac{1}{(\barz_1 - \barz^*_2)}\right].
\end{eqnarray}
The conformal transformation in Eq.~(\ref{opentrans})  along
with the transformation property (\ref{conformalinv}), then yield
\begin{eqnarray}
\langle \partial_{x}\theta(\barw_1,\barw^*_1)
\partial_{x'} \theta(\barw_2, \barw^*_2) \rangle_{\rm obc} &=&
\frac{K}{4} \Bigg\{ \left[\frac{\pi/2L}{\sinh(\pi(\barw_1-\barw_2)/2L) } \right]^2 +
\left[\frac{\pi/2L}{\sinh(\pi(\barw^*_1-\barw^*_2)/2L) } \right]^2 \\
&& + \left[\frac{\pi/2L}{\sinh(\pi(\barw^*_1-\barw_2)/2L) } \right]^2 
+ \left[\frac{\pi/2L}{\sinh(\pi(\barw^*_1-\barw^*_2)/2L) } \right]^2\Bigg\} 
\end{eqnarray}
Upon setting $\barw_1 = - i x$ and $\barw_2 = i x'$, we finally 
obtain
\begin{equation}
\langle \partial_x \theta(x) \partial_{x'} \theta(x') 
\rangle_{\rm obc} = -\frac{K\rho^2_{0}}{2} \left\{
 \left[\frac{1}{d(x-x'| 2L)} \right]^2 +  \left[\frac{1}{d(x+ x'| 2L)}\right]^2
\right\}
\end{equation}

\begin{thebibliography}{40}
\bibitem{notation}
In Ref.~\onlinecite{Haldane81a},
Haldane denotes $\eta = 2 K$. Note, however, that in this reference, 
the term proportional to $(\rho_0 x)^{-2}$ in
the expression for $\langle \rho(x)\rho(0)\rangle$ has the wrong sign.
\end{thebibliography}
\end{document}